\documentclass[a4paper,10pt]{book}
\usepackage[utf8]{inputenc}
\usepackage{amssymb}
\usepackage{amsmath,hyperref}
\usepackage{xcolor}
\usepackage{fancyhdr}
\usepackage[T1]{fontenc} 
\usepackage[pdftex]{graphicx}

\hypersetup{
    pagebackref = true,
    hyperfootnotes = true,
    hyperindex = true,
    linkcolor = black,
    citecolor= black,
    urlcolor = black,
    linkcolor={black!50!black},
    linkbordercolor=black,
    citebordercolor=black,
    urlbordercolor=black,
    citecolor=black,
    urlcolor=black,
}

\title{On Quantum Anomalous Effects In Electrodynamics Of The Early Universe}
\author{Petar Pavlović}

\begin{document}

\begin{titlepage}
\begin{center}

\mbox{}
\vspace{4cm}

{\linespread{1.3} \textbf{\LARGE{On Quantum Anomalous Effects}}}\\
{\linespread{1.3} \textbf{\LARGE{in Electrodynamics}}}\\
{\linespread{1.3} \textbf{\LARGE{of the Early Universe}}}
\vfill

\textbf{Dissertation\\ zur Erlangung des Doktorgrades\\
an der Fakultät für Mathematik,\\ Informmatik und Naturwissenschaften\\ 
Fachbereich Physik\\
der Universität Hamburg}\\
\vspace{3cm}

{\bf vorgelegt von\\}
\vspace{0.5 cm}
{\LARGE \textbf{Petar Pavlović}}\\
\vfill
\textbf{Hamburg\\ 2018}

\end{center}
\end{titlepage}

\newpage\null \frontmatter
\thispagestyle{plain}

\newpage 
\vspace*{8cm}






\newpage
\thispagestyle{plain}
\mbox{}
\newpage

\newpage 
\vspace*{8cm}
\pdfbookmark[0]{Dedication}{dedication} 
{ \flushright \textit{Wenn nicht mehr Zahlen und Figuren\\
Sind Schlüssel aller Kreaturen\\
Wenn die, so singen oder küssen,\\
Mehr als die Tiefgelehrten wissen,\\
Wenn sich die Welt ins freye Leben\\
Und in die Welt wird zurück begeben,\\
Wenn dann sich wieder Licht und Schatten\\
Zu ächter Klarheit werden gatten,\\
Und man in Mährchen und Gedichten\\
Erkennt die wahren Weltgeschichten,\\
Dann fliegt vor Einem geheimen Wort\\
Das ganze verkehrte Wesen fort.}\\ \vspace{0.5cm}
Novalis, Schriften (Historisch-kritische Ausgabe), Bd. 1,1960, S. 344.\\ }

\newpage
\thispagestyle{plain}
\mbox{}
\newpage

\newpage
\thispagestyle{plain} 
\mbox{}

\section*{Abstract}

This dissertation studies the quantum anomalous effects on the description of high energy electrodynamics. We argue that on the temperatures 
comparable to the electroweak scale, characteristic for the early Universe and objects like neutron stars, the  description of electromagnetic fields 
in conductive plasmas needs to be extended to include the effects of chiral anomaly. It is demonstrated that chiral effects can have a significant influence on 
the evolution of magnetic fields, tending to produce exponential amplification, creation of magnetic helicity from initially non-helical fields, and can lead to an inverse energy 
transfer. 
We further discuss the modified magnetohydrodynamic equations around the electroweak transition. The obtained solutions demonstrate that 
the asymmetry between right-handed and left-handed charged fermions of negligible mass typically grows with time when approaching the electroweak crossover from higher temperatures, 
until it undergoes a fast decrease at the transition, and then eventually gets damped at lower temperatures in the broken phase. At the same time, the 
dissipation of magnetic fields gets slower due to the chiral effects. We furthermore report some first analytical attempts in the study of chiral magnetohydrodynamic turbulence. 
Using the analysis of simplified regimes and qualitative arguments, it is shown that anomalous effects  can strongly support turbulent inverse cascade and lead to a faster growth of the correlation length, 
when compared to the evolution predicted by the non-chiral magnetohydrodynamics. Finally, the discussion of relaxation towards minimal energy states in the chiral magnetohydrodynamic turbulence 
is also presented. 

\newpage
\thispagestyle{plain} 
\mbox{}

\chapter*{List of Publications}
\thispagestyle{plain}

This thesis is based on the following publications: \vspace{1cm}

Petar Pavlović, Natacha Leite, Günter Sigl, \textit{Modified Magnetohydrodynamics Around the Electroweak Transition}, JCAP 1606 (2016) no.06, 044 \\ \vspace{0.5cm}

Petar Pavlović, Natacha Leite, Günter Sigl, \textit{Chiral Magnetohydrodynamic Turbulence}, Phys. Rev. D 96, 023504 (2017) \\ \vspace{0.5cm}

\newpage
\thispagestyle{plain} 
\mbox{}

\tableofcontents
\mainmatter
\chapter{Introduction}

Electromagnetism is certainly one of the most fascinating phenomena in Nature, manifesting in the early experiences of every person 
through the effects of lightening and magnetic attraction; at the same time we can say that no other interaction had such profound influence on the development 
of science -- from the introduction of field concept and first unification in physics, to quantum mechanics and gauge field theories -- while
the application of its principles drastically changed our societies and overall living conditions. As we will further demonstrate in this work, the consideration of electromagnetic phenomena in various systems also leads to rich interplay between electromagnetism and such different subjects as cosmology and 
turbulence theory. A particularly interesting finding is that electromagnetic phenomena can also be found in the Universe, where they play a 
significant role in many various systems and processes. On a fundamental level, it is believed that electromagnetism is 
most precisely described by quantum electrodynamics (QED) and electroweak theory. On the large scales those microscopic quantum aspects 
of electromagnetism do not manifest and it is very satisfactorily described by the classical theory of Faraday and Maxwell. 
However, in the context of high energy systems characteristic for the early Universe, quantum phenomena become relevant and they introduce 
important modifications to the classical description, potentially leading to effects which are also manifest on macroscopic scales. So we expect that there will be intermediary regimes in which the quantum effects will manifest as observable macroscopic corrections to the classical electromagnetic theory. This problem will be in the focus of our work, where we will report some of our recent findings and discuss consequences of the quantum chiral anomaly effect on the evolution 
of magnetic fields. We will further argue that the study of turbulence in such conditions needs to be generalized and described as chiral 
magnetohydrodynamic turbulence. These questions are important both for the sake of better understanding the properties of macroscopic electromagnetic 
fields in such high energy regimes, and for obtaining a more precise description of the early Universe physics. It is because of this reasons that the study of chiral magnetic effect
has received an increased attention of the researchers in the last period, and was studied in such different systems as
quark-gluon plasma and heavy-ion collisions, early Universe, neutron stars and core-collapse supernovae.  \\ \\
In the remainder of the Introduction we will first briefly review the history of electromagnetic theory. We believe this is important, since the 
conceptual foundations, as well as their development, determine and explain the notions which are used in the current research, thus opening 
the perspective for the new insights. However, at the same time they can also potentially cause limitations in the description of new phenomena if not critically approached. 
On the other hand, the history 
of every respective field is certainly the best inspiration and motivation for new research. \\ \\
In chapter \ref{quant} we briefly review the fundamentals of quantum field theory of electromagnetic and electroweak interaction and introduce 
the chiral anomaly effect. 
In the following chapter \ref{magne}, we discuss the magnetohydrodynamic description (MHD) of electromagnetism, which is a suitable approximation 
for the application to the cosmological setting. 
In \ref{turbo} we review the fundamentals of hydrodynamic and magnetohydrodynamic turbulence, which will then be applied in the later part 
of the work, where there modifications will be discussed. 
We start the discussion on the consequences of the chiral magnetic effect in chapter \ref{chir} where we discuss some general important properties of chiral MHD equations. We study the application
of chiral magnetohydrodynamics to description of magnetic fields in the early Universe in \ref{chirea}, where we also analyse the solutions 
of the modified MHD equations around the electroweak transition. 
The issue of chiral anomaly turbulence is studied in chapter \ref{chirturb}, were we present some first attempts in the analytical description of this 
problem, which became the subject of research only recently. After reviewing some recent numerical findings, we first study it using 
qualitative models, and then consider the regime of weak anomaly effects in more detail. Finally, we finish that chapter with the consideration
of minimal energy states of chiral MHD turbulence. 
 
\section{Early advances}

According to currently accepted Standard model of particles - which aims to describe the basic structure of matter - the interconnection between constituents of Nature 
is established through four fundamental interactions: electromagnetic, gravitational, strong nuclear and weak nuclear. While the last two
types of interaction were introduced only in the 20th century, gravity and electromagnetism were known as a phenomena since the beginning of humanity.
However, systematic speculation on their nature and causes without any supernatural assumptions occurred much later, as a part of the first 
philosophical speculations on Nature. Although gravity is one of the simplest experiences of everyday life, electricity and magnetism
were the first phenomena to actually be considered as an interaction - in the broadest sense of a specific mechanism through which a direct influence of one object on 
another is manifested. On the other hand, gravity - as described in Aristotle's Physics  \cite{Aristotel}- was mostly considered as a natural tendency of bodies
to be situated at the surface of the Earth according to their nature, rather than a type of interaction between the Earth and bodies 
In the West, the first speculative approaches towards the understanding of electricity and magnetism were also part of the development of early pre-Socratic Greek
philosophy. Hommage to ancient Greek thought will forever stay engraved in the names of this phenomena which are of the Greek origin - \textit{magnet} from
``magnitis lithos'' ($\mu \alpha \gamma \nu \eta \tau \eta \varsigma$ $\lambda \iota \theta o \varsigma)$, meaning a stone from the Greek region Magnesia; and \textit{electron}
coming from the Greek word for amber ($\eta \lambda \varepsilon \kappa \tau \rho \sigma \nu$). As recorded by Aristotle \cite{anima}, Thales of Miletus, in general recognized as the first
philosopher of Western civilization, seems to have claimed that magnet has a soul, which is manifested in its ability to attract iron objects. Although this concept may, on the first sight, seem like being based on the 
supernatural reasoning, to fully appreciate its contribution one needs to understand at least some basic aspects of what the concept of ``soul''
meant for early Greek philosophers. By observing natural phenomena, these early thinkers concluded that any change in the movement of bodies
happens only if bodies are influenced by some external cause. On the other hand, in the case of living beings it could be observed
that constant change happens without any apparent external cause. The term soul simply denoted the substance or mechanism, whatever be its constitution, which acts as a cause of this independent
movement specific for all living beings \footnote{``For a feature of all these theories is the supposition that the production of movement is the most characteristic feature of the soul 
and that while it is through the soul that all other things are moved the soul's movement is produced by itself. And this is based on our seeing nothing move that is not itself moved.''\cite{anima}}. In fact, most of the early philosophers, 
apart from Plato, Pythagoreans and Anaxagoras, shared the view that the soul is a material
substance. From this perspective, the position of Thales could be essentially understood as attributing the property to create movement to a magnet. Thus, any magnet would have an intrinsic property of creating motion (its ``soul''), which 
would exist even in the case it is not temporary manifested. This position is in fact conceptually  
close to the modern notion of magnetic field, introduced by Michael Faraday. In Faraday's view, which we will discuss later, magnetism is not understood as a phenomenon that happens through
action at a distance of two magnetized bodies (i.e. the magnetic force); but it is the intrinsic property of a magnet to change the space around it, which happens even when there is only one magnet present. 
Therefore, rather than a simple superstition, the position
of Thales could be understood as a way of thinking which - of course, in an embryonic form - anticipated the modern notion of the field, developed several
thousands years later. The opposite theory of the origin of magnetism was proposed by Empedocles, another well known pre-Socratic philosopher, primary known for elaborating the concept
of four elements as the basis of reality. While Thales explained magnetism as a manifestation of a more abstract principle, Empedocles understood it as result of a direct mechanical
influences. Following the account of Alexander of Aphrodisias, it seems that Empedocles attributed the attraction of iron and the magnetic stone to effluvium of particles emanating
from both \footnote{``Empedocles says that the iron is borne towards the stone by the effluvia emanating from both and because the pores of the stone are fitted to receive the effluvium of the iron. The effluvium of the 
stone then expels the air from the pores of the iron. Once the air is expelled, the iron itself is carried along by the abundant flow of the effluvium. Again, when the effluvium of the iron moves 
towards the pores of the stone, which are fitted to receive it, the iron begins to move with it.” \cite{guthrie}
}. Similar approaches, based on concrete mechanical interpretations, will remain dominant in explaining the electricity and magnetism during the following periods, until the development
and triumph of the electromagnetic field theory of Faraday and Maxwell in the 19th century. This demonstrates that conceptual
and philosophical origin of physical concepts always influences the practical way in which the research in physics is done, even more
if the research is not critically aware of it.\\ \\
The interest in electromagnetic phenomena did not exist only in ancient Greece. In parallel, first similar investigations
were also happening in China and India.
The first recorded consideration of magnetism in India
was concerned with its medical application--in terms of extracting the iron parts out of the body by the use of a magnet -- as recorded in a 
medical treatise \textit{Sushruta Samhita} \cite{sush}. This signifies that magnetism was well known a long time before that work was finished. In China, the first short mention of magnetism, 
as an attraction of iron objects to loadstones, is attributed to the work \textit{Master of Demon Valley} \cite{demon} where it was modestly used as a metaphore in one passage. The first explanations given by Chinese thinkers were in essence very similar to the already discussed
views of Thales. According to the interpretation of Wang Chong and Guo Pu, some items will attract others due to the compatible interaction between them.
Namely, it was stated the interaction will happen if their \textit{qi} is of the same nature, while the difference in their \textit{qi}
will make mutual influence impossible \cite{china}. The meaning of the notion of \textit{qi} in Chinese philosophy is not a simple issue, and it would
require a separate discussion. For our purposes, however, it can simply be described as a fundamental, all-encompassing substance
with internal dynamics, which manifests in different levels and forms in all concrete things. \footnote{For instance this view
can be clearly seen in the writings of Zhang Zai: ''The Great Void consists of Qi. Qi condenses to become the myriad things.
Things of necessity desintegrate and return to the Great Void...If Qi condenses, its visibility becomes effective and
physical form appears.``\cite{source}} Therefore, if understood from this perspective, the concept of \textit{qi} also appears close to the modern notion of
field. As can be seen, this interpretation treats electricity and magnetism
in unified fashion, assuming that both phenomena have the same origin in the dynamic properties of matter (\textit{qi}). In the West such
perspective will be presented only by the end of the 19th century, with the work of Faraday and Maxwell, as will be discussed later on in this chapter. From the discussed significance
of the notion of \textit{qi}, which stresses the ever changing nature of reality and its elements, it can be observed that the early Chinese view on magnetism is closely connected to the position 
attributed to Thales. However, the difference visible from the discussion above
is that the Chinese concept of \textit{qi} was much more theoretically developed and general, while not having any traces of anthropomorphic concepts, and also well defined
in the body of Chinese philosophy already present before the discussions of Wang Chong and Guo Po. Chinese thinkers have also made some first important steps
in the direction of measuring the strength of magnetic interactions. It was given in terms of
weight of iron pieces that loadstone was capable to support \cite{china}. Therefore, even more than for the explanation attributed to Thales,
this Chinese position is conceptually compatible with our modern views on electromagnetic field. Of course, the obvious shortcoming of these early ideas is
that they were just giving the basic conceptual framework, but not any kind of quantitative explanation - it is for instance not clear how and why
is the different nature of \textit{qi} determined, and by which material mechanisms is the mutual influence manifested. The complete and proper understanding of these questions
needed to wait until the 20th century and discovery of atomic structure of matter, as well as modern understanding of magnetic properties
of materials. 

\section{Observations and experiments}

Since it was not directly possible to empirically verify different theories explaining their origin, further development of ideas related to electricity and magnetism was
for centuries limited only to practical purposes and applications. The most important one was the discovery of Earth's magnetism and compass, which also came as a result
of early Chinese explorations. The first description of compass, and its usage to determine the Earth's magnetic pole, is attributed
to Shen Kuo, who mentioned it in his work Dream Pool Essays \cite{kuo} written in the 11th century. The simple observation that a magnet placed in a bowl of water
declines towards the direction of north and south pole, had an enormous effect on the development of civilization in general. From China this
discovery was quickly taken over by the Arabs who soon introduced it to Europe, and there compass was already in use by the twelfth century - enabling a completely
new level of orientation and safety for sailors. This development opened the way for a theoretical and experimental study, as well as the review of all known properties
of magnetism, undertaken by Petrus Peregrinus of Maricourt. In his letter, \textit{Epstola de magnete}, composed in the 13th century, 
Peregrinus investigated the magnetic lines and points of their intersection - which he called ``poles'' of the magnet, observed that repulsion
or attraction depend on the types of poles of two magnets, and that cutting of a magnet makes each piece to become a new complete magnet \cite{hisper}. Peregrinus also
proposed in his letter a new theory in which he tried to explain that magnetism is a consequence of imparting the power of magnetic poles of heavens 
to the poles of magnetized needle; while in the second part of his letter he explained principles for the construction of several types
of magnetic instruments. 
 \\ \\
In the following several hundred years there were no new conceptual discoveries related to electricity and 
magnetism, and interest in these phenomena was primarily of practical and nautical nature. However, during this period - which also corresponded
to the renaissance, an important change in scientific paradigm occurred -- the new framework of research that stressed the importance of experiments, the need for a synthesis of
speculation and experiment, and the importance of mathematical methods in explaining reality. These new assumptions, strongly rooted in the 
modern foundations of scientific theories, also enabled the new approach in investigation of electric and magnetic effects in the centuries to follow. 
This influence is clearly
visible in the important contribution of William Gilbert of Colchester, which was presented in his famous work \textit{De Magnete} \cite{magnete}, finished in 1600. Gilbert presented all
knowledge of that period related to magnetism, as well as his own research, ideas and experiments. His work is not only important
as the most complete analysis of electricity and magnetism up to that period, as a review of different experiments, and critical application of scientific
method, but also because of the new concepts which were put forth by Gilbert himself. Gilbert proposed that the Earth itself is a giant magnet, showing all
magnetic properties. He supported this thesis with various experiments, such as the downward inclination of magnetic needle and observations
made on spherical magnets. Observing the declination of magnetic needle from the northern direction, Gilbert concluded that the Earth is
not a perfect sphere, but that this declination is caused by non-homogeneous distribution of its mass. In his second book of \textit{De Magnete} Gilbert also studied electricity and compares it with magnetism, 
demonstrating that electrical interaction is not just specific for amber, but that various other materials can also easily be charged. This 
discussion was the first study published on electricity, and it also contains the first instrument for measuring the presence of electric
charge -- \textit{versorium} \cite{hisper}. In his attempts to explain the origin of electricity, Gilbert was inspired by the already mentioned Empedocle's 
theory of effluvia, but - observing that magnetic interaction penetrates thick shields and screens - he did not apply it to magnetism. Gilbert was of the opinion that the whole phenomenon of magnetism has its origin
in Earth's magnetism. Thus, according to Gilbert electricity and magnetism were of different origin - first coming from the effluvia of particles, and the second being the result
of terrestrial nature of magnetic materials.\\ \\
Gilbert's results opened the way for the intensive experimental research that followed during the next two centuries. Apart from gathering 
and organizing observational facts about electricity and magnetism (including the division between
the electric conductors and insulators), the most 
important developments were related to discovery of various electrical apparatus such as the electrostatic generator (attributed to
Otto von Guericke) and the first capacitor - Leyden jar (discovered simultaneously by Ewald Georg von Kleist and Pieter van Musschenbroek) \cite{histor}. 
Soon also followed some very important theoretical advances.
In the second half of the 18th century Benjamin Franklin demonstrated that lightening is of electric origin, and also introduced the important classification
of electricity - based on the mutual repulsion or attraction of charged bodies \cite{franklin}. He labeled these two types of electric charge as 
positive and negative -- which is the classification that is still in use today. Shortly after Franklin, Charles-Augusting de Coulomb made a very important discovery,
which for the first time made a precise quantitative description of electricity and magnetism possible. By inventing torsion balance, 
Coulomb was able to precisely measure weak interactions between charged bodies and magnetic poles, finding that 
the attraction or repulsion is proportional to the inverse square of distance between bodies. Although the same law was describing both the 
electric and magnetic interaction, Coulomb still did not speculate about their common origin \cite{coulomb}. \\ \\
Until the end of the 18th century, essentially only the electrostatic and magnetic phenomena were in the focus of the research. In the case
of static fields, electricity and magnetism appear as
different and not interconnected phenomena -- apart from some formal similarities in the interaction properties. From a fundamental point of view, physical information regarding static fields is exhausted
by the Coulomb law of interaction and the fact that magnetism always implies the existence of two poles in a single piece of material, while electric charges can appear 
as separate. Further development of knowledge regarding electricity and magnetism therefore required the analysis of dynamic fields. 
The main obstacle in this development was the fact that apart from lightening and rapid electric 
discharges - which are very difficult to manipulate experimentally, main sources of electricity were based on friction and static
electricity. This has dramatically changed after a discovery by Luigi Galvani and Alessandro 
Volta. While conducting his experiments on the influence of electric discharges on biological tissues, Galvani noticed that frog's legs 
twitch when connected to electrostatic generator. Galvani attributed his discovery to some new kind of electricity existing in organisms, the 
so called ``animal electricity'' \cite{galvani}. 
Influenced by Galvani's discovery, Volta explained this phenomenon assuming that different metals, placed on the tissue, act as a source of
the electromotive force, and muscles play the role of the conductor between them \cite{volta}. Following this understanding, Volta constructed 
the first electric battery, also called the Voltaic pile - which consisted of zinc and copper as electrodes and sulfuric acid mixed with water
between them. Volta also concluded that the nature of different metals placed in the medium between them, being characterized by different
electric potentials, creates the necessary condition for the current flow. This discovery of battery now finally opened the way for
observations with relatively stable and strong electrical currents, that could last for a long period of time. Now the interest of investigation
focused on electrodynamic effects, leading to discoveries of an large number of physical principles and laws that describe the processes in 
electric circuits, as well as the chemical effects of electrical current and termo-electricity. Of these developments the most important was a 
law establishing the relation between the electric current, resistance and electromotive force, discovered by
Georg Ohm. \\ \\
One of the most important discoveries in the history of electricity and magnetism, which will - in the time span of only few decades - finally lead to the 
establishment of a new paradigm of unified electromagnetic interaction, actually happened by chance. During one of his lectures Hans Christian \O{}rsted noticed
that electric currents can deflect a magnetic needle, which clearly demonstrated that electricity and magnetism are fundamentally connected, 
and not separate phenomena. Further investigation of magnetic effects caused by electric currents was continued by Andr\'{e}-Marie Amp\`{e}re, 
who analysed the properties of interactions between currents \cite{ampere}. That led him to the discovery of a well known law - according to which the attraction (or repulsion)
of two current-carrying conductors is proportional to their current intensities and lengths, and inversely proportional to their distance. 
Amp\`{e}re also proposed a new theory of magnetism, which was based on elementary atomic currents acting as a of cause both electric and magnetic effects. 
The huge importance of this new idea comes from the fact that it was the first developed theory which unified two previously distinct physical interactions
into a single phenomenon, guided by an established set of physical principles. Equally important is the significance of Amp\`{e}re's idea that electricity is a
fundamental phenomenon, and magnetism is just its consequence -- being the result of the interaction between charges set in motion. These new 
revolutionary concepts introduced by Amp\`{e}re are confirmed up to our current days. The asymmetry between electric and magnetic field -
the first being the fundamental one, and the second its special manifestation - determines the mathematical structure of Maxwell's equations,
used for a general description of all electromagnetic effects, and it also led to the establishment of Einstein's special theory of relativity, and therefore
also implicitly to Einstein's general theory of relativity. 

\section{Electromagnetic field theory of Faraday and Maxwell}

In the first half of the 19th century, when Michael Farady was starting his research on electricity and magnetism, the dominant paradigm was based on a purely mechanical reasoning,
which described these phenomena as a fluid, much inspired by Newton's fluid theories. This conceptual view, which could be traced back to Empedocles --as discussed previously, was based only
on the notion of charges, their movements and interactions at a distance between them. After conducting numerous electrodynamic experiments, Faraday - a self-taught experimentalist
who had only elementary knowledge of mathematics - started to develop his own radically different concepts. After several decades of active research these
revolutionary ideas were taken by Maxwell who put them in mathematical form, thus leading to the establishment of a complete theory of electromagnetism.\\ \\
Probably the central concept for Faraday were the lines of force, that could be visualized by iron dust around magnets and wires conducting current \footnote{Faraday defines the 
lines of force in the following manner: ``A line of magnetic force may be defined as that line which is described by a very small magnetic needle, when it is so moved 
in either direction correspondent to its length, that the needle is constantly a  tangent to the line of motion; or it is that line along which, if a  transverse wire be moved 
in either direction, there is no tendency to the formation of any current in the wire, whilst if moved in any other direction there is such a tendency; or it is that line which coincides 
with the direction of the magnecrystallic axis of a crystal of bismuth, which is carried in either direction along it. The direction of these lines about and amongst magnets and 
electric currents, is easily represented and understood, in a general manner, by the ordinary use of iron filings.'' \cite{faraday} }. From charged bodies and their 
actions - which were the only thing of interest of physical community of that period, Faraday turned his attention to the space surrounding them. In his view, this space was enabling the observed
configurations of the lines of force. For Faraday - and that point is actually the birth place of field theory -- the space surrounding magnets and charges is characterized by
``powers'' which were causes of action that leads to different observed electrical and magnetic effects. The lines of forces were then quantitatively measuring the intensity of 
such ``powers'' (in modern jargon - field strengths). Thus, the description of electricity and magnetism is no more given in terms of forces between isolated charged objects, but rather
from the viewpoint of the whole system and special state of the space between its components -- in terms of the field \footnote{Faraday was in his works also explicitly using the term
``field'' and he also gave a precise definition what it meant for him:"Any portion of space traversed by lines of magnetic power, may be taken as such a field, and there is probably 
no space without them. The condition of the field may vary in intensity of power from place to place, either along the lines or across them...`` \cite{faraday}}. Faraday was also very
much influenced by \O{}rsted's experiment which demonstrated that electric currents act as a source of magnetic fields. Deeply believing in the unity of electricity and magnetism,
Faraday invested many years in searching for the opposite process - the creation of electric currents from magnetic fields. After numerous failed attempts Faraday noticed that an electric current
can be induced only by varying the magnetic field, thus discovering electromagnetic induction which finally demonstrated that electricity and magnetism are just various manifestations
of the same physical reality, namely the electromagnetic field. This was a big triumph of Faraday's new concepts and the first example
of a successful unification of previously separated interactions in physics.\\ \\
At the end of his research phase of life, Faraday gave his manuscript, containing his experimental discoveries, directly to then young scientist James Clerk Maxwell, with whom he was in 
regular contact \cite{simpson}. Having a strong mathematical background, accepting Faraday's ideas on the electromagnetic field and induction and taking also into account
other discovered laws of electromagnetism - such as Gauss's and Amp\`{e}re's law - Maxwell started his theoretical investigations of electromagnetism \footnote{For a modern researcher
in theoretical physics Maxwell words, in which he explains the relationship between conceptual/physical and mathematical components of a physical theory, should still be of much interest. Even more
with the existing strong tendency of reducing physical theories to abstract mathematical formalism. Their significance also comes from the fact that they are based on the experience
of the physicist who was involved in the construction of one of the most important physical theories: ''The  first  process  therefore  in  the  effectual  study  of  the  science,  must  be  one  of 
simplification  and  reduction  of  the  results  of  previous  investigation  to a  form  in which  the  mind  can  grasp  them.    The  results  of  this  simplification 
may  take  the form  of  a  purely  mathematical  formula  or  of  a  physical  hypothesis.    In  the  first case we entirely lose sight of the phenomena to be explained; and 
though we may trace  out  the  consequences  of  given  laws,  we  can  never  obtain  more  extended views of the connections of the subject.  If, on the other hand, we adopt a 
physical hypothesis, we see the phenomena only through  a medium, and are liable to that blindness   to   facts   and   rashness   in   assumption   which   a   partial   
explanation encourages. We  must  therefore  discover  some  method  of  investigation  which allows the mind at every step to lay  hold of a  clear physical conception,  
without being  committed  to  any  theory  founded  on  the  physical  science  from  which  that conception  is  borrowed,  so  that  it  is  neither  drawn  aside  from  the  subject  in 
pursuit  of  analytical  subtleties,  nor  carried  beyond  the  truth  by  a favorite hypothesis.`` \cite{simpson} }. This will culminate in a new theory of electromagnetism,
which by its strong conceptual foundations, far-reaching consequences (in terms of both direct physical implications, as well as the influence on the subsequent development
of physics), beauty and elegance, can only be compared to Newton's foundation of mechanics. Maxwell described this theory in his own words in the following manner:''
\textit{The theory I propose may therefore be called a theory of the Electromagnetic Field, because it has to do with the space in the neighbourhood of the electric 
and magnetic bodies, and it may be called a Dynamical Theory, because it assumes that in 
that space there is matter in motion, by which the observed electromagnetic phenomena are produced. The 
electromagnetic field is that part of space which contains and surrounds bodies in electric or magnetic conditions}...\cite{maxwell}. Maxwell's theory was the first field theory -- 
an archetypal example of what is now a standard way to describe physical interactions. In the field theory the old and ill defined concept of force, which also implies a non-physical action
at a distance, is replaced by the holistic notion of the field, located in a space and characterized by a finite time of propagation. Einstein's General theory of relativity is based
on the same conceptual foundations, inherited from Faraday-Maxwell's theory, and the same hold as well for all quantum field theories. Maxwell's equations not only brought the unification of electricity and magnetism
to formal mathematical completeness -- previously experimentally confirmed by Faraday, but from them it also followed that a transformation of electric to magnetic field and vice versa propagates 
as a wave characterized by the speed equal to the speed of light. In this manner, three previously separated branches of physical phenomena - electricity, magnetism and optics - 
were brought into a synthesis. It was the first example of unification in physics, that will later be followed by Weinberg-Salam's theory of electroweak unification and further works
on potential unification of the electroweak interaction with strong interaction. The unification of all interactions, including gravity, still remains the most important open question and greatest
challenge in physics.\\ \\
In original Maxwell's version of the theory, electromagnetic field was described by a set of 20 equations, and it was Heaviside
who at the end of the 19th century showed that they can actually be mathematically reduced to only four equations. 
In the modern notation, SI units and in differential form, Maxwell's equations in the free space read;
\begin{equation}
\nabla \cdot \mathbf{E}= \frac{\rho}{\epsilon_{0}}, 
\end{equation}
\begin{equation}
\nabla \times \mathbf{E}=- \frac{\partial \mathbf{B}}{\partial t},
\label{inductionma}
\end{equation}
\begin{equation}
\nabla \cdot \mathbf{B}=0,
\label{mono}
\end{equation}
\begin{equation}
\nabla \times \mathbf{B}=\mu_{0} \mathbf{J}+ \mu_{0} \epsilon_{0}  \frac{\partial \mathbf{E}}{\partial t},
\label{zadnja}
\end{equation}
where $\mathbf{E}$ and $\mathbf{B}$ are the electric and magnetic field, $\rho$ represents the charge density and $\mathbf{J}$ is the current; while $\epsilon_{0}$
and $\mu_{0}$ are the dielectric and permeability constant of the vacuum. The first equation expresses the fact that in Maxwell's theory the charge distribution is
acting as a source of the electric field, mathematically described as the divergence of the electric field. In this respect, charge - the central
category of earlier mechanicistic theories of electricity - still has an important role in the Maxwell theory. But the central difference is
that electric fields can clearly exist in regions of space where $\rho=0$. Therefore, the electric field has independent existence,
it now represents a fundamental physical category, and not just some auxiliary concept. In other words, the electric field can not be reduced to the concept of electric charge.
Here, one can see the direct influence of ideas previously elaborated by Faraday. Mathematically speaking, this equation can be
easily reformulated to already known Coulomb's law. What is, however, significant here is that this form is more general, leads to 
a completely new physical concept of the electromagnetic field as the central category,
and enables the natural connection with remaining equations. The second equation represents the law of induction discovered by Faraday:  
if one starts with no electric field, and with a magnetic field that is changing in time then an electric field will be induced. 
We see that, similar to a distribution of charges, a time-dependent magnetic field creates an electric field. While a distribution of
charges creates the electric field emanating from a given point in space, acting like its source, a varying magnetic field
produces an electric field with rotation, similar in geometrical character to the magnetic field -- and does not produce a divergence of the induced field.
 In a similar sense, magnetic fields have no source points - there is no point in space from which it would be emerging and causing its divergence to be nonzero, as
can be seen from the third equation. This is often expressed as the statement that magnetic monopols do not exist, and the magnetic field lines
are necessarily closed curves. Finally, the last equation shows how the magnetic field is induced -- both by the electric current (i.e. the moving charges), and 
also due to the changes of the electric field. One sees that the analogy of electric and magnetic field is complete in the absence of charges. On the other hand, the fact
that only electric charges exists, and that they in their movement create magnetic fields, leads to the conclusion that the notion of electric field
is physically fundamental, while the magnetic field is only its consequence. Although in certain practical applications of Maxwell's theory
- for instance in the magnetohydrodynamic approximation - it is more suitable to mathematically treat the magnetic field as a primary quantity 
of interest, this fundamental physical fact should never be overlooked.\\ \\
It is important to understand that the existence of charges, acting as field sources, represents limitation of the field-theory program. Instead of having the full description of the complete system 
only in terms of fields and their dynamics, charges as an additional concept need to be introduced. It is also still not
clear why do certain constituents of matter have the property of being charged; moreover, they are inevitably connected to singularities of the field,
which points to the fact that the physical description is incomplete. Further development of field theories - in the framework of quantum field theory approach - 
will follow the route of overcoming this contradiction between the field and the sources of the field. However, the complete, unified and consistent 
description of matter constituents and their interactions, based only on the idea of field, still waits to be found. 

\chapter{Quantum field theory of electromagnetic interaction}\label{quant}

\section{Relativistic field equations}

The classical paradigm of deterministic and continuous description of processes in Nature was negated by the advent of quantum theory, which showed 
that on the fundamental level, the motion of matter needs to be described probabilistically and involving discontinuous aspects.   
Quantum theory based on the Schr\"{o}dinger equation, on the other hand, still preserved the duality of particles and fields: particles were described as conserved, individual and fundamental objects, with their
probability density distribution given by the solution of the Schr\"{o}dinger wave equation; while the electromagnetic interaction was viewed as a field phenomenon. However, this contradiction -
which is also not consistent with the wave/particle duality principle - will be resolved in the quantum field theory. This issue is also connected to the need for constructing
a relativistic version of the quantum theory - since it is precisely the relativistic regime where the conservation of particle number will be violated. Thus, the introduction of a relativistic
formalism overcomes the basic conceptual distinction between particles and fields, and therefore opens the way for their unified description. The first important step
in this direction was done by Paul Dirac in 1928, who found the relativistic equation describing electrons and spin 1/2 particles in general. Searching for a covariant relativistic version of
the Schrodinger equation, he found that it needs to be of the following form \cite{pes}
\begin{equation}
 (i \gamma^{\mu}\partial_{\mu} - m) \psi=0,
 \label{dirac}
\end{equation}
where $\psi$ is the space-time dependent wave function for the electron of mass $m$, and $\gamma$'s are the Dirac
matrices
\begin{equation}
\mathcal \gamma^{0}=
\left(
\begin{array}{cc}
 \mathbf{1}&0\\
0 & - \mathbf{1} \\
\end{array}\right)\; , 
\end{equation}
\begin{equation}
\mathcal \gamma^{i}=
\left(
\begin{array}{cc}
 0&\sigma_{i}\\
-\sigma_{i} & 0 \\
\end{array}\right)\;
\end{equation}
with $i=1,2,3$, and $\sigma$ being the Pauli's matrices
\begin{equation}
\mathcal \sigma_{1}=
\left(
\begin{array}{cc}
 0&1\\
1 & 0 \\
\end{array}\right)\; , 
\end{equation}
\begin{equation}
\mathcal \sigma_{2}=
\left(
\begin{array}{cc}
0&-i\\
i & 0 \\
\end{array}\right)\; , 
\end{equation}
\begin{equation}
\mathcal \sigma_{3}=
\left(
\begin{array}{cc}
 1&0\\
0 & - 1 \\
\end{array}\right)\; . 
\end{equation}
In the Lagrangian formalism, Dirac's equation results from the following Lagrangian density
\begin{equation}
L=\bar{\psi}(x)(i \gamma^{\mu} \partial_{\mu}-m)\psi(x),
\label{diral}
\end{equation}
with $\bar{\psi}(x)=\psi^{\dagger}(x) \gamma_{0}$, when required that the action
$ S= \int d^{4}xL $ is minimal. 
Solutions of Dirac's equation \eqref{dirac} are spinors of the form
\begin{equation}
\mathcal \psi=
\left(
\begin{array}{cccc}
 \psi_{1}\\
\psi_{2}\\
\psi_{3}\\
\psi_{4}
\\
\end{array}\right)\; , 
\end{equation}
By solving equation \eqref{dirac} with this ansatz, and assuming the plane-wave solution,  
it can be shown that this solution is in fact given by two types of plane waves
\begin{equation}
\psi_{1,2}=u(k)_{1,2}e^{-k \cdot x},
\end{equation}
\begin{equation}
\psi_{3,4}=v(k)_{1,2}e^{k \cdot x}.
\end{equation}
The general solution can then be written as a superposition
\begin{equation}
\psi(x)= \int \frac{d^{3}k}{(2 \pi)^3 2k^{0}}\sum_{r=1,2}[b_{r}(\mathbf{k}) u^{r}(\mathbf{k})e^{-i k \cdot x}
+d_{r}^{*}(\mathbf{k})v^{r}(\mathbf{k})e^{i k \cdot x}],
\label{diractot}
\end{equation}
with coefficients $b_{r}(\mathbf{k})$ and $d_{r}(\mathbf{k})$
\section{Quantum electrodynamics}
The classical field theory of Maxwell and Faraday was based on the dichotomy between fields -- described as continuous distributions
of field strengths in space, and sources -- described as some localized distribution of charges. The dynamics of electromagnetic field
was described by the Maxwell field equations, while the charges appeared as some exterior concept, leading to the singularities of the fields. 
The presence of a need for localized charges in the theory actually shows the limitation of the field theory program - if phenomena of physical interactions
can be described by the field concept, then only the description in terms of fields should be necessary, without any exterior additions. 
Development of the early quantum theory led to the realization that the energy of electromagnetic radiation at a given frequency is discrete
and quantized - thus, that radiation, in the classical field theory described by electromagnetic waves, also manifests localized  
properties. On the other hand, it also became apparent that the structure of matter -- classically considered as a collection of discrete and localized
objects (particles) -- shows the wave-like properties of interference and diffraction. Assuming the non-relativistic limit, the evolution
of matter constituents was then described by the Schr\"{o}dinger wave equation in terms of the wave function, which gives the probability density for
their localization. This revolutionary new conceptual framework of describing matter and radiation is expressed through wave-particle
duality: all physical phenomena can be described complementary as being of wave-like (continuous, non-localized) and "particle"-like (discrete, localized)
nature. The wave-like description emphasizes global properties of the system and propagation of radiation and matter constituents,
while the particle-like description emphasizes the local and peculiar properties of constituents, as well as the act of interaction between matter and radiation. \\ \\
The wave-particle duality opened the way for a significant step further in development of a complete electromagnetic field theory, since
now matter constituents could also be described as matter fields -- bringing both the electromagnetic field and matter constituents
to a unified and consistent theoretical description. However, this required going beyond the artifacts of description related to older
quantum mechanics - such as the conserved particle number, which were connected to Schr\"{o}dinger
equation and its non-relativistic nature. Since the uncertainty relations naturally lead to the uncertainty in energy-time fluctuations,
$\delta E \cdot \delta t \geq \hslash/(2 \pi)$ when $\delta E \sim 2mc^2$ the creation of particle-anti-particle pairs becomes important and in general
the assumption of a constant number of particles will be violated. Thus, the proper description of matter needs to proceed from the fact that
at short distances we have to deal with a multitude of matter and anti-matter constituents of unspecified number. On the other hand,
the fact that all particles of the same species are in quantum physics completely identical, signifies that the concept of particles
as a separate and individual entity needs to be abandoned. In both cases, the proper description of matter constituents is given in terms
of excitations of the field, and not as a separate fundamental concept. The formalism of quantum field theory enabled the realization
of these ideas and the overcoming of the aforementioned limitations of the early quantum theory. \\ \\
Starting from the Hamiltonian formalism of classical mechanics the equations of motions can be presented in the form of Poisson brackets
formalism. They are defined by
\begin{equation}
\{f,g\}=\sum_{i}(\frac{\partial f}{\partial q_{i}}\frac{\partial g}{\partial p_{i}}-\frac{\partial g}{\partial q_{i}}\frac{\partial f}{\partial p_{i}}),  
\end{equation}
leading to a relation between generalized coordinates, $q_{i}$, and conjugate momenta, $p_{i}$ given by
\begin{equation}
\{q_{i},p_{k}\}=\delta_{ik}
\end{equation}

The quantization, or transition from classical to quantum mechanics, can then be formally understood as promoting 
 functions of generalized coordinates and their conjugate momenta to operators defined on a Hilbert space, and replacing the Poisson 
brackets with commutation relations between these operators \cite{qft}. These commutation relations are usually written using a commutator between
two operators, say $O_{1}$ and $O_{2}$, given by $[O_{1},O_{2}]= O_{1}O_{2}-O_{2}O_{1}$. The basic steps in the construction of a quantum field theory are
formally the same. The classical field, which was described by some function of time and position, is first promoted to an operator, $\phi_{a}(\mathbf{x})$,
and then the commutation relations are imposed between this field operator and its conjugate momentum, $\pi^{a}(\mathbf{x})$. Since a field
is defined in every point in space this leads to an infinite number of degrees of freedom. Therefore, quantum mechanical
commutation relations need to be generalized to the continuous case
\begin{equation}
 [\phi_{a}(\mathbf{x}),\phi_{b}(\mathbf{y}]=[\pi^{a}(\mathbf{x}),\pi^{b}(\mathbf{y})]=0
\end{equation}
\begin{equation}
[\phi_{a}(\mathbf{x}),\pi^{b}(\mathbf{y})]=i \delta(\mathbf{x}-\mathbf{y})\delta_{b}^{a},
\end{equation}
in units where $\hslash=1$. \\ \\
In the general solution \eqref{dirac}, the expansion coefficients should be considered as operators. It can then be shown that
$b^{\dagger}_{r}(k)$ can be understood as operator leading to the creation of particles, $b_{r}(k)$ leading to the annihilation of particles, 
$d^{\dagger}_{r}(k)$ leading to the creation of antiparticles, and $d_{r}(k)$ leading to the annihilation of antiparticles. 
The relationship between these operators is then given by the anti-commutation rules, with anti-commutator given by
$[O_{1},O_{2}]_{+}=O_{1}O_{2}+O_{2}O_{1}$:
\begin{equation}
[b_{r}(k),b^{\dagger}_{r'}(k')]_{+}= [d_{r}(k),d^{\dagger}_{r'}(k')]_{+}=(2 \pi)^{3} 2 \omega_{k} \delta_{r, r'}\delta(\mathbf{k}-\mathbf{k'}),
\end{equation}
and all other combinations of commutators are vanishing. The Hamiltonian corresponding to the Lagrangian \eqref{diral} of the Dirac field 
is after the quantization given by
\begin{equation}
H=\int \frac{d^{3}k}{(2 \pi)^{3} 2 \omega_{k}} \omega_{k} \sum_{r=1}^{4}(b^{\dagger}_{r}(k)b_{r}(k)+ d^{\dagger}_{r}(k)d_{r}(k))
\end{equation}
\\ \\
The Dirac filed $\psi(x)$ is not a quantity that can be physically measured, having a similar auxiliary status as the vector potential
in the classical Maxwell's theory, $\mathbf{A}$, defined in such a way that $\mathbf{B}=\nabla \times \mathbf{A}$. Therefore, different field choices can lead to the same evolution equations and the same physical 
consequences of the theory. This fact is described as the gauge freedom of the theory. This can of course also be easily understood
in terms of symmetries - since some transformations of the field leave the theory unchanged. It can be easily checked 
that the Dirac Lagrangian \eqref{diral} remains invariant with respect to the transformation $\psi \rightarrow \psi'=e^{-iq \alpha}\psi$, with
$q$ and $\alpha$ being constants. The natural generalization of this transformation is to let $\alpha$ become a function of coordinates,
and therefore the transformation becomes local. Now, the Lagrangian is no longer invariant, and changes as 
$L \rightarrow L'=L + q\bar{\psi}(x) \gamma^{\mu}\psi \partial_{\mu}\alpha$. This result just shows a simple fact that the Lagrangian 
\eqref{diral} is not consistent with the assumption that local field transformation leaves it invariant. In fact, there is no \textit{a priori} reason why the theory should be locally gauge invariant. 
On the other hand, it is always possible to take the assumption of the local gauge invariance and then to see what change to the equations this 
introduces. It turns out that if this assumption is taken, then from this simple purely mathematical condition important terms in 
the Lagrangian arise, which have a direct physical meaning. Therefore,
it seems that the local gauge invariance is justified \textit{a posteriori}, on the base of its consequences - although it is still
unclear what would be the proper reason for the existence of it as an independent principle. Local gauge invariance can be restored
if ordinary derivatives appearing in \eqref{diral} are replaced by covariant derivatives, defined as
\begin{equation}
D_{\mu}=\partial_{\mu}+ iqA_{\mu}, 
\end{equation}
where $A_{\mu}$ is the four-vector potential which has the property to transform under gauge transformations as 
$A_{\mu}(x) \rightarrow A'_{\mu}(x)+ \partial_{\mu} \alpha(x)$. When introduced into the Dirac Lagrangian 
this change directly leads to the coupling between the matter and the radiation field, which was previously not present:
\begin{equation}
L= \bar{\psi}(x)(i \gamma^{\mu} \partial_{\mu}-m)\psi(x)- q\bar{\psi}\gamma^{\mu}\psi A_{\mu}.
\label{qed}
\end{equation}
Therefore, the introduction of a purely mathematical condition - the local gauge invariance - to a Lagrangian describing only
Dirac field, leads to its generalization with the interaction between Dirac and Maxwell field. This signifies that
the physical properties of matter fields for spin $1/2$ particles can not be properly studied if they are not taken together with 
their relation to the electromagnetic field. This is connected with the fact that an electron matter field acts as a source of 
the Maxwell field, which then influences the original electron. In the typical picture used in quantum electrodynamics, the 
electron necessary creates virtual photons around it that influence its own evolution. Therefore, essentially there is no such thing
as free particles. This modified Lagrangian now contains the contribution from the Dirac field as well as its interaction with the 
electromagnetic field. To have the full Lagrangian leading to complete equations of electrodynamics it is necessary to also add
the contribution of the free electromagnetic field. The Lagrangian then becomes:
\begin{equation}
L= \bar{\psi}(x)(i \gamma^{\mu} \partial_{\mu}-m)\psi(x)- q\bar{\psi}\gamma^{\mu}\psi A_{\mu}-\frac{1}{4}F_{\mu \nu} F^{\mu \nu}, 
\end{equation}
where $F_{\mu \nu}$ is a field strength, defined as $F_{\mu \nu}=\partial_{\mu}A_{\nu} - \partial_{\nu} A_{\mu}$
\section{Electroweak theory}
The success of quantum electrodynamics was visible both in terms of the theoretical developments it opened -- such as the discussed unified description of matter 
and interactions in terms of fields, as well as experimental confirmations of the theory -- for instance in the measurement of the Lamb shift in hydrogen and  the magnetic moment of electron \cite{lamb}. This naturally led to the question if other interactions could also be understood in a similar formalism, and 
described as gauge theories. In attempts to apply these principles to the theory of the weak nuclear interaction, it was realized by Weinberg and Salam 
\cite{weinberg, salam} that weak interaction needs to be understood in unity with the electromagnetic interaction, while extending the symmetry group which 
describes the invariance with respect to local gauge transformations. This discovery again fundamentally changed our notions of electromagnetic 
phenomena, in a very similar sense as the theory of Faraday and Maxwell did, now opening the perspective in which, fundamentally, electromagnetism 
is just a manifestation of electroweak interaction, with the apparent differences between electromagnetic and nuclear phenomena coming from the spontaneous breaking of 
symmetry between them on significantly smaller temperatures. As electric and magnetic fields appear different for static fields, that is when the 
field dynamics is not pronounced, so similarly -- unified electroweak description leads to two different interactions via the symmetry breaking 
Lagrangian term activating on lower energies. This mechanism for the spontaneous symmetry breaking -- known as the Higgs mechanism -- was invoked due to the fact that 
gauge invariant electroweak theory requires massless bosons, while the ones characterizing the weak processes needed to have significant masses because of their short range.
As we will briefly review later in this section, the process in which previously massless bosons acquire masses is precisely this mechanism of symmetry breaking, since 
weak nuclear bosons become massive, while photon remains massless -- which gives rise to two seemingly different interactions. It is believed 
that the electroweak symmetry existed in Nature on temperature scales of $100$ GeV which were reached during the radiation epoch 
of the early Universe. Therefore, an understanding of electromagnetism of the Universe in this stage necessarily needs to proceed from  
electroweak theory. \\ \\
Let us first introduce the concept of chiral states for Dirac fields and review some useful definitions regarding them. Each Dirac spinor 
has two independent degrees of freedom, as discussed in the previous section, and they can be represented in different basis. One special 
and often used representation is given in terms of so called chiral states, which can be left and right-handed. Their definition is
\begin{equation}
\psi_{L}=\frac{1- \gamma_{5}}{2} \psi ,
\end{equation}
\begin{equation}
\psi_{R}=\frac{1 + \gamma_{5}}{2} \psi ,
\end{equation}
so that $\psi= \psi_{L} + \psi_{R}$. Here the gamma matrix  $\gamma_{5}$ was used, which is defined as a product of Dirac gamma matrices 
in the following way $\gamma_{5}\equiv i \gamma_{0} \gamma_{1} \gamma_{2} \gamma_{3}.$ It can be checked that this matrix has the following 
properties: $\gamma_{5}^{2}=1$, $\gamma_{5}^{\dagger} = \gamma_{5}$ and it anti-commutes with other matrices: $[\gamma_{5}, \gamma_{\mu}]_{+}=0$. It should be stressed 
that this concept of chirality was introduced in an abstract and formal manner, and is not conceptually related to the concept of helicity --
which gives the relative orientation of spin and momentum for a matter field. In fact, those two concepts coincide only in the special limit
when the fields are massless. \\ \\ 
In order to describe the weak nuclear processes as given by the field which respects the local gauge invariance, it is necessary to take into account 
the significant difference with respect to electromagnetism: while electromagnetism does not discriminate between left-handed and 
right-handed states, it is experimentally known that weak processes involve parity violation -- that is, they prefer the processes 
involving left-handed fermions. To properly describe this and other properties of interactions between fermionic matter constituents, the 
Standard model organizes leptons and quarks in a three generation structure (first consisting of the electron, electron neutrino and u and d quarks;
the second of muon and muon neutrino, c and s quarks; and finally third consisting of the tau and tau neutrino, t and b quark), while each of the quarks moreover comes 
in three colors. Left-handed matter fields 
are then described as doublets (consisting of a neutrino and the corresponding lepton, or two quarks within the same generation), while their right-handed 
counterparts are described as singlets. If we now, motivated by QED, demand that local gauge invariance holds, this type of organization 
of matter constituents can not be reconciled with the simple symmetry that was applied in the case of electrodynamics. Indeed, the one dimensional 
unitary symmetry group, U(1), that was describing gauge symmetry in the QED case now needs to be extended to the simplest symmetry 
group which has a doublet representation, and that is SU(2). Since in the electroweak theory we also expect the electromagnetic U(1) group 
to be present, the symmetry group is given by $SU(2)_{L}\otimes U(1)$. If we now focus on the quark sector introducing the notation 
within one generation as $\psi_{1}=\begin{pmatrix}u\\d \end{pmatrix}_{L}$, $\psi_{2}=u_{R}$, $\psi_{3}=d_{R}$ and the same for the other generations, we can then
construct a free part of the Lagrangian without the mass term, inspired by the QED Lagrangian \cite{qft, pes, pich} 
\begin{equation}
L= \sum_{j=1}^{3}i \tilde{\psi}_{j}(x) \gamma^{\mu} \partial_{\mu} \psi_{j}(x) 
\end{equation}
and it can be checked that it stays invariant under global gauge transformations:
\begin{equation}
\psi_{1}'=e^{i q_{1} \beta}e^{i \frac{\sigma_{i}}{2} \alpha^{i}}\psi_{1}, 
\end{equation}
\begin{equation}
\psi_{2}=e^{i q_{2} \beta}\psi_{2}, 
\end{equation}
\begin{equation}
\psi_{3}= e^{i q_{3} \beta}\psi_{3}, 
\end{equation}
where index $i$ runs from $i=1$ to $i=3$, and the summation over repeated indices is implied. 
This is essentially similar to the gauge property of the QED Lagrangian, but with a more complicated mathematical structure, coming 
from the existence of left-handed doublets. In analogy, further demanding that gauge invariance now also holds locally, that is when $\beta$ 
and $\alpha$s become function of time and space-coordinates, requires that derivatives are replaced by covariant derivates, given by the 
following structure
\begin{equation}
D_{\mu} \psi_{1}=[\partial_{\mu} +ig \frac{\sigma_{i}}{2}W^{i}_{\mu}+i g' q_{1} B_{\mu}]\psi_{1}, 
\end{equation}
\begin{equation}
D_{\mu} \psi_{2}=[\partial_{\mu}  +i g' q_{2} B_{\mu}]\psi_{2},  
\end{equation}
\begin{equation}
D_{\mu} \psi_{3}=[\partial_{\mu}  +i g' q_{3} B_{\mu}]\psi_{3} .
\end{equation}
In QED there was only one gauge parameter that needed to be introduced to enable the local gauge invariance, and thus there was only 
one object defining the covariant derivative, $A_{\mu}$, which is understood as a gauge boson -- a quantum of the electromagnetic field. In the 
case of the Lagrangian discussed above there we have $\beta$ and $\alpha^{i}$, therefore altogether four gauge parameters. They lead to four gauge bosons: $B_{\mu}$, 
$W^{1}_{\mu}, W^{2}_{\mu}$ and $W^{3}_{\mu}$. Proceeding further as in the QED case, it is possible to define the corresponding field strengths 
\begin{equation}
 B_{\mu \nu}=\partial_{\mu}B_{\nu} - \partial_{\nu}B_{\mu},
\end{equation}
\begin{equation}
W^{i}_{\mu \nu}=\partial_{\mu}W^{i}_{\nu} - \partial_{\nu}W^{i}_{\mu} - g \epsilon^{ijk}W^{j}_{\mu} W^{k}_{\nu}, 
\end{equation}
and then the Lagrangian can be extended to take into account these kinetic terms
\begin{equation}
L= \sum_{j=1}^{3}i \tilde{\psi}_{j}(x) \gamma^{\mu} D_{\mu} \psi_{j}(x) - \frac{1}{4}B^{\mu \nu}B_{\mu \nu} - 
\frac{1}{4}W^{i}_{\mu \nu} W^{\mu \nu}_{i} . 
\end{equation}
This Lagrangian now again leads to the interaction of gauge fields with fermions. It can be shown that $W^{1}_{\mu}$ and 
$W^{2}_{\mu}$ lead to charged interactions, and are thus related to $W^{+}$ and $W^{-}$ bosons of the weak interactions. On the other hand, $W^{3}_{\mu}$ and 
$B^{\mu}$ are related to the photon, $A_{\mu}$, and the neutral weak current, $Z$; yet not directly but via Weinberg angle, $\theta_{W}$:
\begin{equation}
\begin{pmatrix} W^{3}_{\mu}\\B_{\mu} \end{pmatrix}=\begin{pmatrix}\cos \theta_{W}& \sin \theta_{W}\\-\sin \theta_{W} &\cos \theta_{W} \end{pmatrix}\begin{pmatrix} Z_{\mu}\\A_{\mu}\end{pmatrix},
\end{equation}
where the Weinberg angle is determined by the connection between couplings specific for different sectors of the electroweak theory
\begin{equation}
g \sin \theta_{W}=g'\cos \theta_{W}=e
\end{equation}
It can be noticed that the electroweak Lagrangian does not include mass terms. The reason for this is that in the presence of such terms 
gauge invariance can not be achieved, and would be broken -- therefore the electroweak theory demands bosons and fermions of vanishing masses. 
Since this is not the case in Nature, in order for the theory to aim at its modeling, an additional mechanism needs to be introduced for explaining 
such state of things. Such mechanism is described as the spontaneous symmetry breaking induced by the scalar boson proposed by Higgs 
\cite{h}, Brout and Englert \cite{be}, which generates mass terms for weak bosons. 

\section{Chiral anomaly}\label{anomaly}
An anomaly is a property of gauge field theories by which some symmetries that existed classically do not exist anymore after renormalization. 
This property can be understood as a consequence of quantum fluctuations which break classical symmetries. Since according to Noether's theorem \cite{noe1, noe2} 
every symmetry corresponds to a conserved current, the existence of quantum anomalies will lead to the divergence of currents that were classically 
conserved. The anomaly in which we will be interested in this work is the one associated with the electromagnetic field. Our primary question 
of interest will be how does the introduction of such anomaly changes the physical properties and the description of the electromagnetic field 
in systems of interest -- for instance in the early Universe, where the conditions for significant contribution of such quantum phenomena 
should be satisfied. \\ \\
The existence of a quantum anomaly can be observed in the context of quantum electrodynamics, by observing the Lagrangian density \ref{qed} \cite{adler}. 
If we now consider a chiral transformation of the following form
\begin{equation}
\psi \rightarrow e^{i \lambda \gamma_{5}}\psi, 
\end{equation}
where $\lambda$ is a constant, it can be seen that the kinetic term is invariant under this transformation, due to the commutating properties 
of $\gamma_{5}$, while the mass term is not. Thus if one would directly apply Noether's theorem it follows that the divergence of 
the chiral (axial-vector) current $j_{5}^{\mu}=\tilde{\psi} \gamma_{\mu} \gamma_{5} \psi=j^{\mu}_{L} - j^{\mu}_{R}$ corresponds to the contribution of the mass term in 
\eqref{qed}. However, when the corresponding Feynmann diagram with one axial-vector and two vector vertices is computed, while demanding that 
the vector current $j_{\mu}=\tilde{\psi} \gamma_{\mu} \psi$ stays conserved -- which is necessary for maintaining the gauge invariance -- it yields a different result. The divergence of the $j_{5}^{\mu}$ 
current is given not only by the mass term, but it also involves a residual correction term -- not involving mass, but depending on the
field strengths $F_{\mu \nu}$. This means that even in the massless limit there will be a non-conservation of the chiral current, which was 
conserved classically -- this signals the existence of a quantum anomaly. The complete divergence of chiral current in a one-loop 
correction is \cite{ad, belr, hoof} 
\begin{equation}
\partial_{\mu}j_{5}^{\mu}=2im j_{5} + \frac{e^2}{16 \pi^2} \epsilon^{\mu \nu \rho \sigma} F_{\mu \nu} F_{\rho \sigma} .
\label{divan}
\end{equation}
This anomaly is called Adler–Bell–Jackiw anomaly \cite{quantumtheory} and since 
it is related to chiral current, it is also known as chiral anomaly. Its important property is that it is not possible to add any local 
polynomial term which would compensate it, and it is not possible to modify quantum electrodynamics to eliminate the chiral anomaly without 
violating gauge invariance, renormalizability or unitarity \cite{adler}. The first application of chiral anomaly was in the problem of a pion 
decay into photons, where it was shown to lead to an increase of the calculated reaction rate, in accordance with experiments \cite{ad, belr}. 
We will further study the effect of the anomalous correction appearing in \eqref{divan} on the electrodynamics of massless charged fermions
in chapter \ref{chir}, section \ref{introchir}, and we will then apply it to the electrodynamics of the early Universe in the following 
sections.

\chapter{Magnetohydrodynamics}\label{magne}
Although the classical Maxwell equations were derived two centuries ago, and it is known that they should on the fundamental level
be replaced by the quantized description of the electromagnetic field -- given by the quantum electrodynamics and 
electroweak theory -- the study of their implications is still one of the very active and important 
research fields in physics. This is especially manifest in research connected to various astrophysical objects
and cosmological problems. This comes from the fact that many such systems of interest are characterized by the presence of magnetic 
fields -- that can be described purely in classical terms, and which significantly influence their properties and are therefore important for the proper physical description of such systems. 
Looking back on the history of magnetism,
as was shortly presented in the introduction, it must be concluded how remarkable it is that the phenomenon
speculated upon by Thales, Wang Chong and Guo Pu thousands of years ago still occupies the interest of
researchers and hides many unanswered questions. These early researchers of magnetic attraction and iron needle
declination could never have imagined
that magnetism is actually present in such enormous systems as
stars and galaxies. Moreover, it is very often assumed that
magnetic fields were present in the early universe, potentially leading to the creation of currently observed 
astrophysical magnetic fields, but also that they could influence such diverse significant processes as CMB spectrum creation, primordial 
nucleosynthesis, baryon asymmetry of the Universe and cosmological gravitational wave production \cite{vale, zucca, kunze, silk, yamazaki}. The advanced
study of electromagnetic theory is therefore a necessary condition for the understanding of different fundamental astrophysical
and cosmological processes, apart from being an important topic in its own right.
Since all of those systems are clearly macroscopic, the underlying quantum nature
of electromagnetic fields is not essential in their study, and all properties of interest in the considered systems can be fully
analysed within the framework of classical Maxwell's theory. Although suitable for a description of very broad interesting phenomena this assumption 
will most certainly not be valid in the conditions characteristic for the very early Universe, where on very high temperatures the full quantum picture of electromagnetic interactions becomes necessary. 
We can also assume that going enough back in time in the history of the Universe,
yet unknown field properties and processes, based on the expected unification of electroweak, strong nuclear and gravitational interactions, will take 
place and determine the dynamics of matter. These assumed regimes of Nature are still not in the reach of our knowledge, but physics at lower energies, 
around the electroweak scale, is already one of the very important research topics. In the context of such high-energy macroscopic systems,
the description of electromagnetism in terms of Maxwell's equations needs to be supplemented with quantum effects which become significant and lead
to macroscopic consequences -- such systems are therefore a borderline between classical and quantum realms. Such effects, coming from  
field theory considerations, will be the main topic of this work where we will study the consequences of the chiral anomaly effect on 
the evolution of magnetic fields.
\\ \\
In most of the mentioned systems of interest, also
including the early Universe, the electrical conductivity is typically very high, which causes the electric field to decay fast
and therefore to be generally insignificant. Sources of electromagnetic fields in these systems are typically given by
some moving liquids, gas or plasma-- which can be treated as globally electrically neutral. In these cases some very useful approximations
can be made, but in the same time Maxwell's equations need to be considered together with the equation describing the evolution of the moving fluid -- the Navier-Stokes equation. This approach leads to the 
magnetohydrodynamics (MHD), as the study of the electromagnetism of moving fluids. Properly speaking, the description of moving fluids
should be studied in the framework of special relativity, since it is well known that Galileo-Newtonian kinematics and dynamics do not give the
fundamentally proper description of change and movement, but are just an approximation which gives satisfactory results when the considered
velocities are small compared to the velocity of light. Therefore, a more correct description of electromagnetism of moving fluids
would be given in terms of relativistic tensorial equations involving the conservation laws for the stress-energy tensor and the electromagnetic tensor, involving
the four-velocity, rather than ordinary electromagnetic field components and vector-velocity. This description leads to the relativistic
MHD \cite{rel1, rel2, rel3, rel4}. However, in most astrophysical systems, and the same can be argued for
the phases of the early Universe, the characteristic velocities of the system are much smaller than the speed of light, which justifies
non-relativistic MHD as a viable approximation and mathematically simplifies the description of such systems. We will therefore in this
work concentrate only on the non-relativistic MHD description. It should, however, be stressed that such an assumption can not be applied
to some specific problems that involve fluids moving with relativistic velocity -- for instance in accretion flows on to
black holes and neutron stars and relativistic jets and blasts \cite{rela1, rela2, rela3, rela4}. \\  \\
Historically, the physical foundations of magnetohydrodynamics are given by Maxwell's equations and the Navier-Stokes 
equation describing the fluid flow, which were both known by the end of the 19th century. However, a detailed study of the relationship between
electromagnetism and fluid motions was not of any interest before late 1930's and 1940's -- due to the lack of any theoretical
and practical motivation before that period \cite{hismhd1, hismhd2}. Still, it could be stated that the father of electromagnetic theory, Michael Faraday,
was also the father of magnetohydrodynamics. In his experiment Faraday tried to measure the induced voltage across the river Thames, 
which he predicted on the basis of his induction law, generated by the river's relative motion with respect to Earth's magnetic field. 
This predicted voltage difference was however too small
to be measured by the instruments used by Faraday and in the following period there was no significant interest to continue the research
in a similar direction. Interest in the study of magnetohydrodynamics
started to develop in the third decade of the 20th century with the advance of astrophysical studies, which showed that magnetic fields and 
plasmas are characterizing various structures in the observed universe, from solar systems to galaxies. Since those structures
are also in general characterized by moving masses, which could at least to some accuracy be described as plasma, the study of 
magnetohydrodynamics became necessary for their description. 
It was in this context that modern magnetohydrodynamics was developed by the important works of Hannes Alfv\'{e}n, who was also the inventor of this name for the study of 
electromagnetism of conducting fluids \cite{al1,al2}.  A bit later, during the 1950's, MHD also came into the focus of plasma physicists. Today, magnetohydrodynamics is a very active field of research,
being used in such various fields as astrophysics, cosmology, plasma physics and applied science and engineering. \\ \\ 
Finally, let us here try to stress the fact that research on electromagnetism in the cosmological and astrophysical setting, also done under
the MHD approximation, is not just an example of an application of a more fundamental theory, but also a way to search for a better understanding of electromagnetism, including
the investigation of concepts beyond our current knowledge. By analysing various systems from this perspective -- which often includes other interactions
and different physical mechanisms -- we also test the limits of our understanding of electromagnetism. During this work many new concepts, modifications, 
and effects are proposed and introduced, which sometimes can challenge even fundamental notions and predictions. Therefore, this research should not be
considered as separated from the work on the theoretical foundations of electromagnetism and its role in our physical picture of the Universe. 
\\ \\
The usual basic assumptions of the magnetohydrodynamical approach consist of the fluid approximation, electrical neutrality of the plasma, and
the existence of a local relation between electric field and current density (Ohm's law) \cite{spruit}. Under these assumptions, a simplified system
of Maxwell's and the fluid equation can be consistently formulated and solved. The first assumption is actually also used to derive the 
equations of fluid mechanics. In this approach the fluid medium is mathematically treated as a continuum, with well defined physical quantities - such 
as temperature, velocity, pressure, field strengths etc. - around every point of the fluid. Since the real medium is
necessarily discontinuous on some level (at least due to its atomic structure), this assumption is of course only an approximation. However,
its usage is justified as long as the physical scales of interest, describing the fluid and the electromagnetic phenomena we are interested in,
are much larger than the scales where discontinuities become manifest. The electrical conductivity of a fluid is
in physically realistic systems mostly the result of the partial ionization of the plasma. However, the fluid is taken to be globally electrically neutral, which significantly simplifies the mathematical structure of Maxwell's equations since there are no source terms present. This condition is
very often satisfied, for instance in astrophysical settings.
The existence of Ohm's law means that regardless of which of the local dynamic processes influence the properties and evolution of the fluid,
they can all be  averaged to lead to some effective relation between local electric field and current density. As already mentioned, it is often the case
that the conductivity of typical systems of interest will be high, which further simplifies the considered equations. A specially important
idealized case is that of the infinite conductivity or vanishing resistivity. In this regime, the MHD description becomes particularly simple and elegant, with a large 
number of symmetries and conserved quantities. For that reason, special attention is usually given to the study of non-resistive, ideal MHD, and the more
realistic resistive MHD is often introduced as a deviation from this regime. We note that in this chapter and remaining parts of this work 
we will not use SI, but rather Lorentz-Heaviside units. 
\section{Ideal MHD}
Ideal MHD considers the case of a perfectly conducting fluid characterized by the presence of an electromagnetic field. In this case the electric field
has a very simple form. Since the current density is proportional to the electric field and conductivity, $\mathbf{J}= \sigma \mathbf{E}$, the only way to have a finite -- that is physically meaningful -- current 
density in the case of infinite conductivity is if the electric field is vanishing. Therefore, the electric field
in the ideal MHD needs to be vanishing in the reference frame which is at rest with respect to the fluid (that is which is comoving with the fluid).
This will, however, not be the case in systems which are moving with some relative velocity, $\mathbf{v}$, with respect to the fluid (or to say equivalently, in which appears
to the observer that the fluid is moving with some nonzero velocity). In that case, the electric field, $\mathbf{E}$, for an observer measuring it with respect to the
moving fluid, is given by the transformations of the electromagnetic tensor in the respective systems
\begin{equation}
F^{\mu' \nu'}=\frac{dx^{\mu'}}{dx^{\mu}}\frac{dx^{\nu'}}{dx^{\nu}}F^{\mu \nu},
\label{trans}
\end{equation}
where the electromagnetic tensor is in Cartesian coordinates given as
\begin{eqnarray}
F^{\mu\nu}=\left(\begin{array}{cccc}
0&-E_{x}&-E_{y}&-E_{z}\\
E_{x}&0 & -B_{z} & B_{y}\\
E_{y}&B_{z} &0 &-B_{x}\\
E_{z}& -B_{y} & B_{x} &0
\end{array}\right)\label{stress}\; ,
\end{eqnarray} 
and $x^{\mu'}$, $x^{\mu}$ are coordinates of the rest frame of the fluid and the observer. 
Using \eqref{trans} and \eqref{stress} and taking into account that the transformation between the systems is given by Lorentz transformation,
coming from a relative motion between the respective systems with velocity $\mathbf{v}$, and that the electric field 
in the rest frame of the fluid $\mathbf{E}$' is vanishing, in the non-relativistic limit, $v \ll 1$, we obtain
\begin{equation}
\mathbf{E}=- \mathbf{v} \times \mathbf{B}, 
\label{malajed}
\end{equation}
where we used the fact that Lorentz transformation between the coordinates is given by (in the direction of x-axis)
\begin{eqnarray}
\frac{dx^{\mu'}}{dx^{\mu}}=\left(\begin{array}{cccc}
\gamma&-\gamma v&0&0\\
-\gamma v&\gamma &0&0\\
0&0 &1 &0\\
0&0&0&1
\end{array}\right)\; ,
\end{eqnarray} 
with $\gamma=1/\sqrt{1-v^{2}}$. From \eqref{malajed} we therefore see, that the electric field measured by the observer is completely specified by the magnetic field and velocity. Using this equation,
together with the Faraday induction law \eqref{inductionma} leads to the evolution equation of the magnetic field in the non-relativistic ideal MHD approximation
\begin{equation}
\frac{\partial \mathbf{B}}{\partial t}=\nabla \times (\mathbf{v} \times \mathbf{B}). 
\label{idealind}
\end{equation}
While the corresponding original Maxwell equation \eqref{inductionma} was describing the induction of an electric field due to the change in
magnetic field, the new relation in the ideal MHD approximation \eqref{idealind} now describes a rather different picture of change in the magnetic field influenced
by the velocity, and its geometrical relation to the magnetic field. 
Using the equation for the electric field measured by the observer in the ideal MHD limit, \eqref{malajed}, together with Maxwell's equation including current density \eqref{zadnja}
leads to
\begin{equation}
\mathbf{J}=\frac{\partial}{\partial t}(\mathbf{v} \times \mathbf{B})+ \nabla \times \mathbf{B}
\end{equation}
When the velocities are non-relativistic, $v \ll 1$ then the displacement current -- corresponding to the term $\frac{\partial}{\partial t}(\mathbf{v} \times \mathbf{B})$ -- 
can be neglected, since it is being proportional to $v^{2}$ much smaller than the second term in the equation above. This is one of the standard MHD approximations leading to
\begin{equation}
\mathbf{J}= \nabla \times \mathbf{B} . 
\end{equation}
The same relation can also be shown to be valid in the non-ideal MHD case. From this one can deduce a direct consequence, that -- under this approximation -- currents in MHD, similar to magnetic fields in general,
do not have sources or sinks, $\nabla \cdot \mathbf{J}=0$.\\ \\
Finally, in order to have a full description of the considered system of conducting fluid in motion, an equation governing the evolution of the velocity needs to be given. It is given by the
Navier-Stokes equation from fluid mechanics \cite{granger}:
\begin{equation}
 \rho\left[\frac{\partial \textbf{v}}{\partial t} + (\textbf{v} \cdot \nabla) \textbf{v} - \nu \nabla^{2}\textbf{v}\right]= - \nabla p + (\nabla \times \mathbf{B}) \times \mathbf{B}
\label{navier}
 \end{equation}
where  $\rho$ is the matter density, $p$ pressure and $\nu$ kinematic viscosity, and the equation for the Lorentz force,
$\mathbf{F_{L}}=\mathbf{J} \times \mathbf{B}$, has also been used \footnote{Let us note that this expression for Lorentz force, 
giving the force on the electric current in the magnetic field, can be derived similarly as \eqref{malajed} considering the non-relativistic limit 
of transformation \eqref{trans}, thus calculating the force effect of the electric field measured by the observer moving with velocity $\mathbf{v}$ 
with respect to the fluid characterized by vanishing electric field in its rest frame, and magnetic field $\mathbf{B}$}. The movement of the fluid is further constrained by the continuity relation, which is in fact
a generalization of the conservation of mass to the case of a moving fluid.
\begin{equation}
\frac{\partial \rho}{\partial t}+ \nabla(\rho \cdot \textbf{v})=0,
\label{cont}
\end{equation}
Additionally, in order to relate the thermodynamic quantities of pressure, temperature and energy density it is further necessary to specify an equation of state (such as the ideal gas law
for instance). One example is the special, but often very satisfactory regime of the incompressible fluid, $\nabla \cdot \rho=0$. It is of particular interest in the study of MHD turbulence
since it simplifies the study of this complex phenomenon, and is often used in the astrophysical and cosmological contexts \cite{in1, in2, in3, in4}. \\ \\
In the ideal MHD regime there is a strong mathematical similarity between the fluid movement and the evolution of magnetic
fields. This is also manifest in the properties of magnetic field lines, which -- similar to properties of the fluid given by the continuity equation -- 
make a conserved flux. In the general problems of electrodynamics this is clearly not the case, since the magnetic flux -- which measures the local magnetic field strength, being equal 
to its integral over some surface -- will change in time. This means that the total number of magnetic 
field lines passing through some closed surface will in general constantly change. In this case field lines are not a good object to track the evolution of magnetic field - since they
change from one moment to the next.
It is the opposite in the ideal MHD case, where the conserved nature of magnetic flux makes them individually traceable and very useful in the conceptual picture of the field evolution.  
The conservation of magnetic flux in the ideal MHD case can be shown directly by using the definition of magnetic flux over some surface moving along with the fluid, $\mathbf{A}$, $\Phi= \int \mathbf{B} \cdot d\mathbf{A}$ and \eqref{idealind}
\begin{equation}
\frac{d\Phi}{dt}= \int \nabla \times (\mathbf{v} \times \mathbf{B})d\mathbf{A} + \oint (\mathbf{B} \cdot \mathbf{v}) \times d\mathbf{l}=0, 
\end{equation}
where in the last equality the Stokes theorem was used. This simple result of magnetic flux conservation in ideal MHD is also known as Alfv\'{e}n theorem. 
One of the interesting consequences of the conservation of magnetic flux is that the interconnection between the fluid flow and magnetic
field can lead to the amplification of magnetic fields. Conceptually speaking this can be understood as a result of the field flow changing
the comoving surface over which the magnetic flux is defined, which leads to the change in magnetic field in order to compensate this change and keep
the magnetic flux constant. This can be seen in a more mathematical and rigorous fashion as follows. Let us first note that the induction
equation \eqref{idealind} can be written as
\begin{equation}
\frac{\partial \mathbf{B}}{\partial t}= - \mathbf{B} \nabla \cdot \mathbf{v} -(\mathbf{v} \cdot \nabla)\mathbf{B}+ (\mathbf{B} \cdot \nabla)
\mathbf{v}
\end{equation}
Assuming the incompressible case, which by the virtue of the continuity equation implies that $\nabla \cdot \mathbf{v}=0$, and writing the
change of the magnetic field as a total derivative by using the second term on the right hand side, we obtain
\begin{equation}
\frac{d\mathbf{B}}{dt}=(\mathbf{B} \cdot \nabla)\mathbf{v}. 
\end{equation}
As long as $(\mathbf{B} \cdot \nabla)\mathbf{v}>0$ the magnetic field will grow as a consequence of suitable spatial distribution
between the magnetic field at a given moment and the velocity field. As a simple example we can consider the case of a magnetic field and 
velocity having only one vector component different from zero, say ${B}_{x}=B(t)$ and $v_{x}=f(x)$. Then it follows that
$B(t)=B(t=0)e^{f'(x) t}$, with prime denoting the derivative with respect to the argument,
and we see that the magnetic field was in this configuration exponentially amplified if $f'(x) >  0$ and exponentially
damped if $ f'(x) < 0$. \\ \\
Magnetic helicity represents an important and often used additional quantity, due to its conservation in the ideal MHD case,
which we will discuss bellow. 
If $\mathbf{A}$ is a vector potential corresponding to a magnetic field $\mathbf{B}$ then the magnetic helicity density, $h$, is defined
as
\begin{equation}
h=\frac{1}{V}\int_{V} \mathbf{A} \cdot \mathbf{B} dV
\label{heldef}
\end{equation}
A fundamental problem with this definition of helicity comes from the fact that the vector potential $\mathbf{A}$ is not an uniquely defined
quantity, but is dependent on the chosen gauge. Namely, if we make a gauge transformation $\mathbf{A} \rightarrow \mathbf{A} + \nabla \psi $,
where $\psi$ is some scalar function, then the resulting magnetic field, $\mathbf{B}=\nabla \times \mathbf{A}$, will not change. 
Since the magnetic field is connected to actual physical measurements, and $\mathbf{A}$ represents an auxiliary mathematical object, there
is no way do discriminate between all possible forms of $\mathbf{A}$ related by gauge transformations, and therefore helicity 
would not be uniquely defined either. The difference between definitions of helicity based on the vector potentials related by the transformation
$\mathbf{A} \rightarrow \mathbf{A} + \nabla \psi $ is given by
\begin{equation}
\delta h= \frac{1}{V} \int_{V} \nabla(\mathbf{B} \psi)dV 
\end{equation}
Using the Gauss theorem, and transforming this integral into a surface integral, it follows that 
$\delta h=0$ if $\mathbf{B} \cdot d\mathbf{a}=0$. Therefore, magnetic helicity can be constructed as an uniquely defined quantity in the special case in which there is no magnetic field in the direction of a surface which encloses the volume
$V$, over which the integration was defined. Since no component of the magnetic field in this case points in the direction of the boundary,
it can also be stated that the definition of helicity is gauge independent in the case where the magnetic field is completely contained
in the region of the volume $V$. On the other hand, practical difficulties with using the definition of helicity rigorously come from the 
fact that in many systems it may not be possible to find a volume such that $\mathbf{B} \cdot d\mathbf{a}=0$ holds generally. \\ \\
Assuming that it is possible to give an unique  gauge independent definition of helicity, we turn to the question of its conservation in the
case of ideal MHD. This can be demonstrated by direct computation. Starting from the definition of magnetic helicity, and using the fact
$\mathbf{E}=- \nabla \varphi + \partial \mathbf{A}/ \partial t$, with $\varphi$ being the scalar potential of electric field, time change of magnetic helicity can be written as
\begin{equation}
\frac{dh}{dt}= -\frac{2}{V} \int \mathbf{E} \cdot \mathbf{B} dV - \frac{1}{V} \int \nabla \cdot(\mathbf{A} \times \mathbf{B})dV 
\label{helcons}
\end{equation}
Using the Gauss theorem and assuming that $\mathbf{A} \times \mathbf{B}$ vanishes in the direction normal to the surface enclosing the
integration volume we are left only with the first term on the left hand side of the previous equation. However, since in the ideal MHD case,
as discussed earlier, $\mathbf{E}=- \mathbf{v} \times \mathbf{B}$, this term is also zero and helicity is conserved. \\ \\ 
Apart from being conserved in the ideal MHD approximation, another important aspect of magnetic helicity is that it describes
the global topology of magnetic field. So in the case of ideal MHD, where the magnetic helicity is a conserved quantity, the topology of
magnetic field lines will not change - they will not link or twist. This insight of helicity as a measure of global field topology actually developed
from the work of Gauss, who studied the linking of asteroid's orbit to Earth's orbit by means of introducing a quantitative measure of their linking. 
Later, Gauss applied his definition to electromagnetic field lines, inspired by the work on electromagnetic induction by Michael Faraday, and this is how
the topological concept of helicity was introduced in the electromagnetic theory. The Gauss linking number is defines as follows \cite{helicitygau}. Let the first curve be 
parametrized with $\sigma$ and the points on that curve given by $\mathbf{x}(\sigma)$ and let the second curve be parametrized by $\tau$ and its points given by
$\mathbf{y}(\tau)$, and we also define their difference as $\mathbf{r}= \mathbf{y} - \mathbf{x}$. Then the Gauss linking number is
\begin{equation}
L_{1,2}=-\frac{1}{4 \pi}\oint \oint \frac{d \mathbf{x}}{d \sigma} \cdot \frac{\mathbf{r}}{r^{3}} \times \frac{d \mathbf{y}}{d \tau} d \tau d \sigma .
\label{guslink}
\end{equation}
It can easily be shown that this general mathematical definition, giving the measure of linking between arbitrary curves, can be applied to the case
of magnetic field lines, yielding equation \eqref{heldef}. In this respect, the Gauss linking number summed over every pair of magnetic field lines within a volume, gives the
corresponding magnetic helicity.
If there would be a discrete number of field lines, the magnetic helicity could then be calculated as
\begin{equation}
h=\frac{1}{V} \sum_{i,j} L_{i,j} \phi_{i} \phi_{j}, 
\label{sumhel}
\end{equation}
where the sum goes over all possible combinations of field lines. 
Assuming that the magnetic field is located in the closed volume, that is $\mathbf{B} \cdot \mathbf{n}=0$ (where $\mathbf{n}$ is a normal unit vector
at the boundary), remembering the definition of magnetic flux, and using this idea we have
\begin{equation}
h= - \frac{1}{V} \frac{1}{4 \pi} \int \int \mathbf{B(\mathbf{x})} \cdot \frac{\mathbf{r}}{r^{3}} \times \mathbf{B(\mathbf{y})}d^{3}x d^{3}y ,
\label{linking}
\end{equation}
where the integral goes over two volume elements to take into account the contribution of all the field lines. Using the standard expression for the vector potential
\begin{equation}
\mathbf{A(x)} =- \frac{1}{4 \pi} \int \frac{\mathbf{r}}{r^{3}}\times \mathbf{B(y)} d^{3}y,
\end{equation}
together with \eqref{linking} we finally obtain the familiar expression for magnetic helicity density
\begin{equation}
h= \frac{1}{V} \int \mathbf{A} \cdot \mathbf{B} d^{3}x.
\end{equation}
We can then clearly see how the magnetic helicity is thus used as a measure of field helicity if we consider a simplified case of helicity associated to a field
configuration corresponding to only two field lines. 
In the case in which two field lines do not link, the associated linking number \ref{guslink} is just zero, and the corresponding configuration has a zero helicity. In the opposite case,
it is clear from \eqref{sumhel} that their helicity is simply given by $h=(2/V) \phi_{1} \phi_{2}$. 

\section{Resistive MHD}
We now turn to the more realistic case, in which the electric field in the fluid frame, $\mathbf{E}'$, is not vanishing and is related to the electric
current density. The usual assumption in this case is that the relation between the 
current density and electric field is linear (given by the Ohm's law)
\begin{equation}
\mathbf{j}'=\sigma \mathbf{E}',
\end{equation}
where $\sigma$ is the electric conductivity. In the non-relativistic limit this relation leads to the equation for the current
measured in the reference frame of the observer, with respect to whom the fluid moves with velocity $\mathbf{v}$
\begin{equation}
\mathbf{j}= \sigma (\mathbf{E}+ \mathbf{v} \times \mathbf{B}), 
\end{equation}
and using this expression, the induction equation in the case of ideal MHD \eqref{idealind} is now replaced by
\begin{equation}
\frac{\partial \mathbf{B}}{\partial t}= \nabla \times (\mathbf{v} \times \mathbf{B} - \eta \nabla \times \mathbf{B}), 
\end{equation}
where we have introduced the magnetic diffusivity given by $\eta=1/(4 \pi \sigma)$. 
It can be easily checked that now Alfv\'{e}n's theorem is no longer valid and the magnetic flux is not conserved. In fact,
it is given by
\begin{equation}
\frac{d \Phi}{dt}=-\frac{1}{4 \pi \sigma} \oint_{l} \bold{j} \cdot d\bold{l} . 
\end{equation}
In a similar manner, in resistive MHD helicity is also no longer a conserved quantity. Using the induction equation
in the resistive case it follows
\begin{equation}
 \frac{dh}{dt}=-\frac{2}{V} \int d^3{x} \frac{1}{4 \pi \sigma}(\nabla \times \textbf{B})\cdot  \textbf{B}
\end{equation}
Therefore, some very useful concepts that simplify MHD problems considerably in the ideal case -- such as the individuality of the field lines which 
can be followed during the field evolution, and the invariance of the field topology described by helicity -- are no longer valid 
when the effects of finite conductivity are taken into account. However, in many practical applications in cosmology and astrophysics, the conductivity
is in general very high so that the effects of change of magnetic helicity and flux can be ignored. 

\chapter{Electromagnetic fields and turbulence}\label{turbo}
The subject of turbulence represents a difficulty not only in its physical description, but even in its definition -- since no universally 
accepted definition exists, and most of the attempts simply represent a list of its properties, rather than a developed characterization.
For the need of our discussion we will (informally) define it as a highly-irregular motion of fluids, which can not be properly described
in terms of deterministic functions. Since the movement of a fluid is described by the Navier-Stokes equation, it follows that turbulent 
velocity configurations appear as its solutions. In fact, although Navier-Stokes equation \eqref{navier} represents a deterministic differential equation, due to its non-linear character it leads
to solutions which are highly irregular. This mathematically describes the well known phenomenon of turbulence in Nature, which in the same time --despite
its chaotic nature -- manifests structure and some aspects of deterministic global properties, as will be discussed later. Although 
turbulent phenomena are experimentally observed for thousands of years, and the basic laws of fluid dynamics that lead to turbulent behavior
are known for several hundred years, it is still one of the unsolved problems in physics. The chaotic nature of turbulence, and mathematical
complexity of its description, still make the question of finding a complete theoretical model, which could describe the evolution of 
turbulence and its universal properties, to remain unanswered. \\ \\
In the previous section we have discussed how a complete description of the evolution of electromagnetic fields related to 
conductive fluids also needs to take into account the evolution of the fluid's velocity and its interconnection with the electromagnetic field.
This interconnection happens due to the dependence of the field measured by the observer on the respective movements of different sections of the fluid, and on the other hand due to the presence of Lorentz force,
influencing the time evolution of velocity. In the case in which the fluid flow becomes turbulent, which often happens in practice, turbulence
will therefore also influence the spatial distribution and time evolution of electromagnetic field. This leads to the conclusion that the proper
description of electromagnetic fields in fluids is inseparable from the effects of turbulence. In fact, most of the astrophysical 
systems are characterized by the existence of magnetic fields and turbulence -- for instance galactic plasma, solar wind, accretion discs
etc., while in the cosmological setting turbulence related to magnetic field evolution could have played an important role
during the phase transitions in the early Universe. \\ \\
Unfortunately, since there is still no complete theory of turbulence  even in the case of hydrodynamic flows, the problem becomes even greater in the more complex setting in which the electromagnetic field is also present, 
and coupled to the velocity of the fluid. In this situation the only possible strategy is to use existing working hypothesis, approximations
and effective models, together with properties of solutions deduced from observations and numerical simulations. In the same sense in 
which the electrodynamics of conductive fluids is described by the MHD theory, the same set of assumptions applied to the case of electromagnetic
fields appearing in turbulent fluids will lead to the theory of MHD turbulence. 
\section{General properties of turbulence in fluids}
Turbulence, as a chaotic movement of the fluid -- as opposed to its laminar motion -- was surely observed from the beginning of civilizations, but the first systematic description of this phenomenon dates to the
16th century and the work of Leonardo da Vinci. In his sketch book Leonardo da Vinci illustrated some examples of turbulent flow patterns,
and also gave a description of its observed properties, which can be considered as very precise, surprisingly modern in its character, and anticipating the concept of transfer of energy
from larger to smaller turbulent structures \footnote{``the smallest eddies are almost numberless, and large
things are rotated only by large eddies and not by small ones,
and small things are turned by small eddies and large.”, \cite{curdy}}. It was Leonardo who gave the name to this phenomenon,
using the word ''turbulenza'', meaning the gathering of people. \\ \\
The modern foundation of theory of turbulence dates back to the mid 19th century and the contribution of Navier and Stokes, who --
starting from the principles of Newtonian physics -- derived the equation believed to describe all the fluid flows given by \cite{landau}
\begin{equation}
 \rho\left[\frac{\partial \textbf{v}}{\partial t} + (\textbf{v} \cdot \nabla) \textbf{v} - \nu \nabla^{2}\textbf{v}\right]= - \nabla p , 
 \label{hidronav}
\end{equation}
assuming there are no additional forces acting on the fluid -- while in general they will be present in the right hand side of the equation. 
We note that in the hydrodynamic case, on which we focus in this section in order to review the properties of turbulence in the best known setting, there is no Lorentz-force term which appears the equation \eqref{navier}. 
This equation should also be supplemented with the continuity equation \eqref{cont} which ensures that the total mass of the 
considered hydrodynamic system is always conserved, and which in the case of non-compressible flows leads to the condition
\begin{equation}
\nabla \cdot \mathbf{v}=0.  
\label{divg}
\end{equation}
The pressure is not an independent quantity in the equations above, since it can be eliminated by taking the divergence of equation
\eqref{hidronav} and then using \eqref{divg}. It then follows that pressure is given by the following relation
\begin{equation}
\nabla^{2} p =- \rho \nabla \cdot[(\mathbf{v} \cdot \nabla) \mathbf{v}] . 
\end{equation}
We thus see that the physical role of pressure is essentially to maintain the continuity equation, and that it is determined
by the velocity field. It also follows that pressure is defined in each space point by the velocity distribution given over the
whole flow. This points to a non-local nature of solutions to the Navier-Stokes equation in general, in the sense that different scales of physical processes cannot be
simply decoupled from each other, but they are rather strongly and non-linearly coupled, which is also a significant reason for
the difficulties associated to their analysis \cite{conceptual}.  
The Navier-Stokes equation is in general believed to also properly describe turbulent regimes, which come from its nonlinear structure,
visible by the existence of the term $(\mathbf{v} \cdot \nabla) \mathbf{v}$. The non-linear nature of the Navier-Stokes
equation is at the same time the reason-- together with its related non-locality and non-integrability -- why it is so difficult to treat turbulence analytically. For this reason, exact analytical approach  
has enabled only very limited progress in the understanding of turbulence so far.\\ \\
The first important experimental study of turbulence was conducted by O. Reynolds by the end of the 19th century, who analysed the transition
from turbulent to laminar flows \cite{reynolds}. His important finding was that the transition of turbulence can be characterized by one non-dimensional and universal parameter (i.e not depending
on the concrete setting and on the realization of particular turbulent occurrence). This parameter is now called Reynolds number (Re) and it essentially gives a measure of the ratio
between intertial and viscous forces -- the first supporting the occurrence of turbulence, the second one suppressing it due to viscous damping:
\begin{equation}
Re=\frac{\rho v L}{\mu},
\label{reynold}
\end{equation}
where $\rho$ is the fluid density, $\mu$ is the parameter of dynamic viscosity, $v$ typical value of characteristic velocity of the fluid, and $L$ gives the characteristic length scale of the system.
The Reynolds number parameter is also often described in terms of kinematic viscosity, in which it is defined as $\nu=\mu/\rho$. 
Turbulence is experimentally known to develop for very high values of Reynolds number, 
typically of the order of $10^3$ and more. Starting from a belief that turbulence is too complex of a phenomenon to be solved in its totality
and exactly, Reynolds also introduced the statistical approach for the description of turbulence, expressing the flow variables as 
a sum of average and fluctuating parts. This approach was in fact the most common one in dealing with the problem of turbulence historically, and its review can be found in
\cite{battimelli, frans, orszag}. The fluctuating part is then understood as a random variable, with the integral averages of different 
products of fluctuating quantities describing the properties of turbulence -- such as energy, helicity, correlation length etc. After Reynolds this statistical approach was further developed
by Prandtl, who was trying to give a model of eddy viscosity \cite{Prandtl, battimelli}. Probably the most important further developments 
in this approach were done by Taylor during 1930's, who developed modern
mathematical techniques for its statistical description, such as correlators, Fourier transforms and probability density distributions 
\cite{Taylor}. The development of related techniques and results continues through the following decades and still, despite its limitations, represents one 
of the standard treatments.\\ \\
Following this paradigm of statistical description we can take 
$u(\mathbf{x},t)$ to be a function of space and time describing the distribution of some hydrodynamic quantity -- like velocity, pressure, temperature etc. 
Assuming that this function is integrable we can define its time average at the point $\mathbf{x}$ as
\begin{equation}
\bar{u}(\mathbf{x}) \equiv lim_{T \rightarrow \infty} \frac{1}{T} \int_{0}^{\infty}u(\mathbf{x},t) dt
\label{av}
\end{equation}
It clearly follows from this definition that the average is not a function of time, but only has a spatial dependence. Then the Reynolds 
decomposition, as discussed above, is given by
\begin{equation}
u(\mathbf{x},t)=\bar{u}(\mathbf{x}) + u'(\mathbf{x},t),
\label{re}
\end{equation}
where $u'(\mathbf{x},t)$ represents a fluctuating part of the hydrodynamic quantity. The definition of the average \eqref{av} implies that
$\bar{\bar{u}}(\mathbf{x})=\bar{u}(\mathbf{x})$, and therefore it follows that $\bar{u'}(\mathbf{x},t)=0$, which is consistent
with the assumed random and fluctuating nature of this part of the decomposition. \\ \\
We note that besides the time average and space average, which is defined in an equivalent fashion, a third type of average
which is very often used is the ensemble average. In this procedure, the average is obtained over the configurations of turbulent flows 
corresponding to the same set of external conditions, that is to say in the similar sense as the notion of ensemble is defined in statistical physics. If different members
of the ensemble are defined with a parameter $\alpha$, such that their respective weight is given by the probability density $P(\alpha)$,
then the ensemble average of a quantity $u$ is defined as
\begin{equation}
\langle{u}\rangle=\int u(\alpha)P(\alpha) d \alpha 
\label{an}
\end{equation}
In order to compare the predictions of turbulence theory with experiments, the ensemble average needs to be computed as a space-time 
average over the members of the ensemble -- because the average over time is what is actually related to a concrete physical measurement. The 
assumption usually made is that this correspondence is valid, and this can be shown to follow if some further technical conditions are met (for details see
\cite{orszag, frisch}). This assumption of equality between time averaging and assemble averaging is known as the ergodic hypothesis \\ \\
Using the Reynolds decomposition \eqref{re} in the  Navier-Stokes equation \eqref{hidronav} in the incompressible case, can be shown 
to lead to the Reynolds-averaged Navier-Stokes equation \cite{durbin}
\begin{equation}
\rho \bar{v_{j}} \frac{\partial \bar{v_{i}}}{\partial x_{j}}+ \rho \frac{\partial}{\partial x^{j}}\overline{v'_{i}v'_{j}}=- \frac{\partial \bar{p} \delta_{i,j}}
{\partial x_{j}}+ \mu \frac{\partial}{\partial x^{j}}\frac{\partial}{\partial x^{j}}\bar{v_{i}}.
\end{equation}
Assuming the validity of the ergodic hypothesis this equation can also be written in terms of ensemble averaging \eqref{an}, instead
of the time averaging \eqref{av}. 
The contribution $\rho \overline{v'_{i}v'_{j}}$ plays the role of the stress coming from the fluctuating velocity field and is usually
called Reynolds stress term. Since there is no equation that could determine the components of this stress term, it is obvious that
the number of unknowns is bigger than the number of equations -- this issue is known as the closure problem in turbulence (for a detailed
exposition of this problem see \cite{closure}). In order to address
this issue various models proposing additional relationships to close the Reynolds-averaged Navier-Stokes equation. A discussion on some of
these concrete models can be found in \cite{alfonsi}\\ \\
Although a statistical approach to turbulence is traditionally the most common one, we should stress that it can be strongly criticized 
on several grounds. First of all, the method of taking averages of physical quantities by definition erases some parts of the information about the
system. The fine details of flow structure and its evolution are inevitably lost in the process of averaging, where only the general
statistical patterns remain. In the case of turbulence, with its non-linear dynamics and complexity, those fine details
can be of critical importance. This is also connected with the loss of the unique correspondence between the physical system and its 
averaged statistical representation -- since, in general, the same statistical representation will correspond to different physical 
realizations, due to the loosing of information that comes as a consequence of the averaging procedure. On a more fundamental level, the
Navier-Stokes equation is a deterministic and not a stochastic equation, therefore its solution -- although leading to an irregular evolution
-- can not be described by randomly fluctuating quantities. The statistical approach is thus essentially conceptually inconsistent with the
assumption that the physics of turbulence is completely described by the Navier-Stokes equation. Also, from an experimental viewpoint,
the position that turbulence is a completely random process is strongly disfavored by the existence of coherent structures in turbulence
\cite{emmons, fiedler}. \footnote{As put fort by Kraichnan and Chen:"The statistical-mechanical treatment of turbulence is made
questionable by strong nonlinearity and strong disequilibrium that result in the creation of ordered structures imbedded in disorder``\cite{kraich}}. Based on this view, a more fundamental -- as well as mathematically and physically more consistent --
approach to turbulence is the one which uses the techniques of dynamic systems analysis. In this perspective, turbulence can be viewed 
as an example of the well established fact that non-linear deterministic dynamic equations can lead to solutions which are highly irregular
and cannot be described with simple functions. The study of this phenomenon, known as deterministic chaos, started by the end of the 19th 
century, including the important contributions of Poincar\'{e} \cite{poincare}, while the possible connection between deterministic 
chaos and turbulence was for the first time proposed by Lorentz in 1963 \cite{lorentz}. Further important work following this concept was 
done recently afterwards by Ruelle and Takens \cite{ruelle}, and continued with numerous contributions in the following decades \cite{prvi}.
The application of methods of dynamical systems to study turbulence has evolved from the observation that phase-space trajectories 
corresponding to the Navier-Stokes equation tend to topologically approach a strange attractor -- an attractor characterized by a fractal dimension, 
leading to temporally chaotic behavior \cite{ruelle, prvi}. 
However, despite various advances the approach of deterministic chaos has not yet been able to produce a successful concrete model of turbulent flows
\cite{prvi}. Because of this reason, since our focus here with respect to turbulence is its further study in the context of anomalous MHD in the early universe -- where we 
need some simplified models -- a further discussion
of this approach is beyond the scope of this work, while more information can be found in \cite{chaos, chaos2, chaos3}
\section{The Kolmogorov model of turbulence}
Starting from 1941. Andrei Kolmogorov \cite{kolmo1, kolmo2,kolmo3, kolmo4} in a series of papers developed his model of turbulence which is usually considered as the most
successful model, as well as the standard model in the field. In the attempts to understand MHD turbulence, it was also serving 
as a role-model for new theories. Similarly, different attempts to generalize the Kolmogorov model to MHD turbulence have been proposed, some of which we will
discuss in the following section. But all of these proposals to MHD case are tested and empirically verified to a much smaller extent than
the Kolmogorov model for hydrodynamic turbulence, due to the greater difficulties which accompany MHD turbulence observations, as well as
numerical simulations. The Kolmogorov model is mostly considered to be in accord with empirical observations, to the extent of the
applicability of the assumptions and approximations made. However, there also exist some more critical views regarding the confirmation
of Kolmogorov's model, for a discussion see \cite{conceptual}. In any case, one should note that the Kolmogorov model is not a complete
theory of turbulence, that it does not develop the study of turbulence from first principles and general properties
of Navier-Stokes equation, but it rather represents a phenomenological model resting on a large number of assumptions and approximations, which
can not hold precisely. This manifests in the observed deviations from the scaling predicted by the Kolmogorov theory, the
effects of anomalous scaling and intermittency \cite{lohse}, which also signal the need to go beyond the Kolmogorov model. From the physical perspective, the effect
of intermittency  comes from the fact that the self-similarity of turbulence -- the structure of turbulent eddies being invariant on different scales -- does not hold exactly. This is mostly
the result of the small-scale structures, which do not have uniform characteristics --even in the statistical sense, and therefore break the assumed features of Kolmogorov's approach.

An example of a modern discussion of Kolmogorov's approach, discussing in detail the assumptions 
and controversial steps done in the original papers by Kolmogorov, can by find in \cite{frisch}. \\ \\In the Kolmogorov picture, turbulence is viewed as a cascading of
energy through different scales, described by the mean energy dissipation rate, $\epsilon$ - being the energy distributed between different
scales per time per unit mass. The main physical assumption being made is that on scales much smaller than the ones defining the
boundaries and geometry of the physical system, the influence of this large scales and anisotropies they induce can be neglected, so that
turbulence can be in this interval treated as homogeneous and isotropic by its statistical properties (although not homogeneous and isotropic
in the real physical space). In this interval and under this assumption the phenomenon of turbulence is universal, not depending of physical mechanisms and realizations of turbulent flows, including pumping of energy on the higher scales, so that the energy transfer in this interval
can be described with a single universal parameter $\epsilon$. Before describing the Kolmogorov model precisely it is useful to devote
some time to the discussion of different typical length scales that can be used to characterize turbulence, since --as can be seen from what was stated so far - the question of different scales occupies an important place in the Kolmogorov model. \\
While defining the Reynolds number \eqref{reynold}, which describes the transition to turbulent flow, we introduced the length scale $L$ which is given by the geometrical dimension
of the system under consideration -- for instance, it can be given by the diameter of the pipe in which the fluid is flowing. This defines the first scale, which is set by the boundaries of the system itself. It is on this scale that energy will by some mechanism be brought to the fluid, in order
to start turbulent motion, and then transferred to smaller scales. On the opposite side of the length scale spectrum in the system, we expect that
turbulence will at some point necessarily stop, since the molecular forces will become important, leading to high viscosity. The dominance of viscosity will therefore reduce the characteristic Reynolds numbers specific for that interval, thus violating the condition of high Reynolds number necessary for the establishment of turbulence. Kolmogorov assumed that the main process describing this scale is the energy dissipation, and according to this
that it can be uniquely described by viscosity and energy dissipation rate $\epsilon$. Using dimensional analysis, the simplest combination
of parameters leading to units of length is
\begin{equation}
\eta=\left(\frac{\nu^{3}}{\epsilon}\right)^{1/4} . 
\end{equation}
This scale parametrised by $\eta$ is called the Kolmogorov length scale. Another important scale is the large scale which is of the
same order of magnitude as the one described by the dimension of the system $L$, called the integral scale. The integral scale describes the 
narrow range around of which the maximum of energy in turbulent flow is concentrated. Between the integral scale and the Kolmogorow scale
there is a wide range characterizing the typical structures of developed turbulence. In the Kolmogorov's approach it is thus assumed that
viscosity will there typically be small, and turbulence will show universal features -- and for high Reynolds numbers it will only depend on 
$\epsilon$, without any influence of the larger scales. It is also the interval
in which the proposed Kolmogorov's scaling of energy with typical length should be taken as valid in his model. 
We note that, in this picture, the region of this inertial interval is a function of Reynolds number, becoming bigger with its increase, 
and of course vanishing as the turbulent flow becomes laminar. \\ \\
The main assumptions of Kolmogorov are two related statements which are known as the Kolmogorov's universality hypotheses. The first
one assumes that at very high, but finite, Reynolds number all of the small-scale statistical properties are uniquely and universally
given by the length scale, $\epsilon$ and viscosity. The second Kolmogorov's assumption now considers the limit of the infinite Reynolds 
number in the turbulent flow, stating that in this limit all the statistical properties are determined only by the length scale, and $\epsilon$,
therefore that viscosity is not manifesting any influence. Here, under the notion of the small scales, the scales which are smaller than the integral
scale are understood. In his work \cite{frisch} Frisch proposed that for a consistent and more transparent representation of the Kolmogorov
theory there are three additional hypotheses which should be employed. First of them is that in the limit of infinite Reynolds number
all of the symmetries of the Navier-Stokes equation, which were broken during the transition to turbulence, are recovered in a statistical 
sense for small scales. We also note that it is for instance often assumed that turbulence is homogeneous and isotropic in a statistical 
sense - meaning that properties of statistical averages of turbulent variables and their products do not depend on the point in space and 
orientation, but just on the relative differences between points. Second additional hypothesis in \cite{frisch} assumes that in the same limit
turbulent flow manifests self-similarity at small scales, that is that $\delta \mathbf{v}(\mathbf{x},\lambda \mathbf{l})=\lambda^{h} 
\delta \mathbf{v}(\mathbf{x},\mathbf{l})$, where $\lambda$ is a positive real number, and $l$ a scale smaller than the integral scale. The third assumption is that
the energy dissipation rate, $\epsilon$ remains finite and non-vanishing.\\ \\
Starting from the presented framework and described assumptions, the turbulence theory of Kolmogorov derives several important results -- such as the law of scaling between energy and distance. In the case of high Reynolds number in the inertial interval this 
approximate law can be derived from the scaling arguments. Following the already presented assumptions, the mean energy dissipation
rate can be taken as universal in the inertial interval and thus $\epsilon \sim \delta v^{2}/t=v^{3}/l $. It therefore follows that velocity changes 
with scale according to $1/3$ power law (or to say equally that energy follows a $2/3$ scaling power-law). The form in which this law is
expressed in the Fourier space, giving the dependence of energy density on the wave number, is perhaps the most used and well known one\footnote{It
is interesting to note that Kolmogorov actually did not used the Fourier representation in his papers \cite{conceptual}}. Using the Fourier
decomposition of velocity and energy in the already presented scaling relation it follows 
\begin{equation}
E(k)=C_{K} \epsilon^{2/3}k^{-5/3}, 
\end{equation}
where $C_{k}$ is the Kolmogorov constant, which seems to have values in the interval from unity to two, mostly around 1.6 -- 1.7 \cite{biskamp}.

\section{MHD turbulence}
As already mentioned, MHD turbulence arises as a result of the interplay between Maxwell's equations in the MHD framework -- discussed
in the previous chapter --and the Navier-Stokes equation. Electromagnetic effects now influence the movement of a conductive fluid via
the Lorentz force, so that the Navier-Stokes equation takes the form
 \begin{equation}
\rho\left[\frac{\partial \textbf{v}}{\partial t} + (\textbf{v} \cdot \nabla) \textbf{v} - \nu \nabla^{2}\textbf{v}\right]= - \nabla p + \mathbf{J} \times \mathbf{B},
\label{magnav}
\end{equation}
where we remind the reader that $\mathbf{J}$ represents the current density.
Similarly as before we can take the divergence of the previous equation to obtain a Poisson-like equation for the pressure in the incompressible limit
\begin{equation}
\nabla^2 p=-\nabla \cdot \left[ (\mathbf{v} \cdot \nabla)\mathbf{v}- (\mathbf{B} \cdot \nabla)\mathbf{B} \right] 
\end{equation}
We note that since the incompressible limit simplifies equations significantly, and it is also typically satisfied in many different
terrestrial, astrophysical and cosmological situations \cite{verma} we will use it in the remainder of this work. 
In the hydrodynamic turbulence the dissipation of the system
 is given by the viscous term $\nu \nabla^{2}\textbf{v}$, while in the MHD turbulence we can also see that we have a dissipation of magnetic
energy described by the resistive term $\eta \nabla^{2} \mathbf{B}$ as can be seen by the inspection of eq. \eqref{induction}. Since the
ratio of nonlinear versus dissipative term was described by a Reynolds number \eqref{reynold}, we now also need to introduce the equivalent
ratio for the magnetic effects. It is given by the magnetic Reynolds number
\begin{equation}
 Re_{m}=\frac{vL}{\eta}
\end{equation}
For turbulent flows both Reynolds numbers are in general high, typically of an order of magnitude higher than thousand. \\ \\
In the statistical theory approach, discussed in the last section, we decompose the field variables into average and fluctuating parts. 
This will now also be the case with the magnetic field, which we can then write as $\mathbf{B}_{tot}=\mathbf{B}_{0}+\mathbf{B}$. 
The analysis of turbulence is significantly simplified if we assume the existence of additional symmetries related to certain statistical
properties \cite{verma}. One of those symmetries is homogeneity -- which implies that statistical properties only depend on the relative
distance between points, and not on their absolute position. Under this symmetry a correlator between two variables, say velocities,
is given by
\begin{equation}
<v_{i}(\mathbf{x},t)v_{j}(\mathbf{y},t)>= C_{ij}(\mathbf{x}-\mathbf{y},t) . 
\end{equation}
The second often used symmetry assumption is that of statistical stationarity -- this means that average properties depend only on time
differences:
\begin{equation}
<v_{i}(\mathbf{x},t)v_{j}(\mathbf{x},t')>= C_{ij}(\mathbf{x},t-t') 
\end{equation}
Similarly, the system is isotropic if its average properties are not dependent on the absolute position in space, that is if they are invariant
under rotation. It can be shown \cite{bat} that the isotropic correlation of two functions can be written as
\begin{equation}
C_{ij}(\mathbf{r})=C^{1}(r)r_{i}r_{j} + C^{2}(r) \delta_{ij}, 
\end{equation}
with $C^{1,2}$ being the even functions of $r=|\mathbf{x}-\mathbf{y}|$ and $\delta_{ij}$ Kronecker delta function. 
\\ \\
In the study of MHD turbulence it is often useful to pay attention to ideally conserved quantities -- the invariants in the case of ideal MHD turbulence, when
viscosity and resistivity can be neglected. The departure from their conservation comes only from dissipative effects, and therefore represents a useful constraint
on the evolution of the conductive fluid, making some problems much easier. As it is in general valid in mechanics, when dissipative processes are neglected the total energy
density stays conserved \cite{biskamp} (note that in the case of a MHD fluid placed in some field, for instance gravitational field, one should also add the contribution of the
potential energy to the total energy of the system). 
As we discussed in the previous chapter, see for instance equation \eqref{helcons}, magnetic helicity is also an ideally conserved quantity. The third important conserved quantity is the
so called cross-helicity, measuring the coupling between the fluid and magnetic components of turbulence
\begin{equation}
h_{c}=\frac{1}{V} \int \mathbf{v} \cdot \mathbf{B}d^{3} . 
\end{equation}
In a similar way in which we showed that magnetic helicity also has a direct topological interpretation, measuring the linking between the field lines, the cross-helicity
can be topologically viewed as giving the measure of linking between the vortex lines of the velocity field with the flux lines of the magnetic field \cite{cross}. This comes from the
same geometrical logic of definition of helicity, introduced by Gauss, as we discussed in the section on magnetic helicity (\ref{guslink}, \ref{sumhel}). \\ \\
In order to make the analysis of equations mathematically simpler, it is usual to represent them the Fourier space. Thus, 
velocity and magnetic field can be written using the decomposition
\begin{equation}
\mathbf{B}(\mathbf{r},t)=\int \frac{d^{3} k}{(2 \pi)^{3}}e^{i\mathbf{k} \cdot{\mathbf{r}}}\mathbf{B}(\mathbf{k},t) 
\label{fourmag}
\end{equation}
\begin{equation}
\mathbf{v}(\mathbf{r},t)=\int \frac{d^{3} q}{(2 \pi)^{3}}e^{i\mathbf{q} \cdot{\mathbf{r}}}\mathbf{v}(\mathbf{q},t) 
\end{equation}
Using this Fourier decomposition we can see that the absence of magnetic monopoles, \eqref{mono}, implies the condition $\mathbf{k} \cdot \mathbf{B}=0$
while the assumption of incompressibility implies $\mathbf{k} \cdot \mathbf{v}=0$.
Assuming that magnetic fields are statistically homogeneous and isotropic their correlators will be given by \cite{monin}
{ \begin{eqnarray}
\label{correlator}
\langle B_{i}(\mathbf{k},t) B_{j}(\mathbf{q},t)\rangle = \frac{(2 \pi)^3}{2} \delta(\mathbf{k} +\mathbf{q})[(\delta_{ij}-\hat{k}_{i} \hat{k}_{j})S(k,t)
+ i \epsilon_{ijk}\hat{k}_{k}A(k,t), 
\end{eqnarray} 
where $\hat{k}_{i}$ is the unit vector of $\mathbf{k}$, and $S(k,t)$ and  $A(k,t)$ denote the symmetric and anti-symmetric parts
of the correlator, respectively. Using \eqref{correlator} we can write the magnetic energy density -- $\rho_{m}$ and helicity density, $h$, in the volume $V$ as
\begin{equation} \label{magen}
\rho_{m}=\frac{1}{2V} \int d^{3} r\langle \mathbf{B}^{2}(\mathbf{r},t)\rangle = \int d\ln k \, \rho_{k}(t),
\end{equation}
\begin{equation} \label{maghel}
 h=\frac{1}{V} \int d^{3} r\langle \mathbf{A}(\mathbf{r},t) \cdot \mathbf{B}(\mathbf{r},t)\rangle=\int d\ln k \, h_{k}(t),
\end{equation}
where we have introduced the spectral magnetic energy and helicity $\rho_{k}(t)=k^{3}S(k,t)/(2 \pi)^{2}$ and $h_{k}(t)=k^2 A(k,t)/2 \pi^{2}$, respectively. The maximal value for helicity density is achieved if all of the magnetic energy is stored in one circularly polarized mode, and thus 
$\rho_{k}(t)=(k/2)h_{k}(t)$. This configuration of magnetic field is called maximally helical. \\ \\
As usual, the fluid part of turbulence is characterized by the kinetic energy, $\rho_{K}=(1/2V)\int d^{3} r \rho  v^{2} $. The relative importance of kinetic over magnetic effects in turbulence will be measured by the ratio between the respective
energy densities, $\Gamma=\rho_{K}/ \rho_{m}$, which will in general be a function of time. 
Turbulence will develop on scales between the dissipation scale, where the Reynolds number becomes small and turbulence stops due to dissipation processes, and the scale of the largest magnetic eddies. The latter  is modeled by the magnetic correlation length
\begin{equation}
\xi_{m}=\frac{\int   k^{-1} \rho_{k} d \ln k}{\rho_{m}}, 
\label{corr}
\end{equation}
and the kinetic correlation length can be defined in a similar fashion. Non-linear turbulent phenomena are dominant in the inertial interval -- the interval between the scales where injection and dissipation effects become relevant. 
\\ \\
The equations for incompressible MHD in Fourier space
can then be written in the following form, using the decomposition, $\mathbf{B}_{tot}=\mathbf{B}_{0}+\mathbf{B}$, \cite{biskamp}
\begin{align}
(\frac{\partial}{\partial t}-i(\mathbf{B}_{0} \cdot \mathbf{k})+\nu k^2)v_{i}(\mathbf{k},t)= \nonumber
-ik_{i}p_{tot}(\mathbf{k},t)- ik_{j} \int \frac{d \mathbf{p}}{(2 \pi)^3}v_{j}(\mathbf{k} - \mathbf{p})v_{i}(\mathbf{p},t) \\ 
+ ik_{j} \int \frac{d \mathbf{p}}{(2 \pi)^3}B_{j}(\mathbf{k} - \mathbf{p})B_{i}(\mathbf{p},t)
\label{fourmhd}
\end{align}
\begin{align}
(\frac{\partial}{\partial t}-i(\mathbf{B}_{0} \cdot \mathbf{k})+\nu k^2)B_{i}(\mathbf{k},t)= \nonumber
- ik_{j} \int \frac{d \mathbf{p}}{(2 \pi)^3}v_{j}(\mathbf{k} - \mathbf{p})B_{i}(\mathbf{p},t) \\ 
+ ik_{j} \int \frac{d \mathbf{p}}{(2 \pi)^3}B_{j}(\mathbf{k} - \mathbf{p})u_{i}(\mathbf{p},t)
\label{fourmhd2}
\end{align}
In order to further simplify the notation, as well as to make the formal mathematical  parallelism between magnetic and velocity
field more apparent, Elss\"{a}sser variables are often used. In this variables we introduce 
\begin{equation}
\mathbf{z}^{\pm}=\mathbf{v} \pm 
\frac{\mathbf{B}}{\sqrt{4 \pi \rho}}
\end{equation}
and then \eqref{fourmhd} - \eqref{fourmhd} can be expressed in terms of $\mathbf{z}^{\pm}$ as follows
\begin{equation}
(\frac{\partial}{\partial t} \mp i(\mathbf{B}_{0} \cdot \mathbf{k})+\nu_{+} k^2)z_{i}^{\pm}(\mathbf{k},t)+\nu_{-}k^{2}z_{i}^{\mp}(\mathbf{k})
=-ik_{j}P_{im}(\mathbf{k})\int d\mathbf{p}z_{j}^{\mp}(\mathbf{p})z^{\pm}_{m}(\mathbf{k}-\mathbf{p}),
\label{else}
\end{equation}
where $\nu_{\pm}0)=(\nu \pm \eta)/2$ and 
\begin{equation}
P_{im}(\mathbf{k})=\delta_{im}-\frac{k_{i}k_{m}}{k^{2}}. 
\end{equation}
\section{Phenomenological models of MHD turbulence}
Starting from the middle of the last century various models of MHD turbulence were proposed, mostly inspired by the logic of Kolmogorov's approach.
Also like the model of Kolmogorov these attempts focus mostly on obtaining the energy spectrum, i.e. scaling of energy with the characteristic 
scales. All of the popular models are just phenomenological -- trying to obtain the energy spectrum based on some scaling arguments and technical
assumptions, and not on the mathematically rigorous analysis of the equations and their solutions, or basic physical aspects of turbulent flows --
which is very difficult due to the non-linear and complicated nature of the equations for the MHD turbulence. In this respect all of the 
proposed models should be viewed as approximate approaches and even the basic questions of the dependence of energy with time, and its distribution
among different wave modes, should be considered as not yet answered. Moreover, it can not be stated that numerical simulations and observations
have completely and uniquely determined the proper scalings and therefore finally confirmed some specific model to be the best description
of MHD turbulence or ruled out other models. In contrast to hydrodynamic turbulence -- which was the topic of numerous experiments and was studied
in different experimental setups, terrestrial experiments involving MHD turbulence are practically impossible due to the large value
of resistivity of plasmas. Therefore, the solar wind -- which is the most accessible plasma exhibiting MHD turbulence -- is the most important 
source of experimental data \cite{biskamp, sedamdeset, stosedpet, veltri}. Taking into account the general tendency shown in various
measurements of the solar wind in the recent period, it can be stated that they seem to indicate that a Kolmogorov type scaling can also be 
applied to MHD turbulence \cite{verma, biskamp}. Similar conclusions were also reported in the numerical simulations of MHD turbulence, 
although there were also claims of different types of scalings, predicted by alternative models \cite{biskamp, verma2, biskamp2}\\ \\
The first such model of MHD turbulence was proposed in the sixties, by Kraichnan and Iroshnikov \cite{osampetd,sedsed}. They considered the regime
of the weak turbulence, that is -- the configuration in which the mean magnetic field, $\mathbf{B_{0}}$, is much stronger than the fluctuations 
of magnetic field. Iroshnikov and Kraichnan then argued that Alfven time scale, $\tau_{A}=l/v_{A}$ is the time scale basically defining the dynamics of the MHD turbulence, with $v_{A}=B/\sqrt{4 \pi \rho}$. This conclusion came from their observation that, unlike the hydrodynamic turbulence, the MHD
turbulence needs to take into account the interaction of magnetic perturbations, described by Alfven waves. As can be seen from \eqref{else} the non-liner term, essentially leading to turbulent effects, only couples $z^{+}$ to $z^{-}$ and vice versa. Therefore, only the Alfven waves propagating along opposite directions along the mean magnetic field will interact. On the scale $l$ their interaction will typically be characterized by the time of interaction given by $\tau_{A}$. On the other hand, the non- magnetic time scale, characterizing the hydrodynamic component of turbulence, is given by a time in which one eddy is distorted by interaction with a similar eddy, traveling in the opposite direction. On the same scale, for the eddy of size $\delta z_{l}^{+}$ or $\delta z_{l}^{-}$
this time scale is then given by $\tau_{l}^{\pm}=l/\delta z_{l}^{\mp}$. We therefore have two time scales -- $\tau_{A}$ and $\tau_{l}$ -- but typically
$\tau_{A}\ll t_{L}^{\pm}$ \cite{biskamp}. The evolution of turbulence in the inertial interval will according to this reasoning be dominantly determined by the Alfven time scale
$\tau_{A}$. The interaction of two oppositely traveling MHD wave perturbations is given by $\tau_{A}$, as discussed above, and therefore the relative change of 
amplitude, $ \Delta \delta z_{l}$, with respect to its original size will be small, since it is --following the previous discussion -- given by
\begin{equation}
\frac{\Delta \delta z_{l}}{ \delta z_{l}}=\frac{\tau_{A}}{\tau_{l}^{\pm}}, 
\end{equation}
and $\tau_{A}\ll t_{L}^{\pm}$. It will therefore take $N~(\delta z_{l}/\Delta \delta z_{l})^{2}$ elementary interactions to create a change in amplitude of the order unity. Therefore,
the time needed for energy transport on that scale, that would in the hydrodynamic case be described simply by $\tau_{l}$ will now be given by 
$N \cdot \tau_{A} ~ \tau_{l}^{2}/\tau_{A}$. Replacing the hydrodynamic turbulent time scale $\tau_{l}$ in the Kolmogorov assumption of universality of energy transfer in 
the intertial interval, $\epsilon ~ \delta  E_{l}/\tau_{l}$, with this time, $\tau_{l}^{2}/\tau_{A}$  leads to scaling $\epsilon ~\delta z^{4}_{L} \tau_{a}/l^{2}$ which then leads to a
different spectrum, called the Iroshnikov-Kraichnan spectrum for MHD turbulence
\begin{equation}
\rho_{k}=C_{IK} (\epsilon v_{A})^{1/2} k^{-3/2}, 
\end{equation}
where the factor $C_{IK}$ is also taken to be different than the corresponding factor for the Kolmogorov spectrum. 
Soon after the presented conclusions of Iroshnikov and Kraichnan there were other similar arguments proposed, but obtaining different final results. For instance, while
using the approach similar to Iroshnikov and Kraichnan, Marsch \cite{stodes} used the nonlinear time-scale, given by the characteristic
time scale of perturbations given by Elss\"{a}sser variables, $\tau^{\pm}_{nonlin} \approx 1/(k z_{k}^{\pm})$, and thus obtaining the
relation $\rho_{k} \propto k^{-5/3}$, which is the same scalling as proposed by Kolmogorov. Matthaeus and Zhou \cite{dvesto} postulated
that MHD turbulent time scale is given by a combination of Alfenic and nonlinear time scale so that $\tau=1/(kB_{0}+kz_{k}^{\pm})$. In this model
the spectrum would be effectively different for small and large wavenumbers, leading to Kolmogorov and Iroshnikov-Kraichnan spectrum
respectively. 
\chapter{Chiral anomaly and the evolution of magnetic fields}\label{chir}
In the last two chapters we have reviewed the conceptual and mathematical foundations of the macroscopic theory of electromagnetism applied 
to the description of conductive media in the MHD approximation and discussed some of its open problems. The underlying principles of MHD theory 
are purely classical. However, it is well known that such classical description can be taken as a valid approximation only in a certain limit. When moving 
to high energy scales quantum effects are expected to become important, and eventually the proper description of electromagnetism needs 
to be given in terms of the gauge theory of unified electroweak interactions. When approaching this regime it can be shown that 
quantum effects manifest even on macroscopic scales and modify the description given in terms of Maxwell's equations. As discussed in \ref{anomaly}
quantum fluctuations break the classical chiral symmetry for charged massless fermions, causing an anomalous current. This leads to the problem 
of generalizing the MHD description of plasma for the temperatures where these anomalous effects become non-negligible. In this chapter we will 
present the equations of chiral MHD and study some general properties which these modifications have on the evolution of magnetic fields. 
\section{Chiral MHD equations} \label{introchir}
In the second chapter we have discussed how the quantum effect of chiral anomaly lead to the divergence of axial current for charged massless 
particles, introducing
new terms which change the dynamics of electromagnetic fields.
Taking the massless limit of equation \eqref{divan}, and writing the divergence of the chiral current in terms of the electric, $\mathbf{E}$, and magnetic, $\mathbf{B}$, field 
we have
\begin{equation}
\partial_{\mu}j_{5}^{\mu}=\frac{e^{2}}{2 \pi^{2}} \mathbf{E} \cdot \mathbf{B}. 
\label{chempo}
\end{equation}
Introducing the number of left-handed and right-handed particles
\begin{equation}
n_{L, R}=\frac{1}{V} \int d^{3}x \psi^{\dagger}(1 \pm \gamma_{5})\psi,  
\end{equation}
and using the definition of magnetic helicity, this can be written as
\begin{equation}
 \frac{d(n_{L} - n_{R})}{dt}=-\frac{e^{2}}{4 \pi^2}\frac{dh}{dt}
 \label{kurac}
\end{equation}
This microscopic quantum anomaly can, in the presence of an asymmetry between right-handed 
and left-handed particles, lead to a macroscopic current. Namely, in the case of massless charged particles placed in a (hyper)magnetic field, the chiral
magnetic effect leads to the creation of an additional effective current, which is in the direction of the magnetic field and proportional to the chemical potential corresponding to the difference between the number of left and righ-handed
chiral particles: $\mathbf{j_{5}}=-e^{2}/(2 \pi^{2})\mu_{5} \mathbf{B}$. The difference of chemical potentials of left-handed and right-handed
particles, $\mu_{5}=(\mu_{L}-\mu_{R})/2$, follows an additional evolution equation and is determined by the change of magnetic helicity, corresponding to 
\eqref{kurac}. This can be shown in a rather simple semi-classical picture \cite{kharzeev}. Taking the Dirac sea of massless fermions when no external electromagnetic field 
is applied, the number of chiral particles is conserved. If electric and magnetic fields are now switched on parallel to each other (and we choose their direction to be along the z-axis), a change 
of chiral states will happen, which will lead to an electric current. Note that now, when the fermions are massless, chirality corresponds to 
helicity (projection of spin on momentum, not to be confused with electrodynamic notion of magnetic helicity). The external magnetic field will tend to orient the spins of negatively charged fermions in an anti-parallel direction with respect to the field (and parallel for positively charged fermions). This tendency exists because this type of orientation corresponds to the
minimal energy level in the interaction of leptons with the external field, $W= -{\boldsymbol \mu_m} \cdot\bf{ B}$, where ${\boldsymbol \mu_m} = gq{\bf s}/(2m_e)$ is the electron ($q=-e$) or positron ($q=e$) spin magnetic moment, with $g$ the spin factor. The considered fermions will
experience a force coming from the electric field,  $q \mathbf{E}$, and will therefore tend to have a positive projection 
of spin on momentum -- i.e. right-handed helicity for $q=-e$ (and reverse for positively charged fermions). 
The consequence of this will be that after some time, $t$, the momentum of left-handed fermions will decrease by $p_{L}=-e E t$, and the 
one corresponding to the right-handed fermions will increase by $p_{R}=e E t$. The density of states along the direction of the electric and magnetic fields will thus be $d N_{R}/d z=p_{R}/(2 \pi)$. 
Also, fermions populate Landau levels in the magnetic field with density
$d^{2}N_{R}/d x d y=eB/(2 \pi)$ and when the respective increase and decrease in the number of left and right chiral fermions is 
taken into account it leads to
\begin{equation}
 \frac{d(n_{L}-n_{R})}{d t}=- \frac{e^2}{2 \pi^2 V} \int \textbf{E} \cdot \textbf{B}\, d^{3}\mathbf{r}.
\end{equation}
When the magnetic helicity, as defined in \eqref{magne}, and the chemical potential corresponding to the number of left-handed and right-handed particles is introduced, $\mu_{L,R}= 6 n_{L,R}/T^{2}$, by using \eqref{chempo}, and considering the effective current corresponding to such a change in chemical potential it can be shown to lead to
$\mathbf{j_{5}}=-e^{2}/(2 \pi^{2})\mu_{5} \mathbf{B}$. In the last decades this effect was analysed by various authors 
\cite{vile,redlich, fukushima, pedrini, joyce}, including the application to quark-gluon plasma \cite{35, 36, hirono}. \\ \\
The example of charged massless fermions is given by leptons before the electroweak transition. Therefore, in the setting taking place before the electroweak 
transition in the early Universe, the MHD description of the cosmological plasma should include contributions coming from the  
chiral magnetic effect. Strictly speaking, the chiral magnetic effect rests on the fact that particles are massless. However, at high temperatures characterizing the early Universe, all the processes involving electron mass will be 
highly suppressed. In these conditions one can approximately ignore the difference between helicity and chirality operators, and expect
the existence of a contribution coming from the chiral magnetic effect even after the electroweak transition. The corrections coming from the finite 
electron mass can then be added perturbativelly, in the form of reactions which flip the chiralities of the considered particles, and which
are related to their mass. This approach was followed in the literature, where the modifications of chiral anomaly were investigated
also after the electroweak transition \cite{Boyarsky}, showing that a chiral asymmetry, initially present around the electroweak scale, 
could survive down to the MeV scale. In the recent period this investigation was followed by other works \cite{hiro, pa, gorbar}, where 
the effect of chiral anomaly on the evolution of magnetic fields was further discussed, as well as its potential role on magnetogenesis
and leptogenesis \cite{sha, gi, dvo, sem, kamada, kamada2}. Although in this work we are focused on the question of cosmological magnetic fields, it should also be mentioned that some astrophysical
 objects, for instance such as core collapse supernovae, reach temperatures comparable to the electroweak scale, thus making the study
 of chiral magnetic effect of interest also in this framework. This question was recently addressed in the context of magnetars,
 neutron stars and quark stars -- where it was shown that chiral anomaly can lead to magnetic field enhancement \cite{39,40,41,42,43,44,45}. 
Taking into consideration results of the various recent studies we cited above, we conclude that the proper description of
magnetic fields and their evolution, on temperature scales comparable to the electroweak scale -- which are expected to be reached 
in the framework of early Universe and neutron stars -- needs to take into account the effect of the chiral anomaly. Therefore, the standard
MHD description should in these regimes be replaced by the chiral MHD description of plasma. In the following sections we will
present its main properties and consequences, and the main modifications with respect to the standard MHD turbulence. \\ \\
In the presence of a non-vanishing $\mu_{5}$, the usual MHD equations -- consisting of Maxwell, Navier-Stokes, and continuity equations in the resistive MHD approximation \cite{Giovannini2} -- should 
now also include the contribution 
from the effective chiral current, and they read 
\begin{equation}
 \nabla \times \textbf{B} = 4 \pi\left[ \sigma (\textbf{E} + \textbf{v} \times \textbf{B})- \frac{e^{2}}{2 \pi^{2}}\mu_{5} \textbf{B} \right],
\label{Maxwell}
\end{equation}
\begin{equation}
 \partial_t\textbf{B}= - \nabla \times \textbf{E},
 \label{induction}
\end{equation}
\begin{equation}
 \rho\left[\partial_t\textbf{v} + (\textbf{v} \cdot \nabla) \textbf{v} - \nu \nabla^{2}\textbf{v}\right]= - \nabla p + 4\pi\sigma[ \textbf{E}\times \textbf{B} + (\textbf{v} \times \textbf{B}) \times \textbf{B}],
\label{navsto}
 \end{equation}
\begin{equation}
\partial_t\rho+ \nabla(\rho \cdot \textbf{v})=0,
\label{c}
\end{equation}
where $\sigma$ is the electrical conductivity, $\rho$ the energy density and $\nu$ the kinematic viscosity. Eq. \eqref{chempo}, describing the evolution of the chemical potential, must also be added to this system. 
In order to take into account processes which flip chirality, and therefore tend to erase the chiral chemical potential, $\mu_{5}$,
we also need to add the term given by the rates of such processes, $\Gamma_{f}$. It is also possible that in the considered system some processes
which create the difference of left and right chiral particles are present -- potentially connected to some physical mechanism beyond the 
Standard model. In this case eq. \eqref{chempo} should also include the source term for the chiral asymmetry, $\Pi_{sr}$.
For instance, in the core of a neutron star it follows $\Pi_{sr}=\Gamma_{f} \mu_{5}^{b}$ where $\mu_{5}^{b}$ is the equilibrium value
of the chiral potential of the background medium in the absence of magnetic helicity \cite{43}. Then the generalized version of Eq. \eqref{chempo}
reads
\begin{equation}
\frac{d \mu_{5}}{d t}=\frac{1}{T^{2}} \frac{3e^2}{4 \pi^2}\frac{d h}{d t} - \Gamma_{f}\mu_{5} + \Pi_{sr}. 
\label{chempomod}
\end{equation} 
We note that in the equations above, as well as in the remainder of this work, we assume that the chiral potential, $\mu_{5}$, is only a function of time
and does not depend on spatial coordinates. This is clearly not the most general case, but this assumption will be justified, at least
as an approximation, as long as its spatial variation is negligible. The generalization of MHD equations to the case of a space-dependent chiral
chemical potential was considered in \cite{boyflu}, where it was shown that it leads to the appearance of new terms describing the
effects of inhomogeneities. The  approximation of space independent chiral chemical potential  was  used  in  almost  all  theoretical  
studies  of  the  chiral  magnetic  effect, but the question of how justified this approximation remains in the realistic descriptions
of the early Universe, and what could be the effects of inhomogeneities, is not a straightforward one. Due to the complicated mathematical
nature of the chiral MHD equations, a natural way to discuss this issue is to compare the solutions of numerical simulations related to a space-dependent chiral potential to the ones obtained
with the assumption of a space independent chemical potential. This was done in references \cite{gorbar,bui} and both studies
concluded that inhomogeneities of the chiral asymmetry
have a negligible role for the main aspects of the evolution of the primordial plasma. 
\section{General properties of chiral MHD equations}
Before discussing the consequences of the chiral anomaly on the evolution of the magnetic fields in the early Universe and its influence
on the properties of MHD turbulence, we will first discuss general properties of the chiral MHD equations \eqref{Maxwell} - \eqref{chempomod}. 
It can first be noted that the anomaly contribution significantly complicates the electrodynamic equations. The new variable
of chiral asymmetry chemical potential, $\mu_{5}$, is determined by the change of magnetic helicity in \eqref{chempomod}, which is not
a simple function of the magnetic field - i.e. it is not a main variable appearing in the MHD equations - but it is given as a scalar product of the magnetic field with the vector potential,
$\mathbf{A}$. As we will see in the following sections, the induction equation \eqref{induction} combined with \eqref{Maxwell} can be 
mathematically simplified if the Fourier decomposition \eqref{fourmag} is used. Then the induction equation can be separated into independent
equations giving the time change of magnetic energy and helicity for each value of $k$, which are all modified by the chiral effective current contribution. However, since the evolution
of $\mu_{5}$ is given by the total magnetic helicity, the corresponding system will be a non-trivial differo-integral system of equations -- and the number
of those equations will clearly be infinite for the case of a continuous spectrum. Therefore it is not possible to solve this system analytically 
in general and the main tools for studying the physical consequences of the chiral MHD will be numerical simulations and various
sets of suitable approximations. The second difficulty is given by the description of the coupling between chiraly modified magnetic field and velocity.
Although the anomaly contribution does not enter the Navier-Stokes \eqref{navsto} equation explicitly, due to parallel orientation of
the chiral current with respect to the magnetic field, it influences the velocity indirectly, via changing the magnetic field in 
\eqref{Maxwell}, which then enters in \eqref{navsto}. On the other hand, the effects of the velocity field on the evolution of magnetic field appear directly
via the corresponding term in \eqref{Maxwell}. As we have discussed in more details in the third chapter, the phenomenon of turbulence
remains to be one of the open problems of classical mechanics, due to the highly non-linear nature of the Navier-Stokes equation, and 
related non-deterministic nature of turbulence itself. Even in this case of hydrodynamic turbulence there is no general analytical
strategy which could lead to its proper description. These issues becomes even more apparent in the case of MHD turbulence,
where some of very fundamental questions -- such as the relationship between magnetic and velocity field, the role of helicity,
proper scaling and the time dependence of MHD quantities \cite{biskamp, lazarian} are yet not properly understood. When the above mentioned mathematical difficulties 
of the modified electrodynamic equations are taken into account it is clear that the issue of chiral MHD turbulence can not be solved
in a straightforward manner. With this in mind we will -- in this and the following section -- first turn our attention 
to the problems of chiral MHD in the case when the velocity field is negligible, $\mathbf{v}=0$, in order to first study the consequences of the anomaly
effect in a simpler setting. After that, in the following chapter, we will discuss the chiral MHD turbulence. \\ \\
In order to better understand the character of the tendencies introduced by the addition of the chiral anomaly effect, we first consider the special
regime of the strong chiral anomaly, where the corresponding term associated with the chiral asymmetry chemical potential, 
$\frac{e^{2}}{2 \pi^{2}}\mu_{5} \textbf{B}$ becomes dominant on the right-hand side of \eqref{Maxwell}. In this case the curl vector operator 
acting on the magnetic field, $\nabla \times \mathbf{B}$, is clearly in the direction of the magnetic field. Since 
$\mathbf{J}=\nabla \times \mathbf{B}$, and the Lorentz force is given by $F_{L}=\mathbf{J} \times \mathbf{B}$, this regime clearly leads
to a force-free field configuration. If there is a velocity field present, then in this regime the Navier-Stokes equation reduces to
\begin{equation}
 \rho\left[\partial_t\textbf{v} + (\textbf{v} \cdot \nabla) \textbf{v} - \nu \nabla^{2}\textbf{v}\right]= - \nabla p  . 
 \end{equation}
 If the contributions of MHD effects on the pressure variations can be ignored, we see that in the case of the dominant chiral anomaly effect, 
 the evolution of velocity and magnetic fields tend to decouple, since the former is then influenced only by viscosity and the non-linear term. 
Using the Fourier decomposition \eqref{fourmag} and definitions of magnetic energy and helicity densities in k-space given in \eqref{magen} and \eqref{maghel}, 
the equations \eqref{Maxwell} and \eqref{induction} in this case of dominant chiral asymmetry lead to
\begin{equation}
\frac{d \rho_{k}}{d t}= - \frac{e^{2}}{16 \pi^{2}}\frac{k^{2} \mu_{5}}{\sigma} h_{k} ,
\label{domch}
\end{equation}
\begin{equation}
\frac{d h_{k}}{d t}=- \frac{e^{2}}{4 \pi^2} \frac{\mu_{5}}{\sigma} \rho_{k}. 
\label{domch2}
\end{equation}
By dividing the \eqref{domch} with \eqref{domch2} we conclude that the condition for maximally helical fields
\begin{equation}
\rho_{k}=\frac{k}{2}h_{k} 
\label{maxhel}
\end{equation}
needs to be satisfied in this case. 
We note that $\mu_{5}$ appearing in the equations above is still a general function of time, which depends on the solution of equation 
\eqref{chempomod}. In the special case when $\mu_{5}$ can be treated as slowly changing in time -- which will be the case if the change of helicity and the 
contribution of the source term compensates 
the effects of chirality flipping processes, or if both of them are negligible -- these equations have a simple solution,
which can be easily found (decoupling the system by taking a time derivative of one equation and using the remaining one) to be represented with exponential
functions 
\begin{equation}
\rho_{k}=A_{k}e^{\omega_{k} t} + B_{k}e^{-\omega_{k} t}, 
\end{equation}
with the equivalent solution for $h_{k}$, where
\begin{equation}
\omega_{k}=\frac{e^{2}}{8 \pi^{2} \sigma} \mu_{5}{k},
\end{equation}
and we see that the smallest scales (largest k's) will have the fastest growth. 
Using this simplified model, based on approximations suitable
for the case where the chiral anomaly effect is the dominant process in the evolution of magnetic fields, we have shown some basic tendencies
that the chiral effect tends to introduce in the evolution of magnetic fields. We can conclude that the anomaly will tend to organize 
magnetic fields towards a force-free maximally-helical field configuration. Moreover, it can also lead to exponential growth of magnetic fields,
and in a suitable setting it can therefore be used as a potential mechanism of magnetic field enhancement in astrophysics and cosmology. 
Those three consequences were indeed the main reason for the increased interest in physical applications of the chiral anomaly effect 
in the recent period. Apart from those consequences which are perhaps the most direct, we will in this work also discuss other interesting effects
related to the chiral anomaly -- such as the creation of helicity, and supporting the inverse cascade effect, which will be of special interest
in the study of interdependence between the anomaly and MHD turbulence. 
\section{Chiral anomaly effect and helicity}\label{heliind}
Returning to the general case of the chiral MHD equations, after applying the Fourier decomposition to \eqref{Maxwell} and \eqref{induction},
the chiral induction equation leads to the following equations for the time change of energy and helicity modes:
\begin{equation}
\frac{d \rho_{k}}{d \tau}=-\frac{2k^2}{\sigma} \rho_{k} - \frac{e^{2}}{16 \pi^{2}}\frac{k^{2} \mu_{5}}{\sigma} h_{k} ,
\label{chireq1}
\end{equation}
\begin{equation}
\frac{d h_{k}}{d \tau}= - \frac{2k^{2}}{\sigma}h_{k} - \frac{e^{2}}{4 \pi^2} \frac{\mu_{5}}{\sigma} \rho_{k}. 
\label{chireq2}
\end{equation}
By inspection of these equations we can see that even if the magnetic field is initially characterized by a vanishing magnetic helicity,
the chiral anomaly effect will lead to the creation of finite helicity. This is in contrast with the evolution of the standard, that is non-chiral,
magnetic fields, where initially non-helical fields will stay non-helical. Assuming there is no helicity initially,
energy modes just decay as $\rho_{k}=\rho_{0} \cdot \exp(-2k^2 \tau/ \sigma)$. However, due to the anomaly term, this decay of magnetic energy 
leads to a non-vanishing time change of helicity according to
\begin{equation}
 \frac{d h_{k}}{d \tau}= - \frac{e^{2}}{4 \pi^2} \frac{\mu_{5}}{\sigma} \rho_{0}e^{(-2k^2 \tau/ \sigma)},
\end{equation}
which will lead to the growth of helicity density under the assumption that $\mu_{5}$ has an opposite sign (if the sign is the same, then 
the induced helicity will have a negative value). Therefore, as long as the created helicity stays smaller than the corresponding energy
term in \eqref{chireq2}, the chiral anomaly effect will lead to the induction of a total helicity energy density
\begin{equation}
|h|= \frac{\rho_{0} e^{2}}{4 \pi^2 \sigma} |\int d\tau \int d\ln k \mu_{5}(\tau) e^{(-2k^2 \tau/\sigma)}|.
\end{equation}
The helicity in the right-hand side of \eqref{chireq1} and \eqref{chireq2} will approximately follow this 
tendency of growth as long as the term containing it stays much smaller than the energy density term.
When this is not the case, one needs to solve the full set of coupled equations for the energy
and helicity densities. If we assume that a chiral MHD system starts from a state of vanishing helicity, we expect to have the creation of helicity
and its fast growth, until the term containing it reaches a level comparable to the magnetic energy term,
with the following less dramatic evolution closely related to the evolution of magnetic energy. In this sense, we expect that the creation
of helicity is a fast (compared to the characteristic time scales of the system) and transient phenomena, leading to a balance of helicity
and magnetic energy density. We will soon see that this conclusion is also confirmed in numerical simulations. \\ \\
The fact that chiral anomaly naturally leads to the creation of helicity from initially non-helical field configurations could have important
consequences for the understanding of magnetogenesis models and evolution of magnetic fields in the early Universe. 
Probably the most direct consequence of the presence of helicity in cosmological magnetic
fields is that it can lead to an increase of the correlation length. One of the crucial problems in a large class
of models which assume a cosmological origin of magnetic fields is the small correlation length of the created
fields which, if not produced during inflation, needs to be smaller than the Hubble radius at
the period of their creation. This result is in strong contradiction with the correlation lengths of observed
fields, which can be even of the order of Mpc \cite{Giovannini2, urban, newneronov}.  
This does not have to directly mean that those magnetogenesis models are not a viable option. Even if the initial correlation
length is too small for magnetic fields created during the electroweak and QCD transition, there is the possibility of its growth
during the subsequent evolution of magnetic fields. For instance, magnetic helicity in the presence
of turbulence is known to lead to the development of magnetic structures at progressively
larger scales as a consequence of the inverse cascade that occurs.  These processes, related to MHD turbulence could
then lead to a considerable growth of correlation length \cite{olesen, campanelli}. The effect of the chiral anomaly could therefore also
influence the growth of the magnetic field correlation length if turbulence develops in those early stages of cosmological evolution. 
We will consider these questions in detail in chapter \ref{chirturb}, when we will discuss the problem of
chiral MHD turbulence. 
Moreover, the presence of non-vanishing helicity in the context of primordial magnetic fields could also change constraints on the amplitude of primordial magnetic fields
from gravitational wave production \cite{caprini}. \\ \\
As in the considered case of induced helicity, we can use a similar logic to analyse the case where, conversely, the initial magnetic fields are characterized by a non-vanishing helicity,
but vanishing initial chiral chemical potential. Here we assume there are no particle processes which could source the chiral asymmetry, $\Pi_{sr}=0$.  In the case of MHD described by vanishing velocity, magnetic
helicity will at first just decay if no chiral asymmetry is present, as can be seen from \eqref{chireq1} and \eqref{chireq2}. However, this change
will then lead to a non-vanishing change of chiral chemical potential, as predicted by \eqref{chempomod}. We conclude that the chiral anomaly effect
will modify the MHD description of turbulence in the case that initially there is either a non-vanishing magnetic helicity or non-vanishing
chiral asymmetry present. Only if $\mu_{5}=0$ and $h=0$ will the turbulence follow the standard MHD description. \\ \\
Another interesting and important aspect of the influence of the chiral anomaly on the evolution of magnetic fields is that it can lead
to an inverse transfer of energy -- that is, the transfer of energy from small to larger scales of the system, even if $\mathbf{v}=0$. We will discuss this question with special attention later, when we will investigate
it from the point of view of inverse cascades in chiral MHD turbulence. Here we will just demonstrate it with a simple argument for the
$\mathbf{v=0}$ case. Let us first for simplicity concentrate on the maximally-helical configuration of magnetic fields. Furthermore,
let us start by considering a specific regime of stationary chiral potential, $\mu_{5}^{stat}$, while $\Pi_{sr}=0$. In general, in order that this solution is obtained, the change
of magnetic helicity needs to compensate the loss coming from the spin-flipping reactions, 
$\delta h \propto \mu_{5}^{stat}/\int \Gamma_{f}(\tau)d \tau$. One special realistic case of a stationary regime is the case when the total 
helicity of the system reaches saturation with negligible spin-flip contributions. This regime could also be a reasonable approximation
also for some period of the evolution of chiral magnetic fields around the electroweak transition \cite{pa}. In this case we can obtain
the exact solution for the helicity modes
\begin{equation}
h_{k}=h_{k}^{0}=exp\left[-(\tau - \tau_{0}) \cdot(\frac{2k^{2}}{\sigma}+\frac{e^{2}}{16 \pi^{2}}k \mu_{5}^{stat})\right] . 
\end{equation}
We can see that, initially assuming that $h_{k}$ and $\mu_{5}^{stat}$ have opposite signs, then the stationary solution --
which also corresponds to the mode carrying maximal helicity (which is, in the maximally-helical case obviously the mode also carrying the
maximal energy) -- is reached for
\begin{equation}
k_{5}= \frac{e^{2} \sigma}{16 \pi^{2}}\frac{|\mu_{5}|}{2} .
\label{k5}
\end{equation}
According to this equation, concrete energy and helicity modes characteristic for different scales, will have a different type of evolution, depending on the scale $k$. All modes for which $k<k_{5}$ will be growing, while all modes for which $k>k_{5}$ will be decaying, and 
$k=k_{5}$ gives the stationary case. Since the wave-number, $k$, is giving the inverse length of the considered scale, this actually corresponds
to the transfer of helicity and energy from smaller to larger length scales, which is here induced without turbulence and other effects,
but just due to the chiral anomaly. This analytical solution will no longer be valid in the case when $\mu_{5}$ is no longer stationary,
since for the time-dependent chiral potential the equation can not be integrated in general to yield an analytical solution. However, the 
qualitative analysis remains, apart from the fact that the stationary mode, $k_{5}$, still determined by \eqref{k5}, should now be considered as a function of time, and therefore the 
character of some given mode $k$ (growing, decaying or stationary) will generally also change in time.
\section{Energy conservation and thermodynamics of the chiral magnetic effect}
We have seen that the chiral magnetic effect can lead to an exponential amplification of magnetic fields, which also means of the magnetic energy. In every 
closed system, the total energy needs to be conserved, and the same also remains true when dealing with chiral 
MHD phenomena. The amount of energy which gets transferred to the magnetic energy is initially stored in the medium which contains chiral particles, and is related to the 
chiral asymmetry potential, $\mu_{5}$. This energy initially stored in the chiral asymmetry corresponds to the energy needed to put a difference of $(n_{L} - n_{R})/2$ chiral particles 
into the system. When $\Gamma_{f}=0$ and $\Pi_{sr}=0$ -- in other words, when the considered chiral MHD system represents a closed system -- the total energy of the system stays constant and can only be redistributed 
between the magnetohydrodynamic part and the chiral part. We will demonstrate this in more detail in section \ref{minencon}, where we will also 
show that reactions causing the change of chirality act similarly to classical dissipation processes, such as viscous and resistive damping, transforming 
the energy into random motion of particles. Thus, when $\Gamma_{f}\neq 0$ and $\Pi_{sr}=0$ the total energy, consisting of energy stored in the magnetic field, motion of the fluid and the chiral asymmetry,
will decrease due to these dissipation processes (see \ref{minencon})
\begin{equation}
\frac{d \rho_{tot}}{dt} =  - \frac{1}{V} \int\left[\frac{(\nabla \times \mathbf{B})^{2}}{\sigma} + \nu \cdot \omega^{2}\right] d^{3}r -\frac{T^{2}}{3} \mu_{5}^{2} \Gamma_{f}.
\end{equation}
If we now also include the source term, describing the contribution of processes creating the asymmetry of chiral particles, the change in 
energy would read
\begin{equation}
\frac{d \rho_{tot}}{dt} =  - \frac{1}{V} \int\left[\frac{(\nabla \times \mathbf{B})^{2}}{\sigma} + \nu \cdot \omega^{2}\right] d^{3}r -\frac{T^{2}}{3} \mu_{5}^{2} \Gamma_{f} + \frac{T^{2}}{3} \mu_{5} \Pi_{sr} .
\label{sourceen}
\end{equation}
Therefore, if $\Pi_{sr}$ is large enough the term containing it can become dominant and lead to the increase of total energy in time, $d \rho_{tot}/dt>0$. This situation corresponds to the pumping of energy, 
into the system by some exterior mechanism. Of course, what we will understand as the system and what as its surroundings is arbitrary and dependent 
on the definition which should be chosen suitably for the problem of interest. However, it should be stressed that if we are interested in studying problems 
related to magnetic field amplification by the chiral anomaly, then the system needs to be chosen in such a way that the mechanisms acting as 
a source of chiral asymmetry are also included in its definition -- so that the backreaction on those mechanisms is also taken into account. Otherwise --if the sourcing processes are just introduced as an exterior mechanism, without 
considering their full dynamics and backreaction on them -- the problem of 
field amplification can not really be solved.  In that case this problem is just hidden and implicitly replaced with the problem of where is the energy transferred into 
the chiral MHD system actually coming from, and what is the complete evolution describing its change. \\ \\
The related questions have recently played a significant role in discussions regarding the field amplification in magnetars. It was proposed 
in \cite{dvorneutron1, dvorneutron2, dvorneutron3} that the study of chiral MHD in neutron stars should be supplemented with an additional 
contribution coming from the electroweak interaction of chiral electrons with background nucleons. The difference of interactions for left 
and right electrons is given by the potential $V_{5}=(V_{L} - V_{R})/2 \approx G_{f} n_{n}/2\sqrt{2}$, where $n_{n}$ is the neutron density 
and $G_{f}$ Fermi constant. As discussed in \cite{dvorneutron1, dvorneutron2, dvorneutron3} this effective interaction potential, described 
by a constant term, changes the 
total potential, acting as an addition to the chiral potential, $\mu_{5} \longrightarrow \mu_{5} + V_{5}$, in the effective chiral current, $\mathbf{j_{5}}$. In this sense we see that electroweak 
interactions between chiral electrons and nucleons do not act a source term, $\Pi_{sr}$, since -- due to the fact that $V_{5}$ is a constant 
term -- they do not change the time derivative of $\mu_{5}'=\mu_{5} + V_{5}$. Including this contribution, the equations now read \cite{dvorneutron1, dvorneutron2, dvorneutron3}
\begin{equation}
\frac{d \rho_{k}}{d \tau}=-\frac{2k^2}{\sigma} \rho_{k} - \frac{e^{2}}{16 \pi^{2}}\frac{k^{2} (\mu_{5}+V_{5})}{\sigma} h_{k} ,
\end{equation}
\begin{equation}
\frac{d h_{k}}{d \tau}= - \frac{2k^{2}}{\sigma}h_{k} - \frac{e^{2}}{4 \pi^2} \frac{\mu_{5}+V_{5}}{\sigma} \rho_{k},
\end{equation}
so that the system of equations is equivalent to \eqref{chireq1} and \eqref{chireq2} with a constant potential shift, $\mu_{5} \longrightarrow \mu_{5} + V_{5}$. 
Since $\Pi_{sr}=0$, and we expect that the evolution of the system can not depend on a constant energy change, one would assume that there is no 
change introduced with respect to \eqref{chireq1} and \eqref{chireq2}. This reasoning seems to be supported by the fact that in general, laws of 
physics do not depend on a choice of energy zero-point, but only on relative energy differences. It was however claimed that similar transformation does not take 
place in the evolution equation for $\mu_{5}$, but that it stays unaltered: 
\begin{equation}
\frac{d \mu_{5}}{d t}=\frac{1}{T^{2}} \frac{3e^2}{4 \pi^2}\frac{d h}{d t} - \Gamma_{f}\mu_{5}.
\label{pickas}
\end{equation} 
Since the chirality-flipping term is introduced phenomenologically, this type of dependence would mean that chirality flips tend to push the level 
of $\mu_{5}$ to zero, as in the case when there is no $V_{5}$. This seems to contradict the expectation that the equilibrium energy level 
of $\mu_{5}$ is now no longer zero due to the contribution from electroweak interactions, but that it should now be $\mu_{5}=-V_{5}$. This issue can also be 
understood by a simple thermodynamical analogy. If we consider a simple thermodynamical system of gas stored in a container with one 
movable wall, with a perpendicular constant external force applied to it -- then the equilibrium position of the wall is clearly changed. 
As in this simple analogy, one would not expect any change in the internal dynamics of the system due to the constant contribution of 
$V_{5}$, and moreover it would follow that the equilibrium position towards which the flipping rates suppress $\mu_{5}$ is also changed and 
is no longer $\mu_{5}=0$. 
\\ \\
This different description 
of the total chemical potential in different equations can lead to significant physical differences. We can formally see what happens if we 
write the equations proposed in \cite{dvorneutron1, dvorneutron2, dvorneutron3} in terms of $\mu_{5}'$ using $\mu_{5}=\mu_{5}'-V_{5}$.  
The equations \eqref{chireq1}-\eqref{chireq2} clearly remain invariant. However the equation for the evolution of the chiral potential \ref{pickas} changes and acquires 
a source term
\begin{equation}
\frac{d \mu_{5}'}{d t}=\frac{1}{T^{2}} \frac{3e^2}{4 \pi^2}\frac{d h}{d t} - \Gamma_{f}\mu_{5}' + \Pi_{sr}, 
\label{dvorso}
\end{equation} 
with $\Pi_{sr}= \Gamma_{f} V_{5}$. If $\Pi_{sr}$ is a dominant term in the equation, the energy will formally 
grow indefinitely as can be seen from \eqref{sourceen}. It is exactly this property of the evolution described by such equations which 
motivated the consideration of $V_{5}$ in the context of neutron stars, since the goal was to somehow strongly amplify the magnetic fields 
to explain the formation of magnetars. Of course, the infinite growth of the magnetic energy is clearly non-physical, and the energy 
eventually needs to come from the energy stored in the thermal bath of particles in the neutron star, which is of course finite. In order 
to fix this problem an \textit{ad hoc} quenching factor was added by hand in \cite{dvorneutron3}, so that the total potential reads 
$(\mu_{5}+V_{5})[1+(B/B_{eq})^{2}]^{-1}$, where $B_{eq}$ would be a maximal value of magnetic field. \\ \\
From the discussion above we can see that such procedure in explaining the magnetic fields in neutron stars is problematic, and can 
not be considered as an viable approach, for two reasons. 
First, the source term in \eqref{dvorso} is not given by realistically considering actual processes which would seed the growth of chiral 
asymmetry, but just from a formal constant shift of the chemical potential, coming from the electroweak interactions with nucleons, but which are not 
consistently taken into account in \eqref{dvorso}. The second reason is that an effective source term introduced in such fashion is not completely 
described as a part of the system, but its evolution needs to be adjusted artificially so that its effects stop after its role of the
magnetic field amplification has been fulfilled. 
\section{Conclusions}
In this chapter we have analysed some general properties of the chiral magnetic effect and the chiral MHD equations, which come as a generalization 
of the standard MHD equations for high enough temperatures. After discussing the effect of chiral anomaly on charged massless fermions 
in the exterior magnetic field, characterized by a non-vanishing helicity, we briefly reviewed the recent progress in the study of this phenomenon
in various physical systems. The mathematical properties of the chiral MHD equations were then discussed, and in order to study their general consequences 
analytically we considered the simplified case of the strong anomaly regime, neglecting all contributions coming from velocity. We have thus 
demonstrated that the chiral anomaly effects tends to lead to maximally helical field configurations and an exponential amplification of magnetic 
fields where the modes corresponding to the smallest scales have the fastest growth. The consequences of the chiral anomaly on the evolution 
of helicity were then discussed analytically, where it was demonstrated how it leads to the creation of helical magnetic fields from fields 
which are initially non-helical, and also how it naturally leads to an inverse transfer of energy from small to large scales. At the end 
we also discussed some problems related to energy considerations in the context of the chiral MHD equations, discussing how the total energy 
needs to be conserved and how the chiral fermions related to the anomaly effect need to be treated consistently as a part of a system reaching the thermodynamic equilibrium 
with other particles. 
\chapter{Chiral MHD in the early universe}\label{chirea}
\section{Magnetic fields in the early Universe}
The very high conductivity of the cosmic medium, both in our epoch and the early Universe \cite{pro,baym}, makes the magnetohydrodynamic approximation 
suitable for the description of electrodynamic processes in these regimes. The electric field can therefore be approximately considered as vanishing -- in accordance with the previous discussion
in chapter \ref{magne} -- and the understanding of electromagnetism in the context of astrophysical objects and Universe in its totality is given predominantly in
terms of magnetic fields. This theoretical consideration, however, does not answer the question on which scales, and with which
field strengths do magnetic fields exist in the Universe -- it could for instance be possible that they are also mostly negligible, especially
on larger cosmological scales, and existing only around stars and planets due to their internal physical processes. This question can 
be answered only by the observations of magnetic fields on various scales of the observable Universe and further empirical investigation of their
properties. The results of these observations speak in favor of a very interesting picture.
Magnetic fields have been observed on almost all scales which were probed:  from stars and planets, to clusters of galaxies and are even inferred to exist in the voids of the inter-
galactic medium \cite{jedan,dva,tri,cetiri,pet, sest, sedam, osam, devet}. Moreover, various measurements of these fields, conducted using the Faraday rotation
and Zeeman splitting methods, imply that typical strengths of magnetic fields in galaxy clusters are
of the order of $10^{-6}$ Gauss, with correlation scales of the order of tens of kpc, and that 
inter-galactic voids can be supposedly characterized by magnetic fields of the order of $10^{-16}$ Gauss. The question which naturally occurs
is which processes and mechanisms could create magnetic fields of such strengths and correlation lengths. A large number of studies which
tried to answer this question assumes that the origin of the observed magnetic fields is astrophysical -- proposing a scenario in which small seed magnetic
fields are generated by galactic currents of charged particles, and then subsequently amplified by a dynamo process \cite{brasubra}. 
However, the already mentioned inferred existence of magnetic fields in the voids of inter-galactic medium seems to suggest that the creation of magnetic fields
was also happening outside the galactic setting. The alternative solution to this question assumes that the origin of those observed magnetic fields
is cosmological, and that cosmologically created seed fields in galaxies could have been later amplified by a dynamo processes. This types of scenarios
require further elaboration on specific types of mechanisms and processes which would then create
the fields of observed strength and correlation lengths in the early Universe. The reason for this is that the cosmological evolution, as predicted by Einstein's general relativity
and standad cosmological $\Lambda$CDM model, cannot explain the creation and amplification of magnetic fields on its own. The conservation
of the stress-energy tensor on the standard FRWL (Friedmann-Robertson-Walker-Lemaître) geometry which describes the largest scales of the Universe requires that magnetic energy, $\rho_{B}$, just decreases, as
$\rho_{B} \sim a^{-4}$, where $a(t)$ is the scale factor of the Universe. The scale factor, $a(t)$, defines the length element, $ds^{2}$, of the FRWL space as
\begin{equation}
ds^{2}= g_{\mu \nu}dx^{\mu} dx^{\nu}=-dt^2 + a^{2}(dx^2 + dy^2 +dz^2), 
\label{fri}
\end{equation}
where we have assumed that the Universe is spatially flat. The description of cosmological dynamics is in the general theory of relativity 
given by solutions of the Einstein equation when applied to geometry described by \eqref{fri}. 
Therefore we first very briefly review some basic mathematical elements which lead to such description (for more details see 
\cite{wald, wein, thorn, caroll}). The Einstein equations read
\begin{equation}
R_{\mu \nu} - \frac{1}{2}R g_{\mu \nu}= k T_{\mu \nu},
\label{ein}
\end{equation}
where $R_{\mu \nu}$ and $R$ are the Ricci tensor and the Ricci scalar respectively, which are determined by the space-time geometry under consideration,
and $T_{\mu \nu}$ is the stress-energy tensor, which models the effects of fields and masses on the considered spacetime. In the case 
of an ideal cosmological fluid in the co-moving frame it can be written as 
\begin{eqnarray}
T_{\mu\nu}=\left(\begin{array}{cccc}
\rho(t)&0&0&0\\
0&-p(t) & 0 & 0\\
0&0 &-p(t) &0\\
0& 0 & 0 & -p(t)
\end{array}\right)\label{energymomentum}\; ,
\end{eqnarray} 
The total energy density, $\rho$ and the pressure density, $p$, are assumed to be related by simple equations of state $p=w \rho$, where 
$w=0$ for solid matter (dust) and $w=1/3$ for radiation.  
The stress-energy tensor needs to be conserved
\begin{equation}
\nabla_{\mu} T^{\mu \nu}=0, 
\label{consse}
\end{equation}
where $\nabla_{\mu}$ is the covariant derivative defined on a given space-time and determined by its geometry and it represents the contribution 
of space-time effects to the standard derivative defined on flat space. For an arbitrary vector $V^{\mu}$ it is given by
$\nabla_{\mu} V^{\alpha}=\partial_{\mu} + \Gamma^{\alpha}_{\mu \beta}V^{\beta}$, where as usual repeated indices imply a contraction. This 
definition can be straightforwardly extended to covariant derivatives of tensors of arbitrary rank. 
Here $\Gamma^{\alpha}_{\mu \beta}$ is called the connection and it describes the change in the definition of derivative depending on specific geometry. 
The most common connection is called Christoffel symbol and it is defined as
\begin{equation}
\Gamma_{ \mu \nu}^{\lambda} = \frac{1}{2}g^{\lambda \tau} \left(g_{ \tau \mu, \nu}+g_{\tau \nu, \mu}- g_{\mu \nu,\tau}\right)\,;\quad  g_{\alpha \beta, \delta} \equiv \frac{\partial g_{ \alpha \beta}}{\partial x^{\delta}}\,, 
\end{equation}
Where the metric tensor, $g_{\mu \nu}$ defines the length element for a specific geometry, $ds^{2}=g_{\mu \nu} dx^{\mu} dx^{\nu}$.
For the Universe described by the metric \eqref{fri}, the Ricci scalar is determined as $R(t)=6H'(t) + 12 H(t)^2$.

The evolution of the Universe, and different matter and radiation components which are its parts, is given by the Friedmann equations --
which are derived from Einstein's equation \eqref{ein} when applied to the spacetime corresponding to \eqref{fri}:
\begin{equation}
H^2= \frac{k}{3}\rho_{tot} 
\label{fridva}
\end{equation}
\begin{equation}
\dot{H}+H^2= - \frac{k}{6}(\rho_{tot} + 3p_{tot}), 
\label{fritri}
\end{equation}
where $\rho_{tot}$ is the total contribution of the energy-density of the Universe, consisting of matter, radiation and (effective) 
dark energy contributions, $p_{tot}$ is the total pressure contribution, $k=8 \pi G$, and $c=1$. 
On the other hand, the conservation of the stress-energy tensor \eqref{consse} applied to the FRWL spacetime\eqref{fri} dictates that the energy density needs to evolve as
\begin{equation}
\rho(t)=\rho_{0}(\frac{a}{a_{0}})^{-3(1+w)},
\end{equation}
where $\rho_{0}$ and $a_{0}$ are the values of energy density and scale factor today. 
By solving the Friedmann equations \eqref{fridva}, \eqref{fritri} during the radiation and matter dominated epoch it follows that the scale factor evolves as $a(t)\sim t^n$, 
with $n=1/2$ for radiation domination and $n=2/3$ for matter domination. Therefore, it is obvious that magnetic fields can only decrease if they are only
determined by the evolution of the Universe as predicted by the standard cosmological model. In cosmological scenarios mostly investigated in the literature
it was usual to consider the setting in which this simple evolution is modified by direct couplings between the electromagnetic quantities and various 
 scalar fields (which are mostly not well motivated) or between the electromagnetic and gravitational sector -- mostly in the context of inflation
 \cite{sobol, sloth, guo, konstantinos, hector, leocamp, soda, bonvin}. The second popular option is not to consider modifications of the usual evolution of electromagnetic fields and the Universe,
 but to analyse the possibilities for magnetogenesis during the cosmological phase transitions, such as the electroweak transition and QCD phase transition, expected to happen in the early Universe.
 Cosmological phase transitions could represent the regimes which are out of thermodynamic equilibrium, and which could also lead to a significant charge separation, and therefore
 to a suitable setting for the explanation of magnetic field creation. The problem with this alternative is that in order to be used
 for the explanation of magnetogenesis, these phase transitions should be of first order, which is characterized by the development of bubbles of the new phase
 in the old phase - which is not the case if the parameters of the Standard model of particles are assumed as valid. In fact,
 unless some extensions of the Standard model are assumed both the electroweak and QCD transitions seem to be of some higher order. \\ \\
 What is common to all these various models is the assumption that magnetic fields were present even during early stages of the evolution of
 the Universe. This assumption becomes even more natural if we take into account the already mentioned presence of magnetic fields on various 
 scales of the Universe today. Moreover, there are no reasons that would prevent the existence of magnetic fields in the early history of
 the Universe, and even the simple scaling relation for the magnetic energy evolution in the FRWL universe discussed above (which in fact still remains
 the only relatively tested and universally accepted relation regarding the cosmological evolution of magnetic fields) suggests that 
 magnetic fields of considerable strength were characteristic for that period. Therefore, this will also be the central assumption in the following
 parts of this work.\\ \\
 The assumption of the existence of magnetic fields in the early Universe naturally leads to the question of their possible description 
 and properties, as well as their influence on other physical processes. The most essential property for this description is that the early
 Universe is characterized by high conductivities \cite{baym}. It then follows that electric fields are small and the MHD description of magnetic fields in the early
 Universe plasma, as described in the chapter on magnetohydrodynamics \ref{magne}, can be considered as a good approximation. In this description, the
 Maxwell equations are also simplified by the requirement of global neutrality of the plasma, i.e. $\nabla \cdot \mathbf{E}=0$ and 
 $\nabla \cdot \mathbf{J}=0$, and the displacement current is also neglected. The effect of magnetic fields on various physical processes
 in the early Universe further constrains possible properties of primordial magnetic fields. One of these effects comes from the
 fact that cosmological magnetic fields act as a source of the cosmic microwave background (CMB) and create
characteristic anisotropy patterns \cite{zeldovich, grish, gasperini}. This constraints the magnitude of cosmological magnetic fields in order to be 
consistent with CMB observations \cite{shira, trivedi, pao, shaw}. The second important effect of magnetic fields on the cosmological
evolution comes from the fact that the energy density of magnetic fields contributes to the total energy density of radiation in the Universe. 
The contribution of magnetic fields therefore enters into the Friedmann equation and determines the expansion rate of the Universe. 
In fact, magnetic fields can not become the dominant energy contribution since, as can be checked by direct computation, an electromagnetically 
dominated universe is not compatible with the symmetric structure of the FRWL spacetime. 
The contribution of magnetic energy to the total energy density of the Universe can have very important consequences on primordial nucleosynthesis, since the expansion rate determines the formation of light 
nuclei and therefore the abundance of chemical elements in the universe, which can be measured with satisfactory 
precision \cite{pri1, pri2, pri3}. This effect can again be used to bound possible values of cosmological magnetic fields to the ones
consistent with the observations \cite{magb1, magb2, magb3}. The results of the cited studies, which analysed the effects of cosmological
magnetic fields on the CMB and primordial nucleosynthesis, can be used to demonstrate that the incompressible MHD approximation, 
that is taking $\nabla \rho=0$ and therefore $\nabla \cdot \mathbf{v}=0$, can be used as a justified approximation in the case of the
early Universe. This can be shown using the fact that fluids can be treated as incompressible if there are no pressure variations, which
could cause a change in the fluid density. The cause of these variations can only come from local influences of magnetic fields, since
the cosmological fluid -- in order to be consistent with FRWL cosmology, needs to be homogeneous and isotropic. For small enough magnetic
fields, the ratio between magnetic pressure and fluid radiation pressure, $B^{2}/(8 \pi p)$ is negligible and pressure variations can
therefore be ignored. It can be argued that $B^{2}/(8 \pi p) \approx 10^{-7}$ and the early Universe plasma can accordingly
be treated as an incompressible fluid to a good approximation \cite{inkomp}. \\ \\
It should be stressed that special care must be taken when approaching the question of the mathematical formulation of electrodynamics
in the cosmological context. As the dynamics of the Universe is explained in the framework of general relativity, which replaces
the old notion of gravitational force with the free movement of test particles on the curved spacetime, the description of electrodynamics
now also needs to be given on a curved spacetime. The formulation of electrodynamics on curved spacetime is guided by one of the fundamental
principles of general relativity -- the principle of general covariance. Since physical laws are assumed to be the same in every system,
the mathematical tensorial structure of equations needs to be unchanged during the transition from one coordinate system to another. This is
possible if the derivatives appearing in Maxwell's equations are replaced by covariant derivatives -- which now take into account
the effects of curvature, since they by their definition do not change the tensorial structure of equations \cite{wald}. 
However, the difficulty lies in the fact that --since the field equations are given in terms of tensor quantities on curved spacetime -- 
the fundamental quantity describing the electromagnetic field is now the electromagnetic tensor, $F^{\mu \nu}$, and not the electric
and magnetic field strengths. This important fact nicely demonstrates the necessary unity of electromagnetic phenomena, described by
a single mathematical object, and also the effects of curved spacetime on the description of electrodynamics (leading to the transformation
of the electromagnetic tensor with respect to the one defined on a flat spacetime). However, for practical needs -- such as in the observation-oriented
analysis and formulation of a simple
system of MHD equations under the assumptions stated above -- it would be suitable to express the equations
in terms of electric and magnetic field rather than the electromagnetic tensor. In order to achieve this it is necessary
to decompose the electromagnetic tensor on the four-vector of electric and magnetic fields, with respect to the observer's geodesic. 
This decomposition was formulated \cite{ellis,barrow, tsagas}, isolating a time direction in some arbitrary spacetime by using the projection
of all physical quantities to the hypersurface orhtogonal to the observer's four velocity, defined as $u^{\mu}=dx^{\mu}/d \lambda$ where 
$\lambda$ is the affine parameter of the considered geodesic, and the four-velocity is normalized: $u^{\mu} u_{\mu}=-1$. Then the electromagnetic
tensor can be decomposed in terms of the four-vector electric and magnetic fields, $E_{\mu}$ and $B_{\mu}$:
\begin{equation}
E_{\mu}=F_{\mu \nu} u^{\nu} 
\end{equation}
\begin{equation}
B_{\mu}=\frac{1}{2} \epsilon_{\mu \nu \rho \lambda}u^{\nu} F^{\rho \lambda},  
\end{equation}
where $\epsilon_{\mu \nu \rho \lambda}$ is the four-dimensional Levi-Civita tensor. In order to consider cosmological magnetic
fields, one can apply this formalism to the case of a spatially flat FRWL universe, where the geodesics of observers are defined
by the co-moving four velocity $u^{\mu}=(1,0,0,0)$. By decomposing Maxwell's equations and defining the electric and magnetic
field three-vectors as spatial components of the respective four-vectors, it can be shown that in the case
of the FRWL spacetime they resemble the flat spacetime Maxwell's equations but with the presence of the scale factor $a(t)$ \cite{kanda}. 
In fact, if the time appearing in the equations is replaced by conformal time, $d \tau/ da$, and all the physical quantities 
are scalled with the conformal factor, $a(t)$, such that $\mathbf{B} \rightarrow a(t)^{2} \mathbf{B}$, 
$\mathbf{E} \rightarrow a(t)^{2} \mathbf{B}$, $\rho \rightarrow a(t)^{3} \rho$, $\mathbf{J} \rightarrow a(t)^{2} \mathbf{J}$, 
$\sigma \rightarrow a(t) \sigma$, $\mu_{5} \rightarrow a(t) \mu_{5}$ etc. then Maxwell's equations have the same form as on the flat
spacetime \cite{kanda,banerjee}. In the remainder of this work we will use this fact and present the equations in their conformal form, 
such that they are suitable for the description of both flat and FRWL spacetime. However, it should be understood that this procedure 
is valid only for the specific case of non-relativistic MHD on the FRWL spacetime. Although the FRWL spacetime is up to now confirmed as a 
good description of the observed Universe on large scales, it is not excluded that some other description could be more valid -- such as
anisotropic models of the universe \cite{ani1,ani2,ani3}, especially in early periods of the cosmological evolution. In the general case,
the above stated conclusion will no longer be valid, and the study of electrodynamics on more general spacetime in terms of
electric and magnetic field needs to proceed from a particular decomposition of the electromagnetic tensor on that spacetime.  
  \section{Chiral MHD around electroweak transition}\label{chirmhd}
In this section our objective will be to investigate the evolution of chiral MHD variables, discussed in the previous sections, in 
the specific framework of the expanding Universe during the period in which it crosses from the symmetric phase -- where electromagnetic and weak interactions
are brought into a synthesis described by the unified electroweak interaction, to the broken phase -- where this is no longer true and 
where the electromagnetic and weak interactions appear as completely unrelated. Since the strategy to study a concrete system has been adopted in this place, we will 
also numerically solve the chiral MHD equations, and compare the obtained solutions with analytical expectations and estimates. Some of the important questions we will ask here
are:  can the chiral magnetic effect lead to a significant enhancement of primordial magnetic fields
around the electroweak transition period, and what is the effect of the electroweak transition on the evolution of chiral asymmetry,
magnetic energy and magnetic helicity? Here we will mostly follow the discussion and results first presented by us in \cite{pa}.
\subsection{Electroweak region MHD}
When temperatures are higher than the temperature of the electroweak transition -- which is on the scale of hundred GeV, the symmetry group
$SU_{L}(2)\otimes U_{Y}(1)$ is believed to be restored, as we discussed previously in the second chapter of this work. 
Electromagnetic and weak nuclear phenomena, which on the usual temperatures characterizing the Universe today appear as completely
separated, with different qualitative and quantitative properties, can on these temperatures existing in the early Universe be described
in the same theoretical framework.
For lower temperatures, which are subsequently reached after the Universe continues to expand and cool down, the symmetry is broken down to $U_{EM}(1)$ group, which corresponds to the existence of ordinary electric and magnetic fields. 
As we have already discussed in the beginning of this chapter, since electrical fields decay due to the high conductivity of the early Universe, the only long-range fields that survive are magnetic ones.
In a similar sense, in the region of the electroweak symmetry, long-range non-Abelian fields decay due to their self-interactions \cite{jaffe}, and the only non-screened 
modes correspond to the $U(1)_{Y}$ hypercharge group \cite{gi}. \\ \\  
Assuming that the electroweak plasma is in a complete equilibrium it can be described by $n_{f}$ chemical potentials, related to the number
of conserved global charges \cite{Giovannini2}
\begin{equation}
 N_{i}=L_{i} - \frac{B}{n_{f}}\,,
\end{equation}
where $L_{i}$ is the lepton number of the \textit{i}-th generation, $n_{f}$ the number of fermionic generations and $B$ the baryon number, which holds strictly only when there is no neutrino mixing. In the absence of chirality flips, i.e. at higher 
temperatures where these processes are out of equilibrium, the number of right-handed electrons is perturbatively conserved and 
one can thus define the corresponding chemical potential $\mu_{R}$. When dealing with lower temperatures, where chirality flipping 
processes -- leading to the non-conservation of chiral states -- need to 
be taken into account, one can then perturbatively add the rate of chirality flipping processes to the equations. Moreover, even in the absence of chirality flips, the number of right-handed 
electrons is not exactly conserved because of the aforementioned Abelian anomaly \cite{gi}. 
Taking this into account, it is possible to obtain MHD equations, giving the evolution of hypermagnetic fields, which
are valid in the electroweak region, before the symmetry was broken. Apart from replacing magnetic fields
with hypermagnetic fields, the Navier-Stokes and continuity
equations stay unaltered. Since the addition of velocity further complicates the equations, and introduces a difficult problem of coupling the field evolution with turbulence, 
we will in the present section assume that characteristic values for the bulk velocity of cosmological plasma are negligible. The first attempts to study the interconnection 
between chiral and velocity effects -- and to analyse the chiral MHD in the turbulent regime -- will be presented in chapter \ref{chirturb}. Limiting our attention
to a velocity-free regime, the full set of MHD equations in the symmetric phase reads \cite{Giovannini2, gi}
\begin{equation}
 \nabla \times\textbf{ B}^{Y}
 = \sigma\textbf{ E}^{Y}- \frac{g'^{2}}{\pi^{2}}\mu_{R} \textbf{B}^{Y}
\label{hmaxwell}
\end{equation}
\begin{equation}
 \partial_\tau\textbf{B}^{Y}
= - \nabla \times \textbf{E}^{Y},
\end{equation}
\begin{equation}
 \nabla \cdot\textbf{ B}^{Y}=0,
\end{equation}
\begin{equation}
 \nabla \cdot \textbf{E}^{Y}=0,
\label{hmaxwell4}
\end{equation}
to which the anomaly equation should also be added
\begin{equation}
\frac{d n_{R}}{d \tau}= \frac{g'^{2}}{4 \pi^{2}}\frac{d h^{Y}}{d \tau} - \Gamma_{s}n_{R},
\end{equation}
where $\Gamma_{s}$ is the chirality flipping rate in the symmetric region, i.e. before the eletroweak transition, and we have introduced the hyper-helicity, $
H^{Y}$, defined analogously to the ordinary one. We can see that this system represents the Maxwell equations for hyperfields, which
also take into account contributions of the effective chiral current, coming from the change in the number of right-handed leptons. We note that here no process producing chiral asymmetry 
are assumed to be active during the electroweak transition, that is $\Pi_{sr}=0$. This is related to the fact that, according to our knowledge, there are no such processes characteristic 
for the electroweak scale which would naturally occur in the context of minimal Standard model. Our assumption is therefore, that for hypermagnetic fields on temperatures slightly above 
the electroweak transition there is some initially non-vanishing chiral asymmetry present -- which could have been produced in the earlier stages of cosmological evolution or that hypermagnetic fields 
are characterized by a non-vanishing magnetic helicity. As we discussed in section \ref{heliind}, if any of those two conditions is satisfied, the evolution of magnetic fields 
will be given by the equations of chiral MHD presented above. \\ \\
It is important to note that in the equations and discussion above we have considered only the
chiral coupling with right-handed leptons, and defined only $\mu_{R}$ and $n_{R}$. If we would also assume the
existence of $\mu_{L}$ in the symmetric phase and assume the presence of left-handed anomalies for the left-handed doublets (such approach being followed in \cite{dvo}),
then the anomaly would couple to left- and right-handed leptons separately instead of to their difference, $\mu_{5}$. 
However, there were a few reasons why this
approach was not adopted here. The main issue is that introducing $\mu_{L}$
would actually violate the equilibrium of five chemical potentials for five conserved charges. 
In this case it would also be necessary to take  into account sphaleron processes, which couple only to left-handed particles, and violate lepton and baryon number. 
This would imply further mathematical difficulties and increase the complexity of the considered equations, and it would make the main 
goal of our study -- the analysis of chiral MHD processes -- much more difficult to reach. 
Therefore, we have followed the approach which assumes that in the symmetric phase only $\mu_{R}$ is non-vanishing, as also assumed in \cite{gi}.
This approach is in accord with scenarios where the baryon asymmetry of the universe comes from leptogenesis and is stored in right-handed 
electrons before the electroweak transition, in which there is no asymmetry between left-handed particles \cite{Davidson} 

The way in which the chemical 
potential depends on the right-handed number density is not trivial and it is related to the number of fermionic generations, Higgs doublets and other features specific to the elementary particles model. In the minimal Standard Model the evolution equation for the chemical potential of right-handed electrons is given by
\cite{gi}
\begin{equation}
\frac{d \mu_{R}}{d \tau}=   \frac{g'^{2}}{8 \pi^{2}}\frac{783}{88}\frac{d h^{Y}}{d \tau} - \Gamma_{s} \mu_{R} . 
\label{muR}
\end{equation}
\subsection{Chirality flipping rates in the symmetric phase }
In order to solve the chiral MHD equations in the symmetric and broken phase we need to determine which processes are giving the
dominant contribution to the flipping rates, $\Gamma_{f}$, in both phases. Since those rates are, as visible from \eqref{chempomod}, having the effect
of suppressing the chiral anomaly influence and they tend to exponentially damp the chiral asymmetry, the importance of modifications to
MHD equations will critically depend on their values and time evolution. In fact, it is essentially the difference of the respective
flipping rates  which leads the effects of the electroweak transition to manifest in the difference of the solutions in both phases. 
This follows from the fact that, independently of the order of the electroweak transition, the difference between the relevant
flipping processes in both phases causes a change in the evolution of $\mu_{5}$, and since $\mu_{5}$ is directly coupled to the
(hyper)magnetic field and the helicity density in the modified MHD equations, this change then introduces a potential difference in the 
evolution of all quantities of interest. \\ \\
We first discuss the dominant processes in the symmetric phase, which have the tendency to change the spin orientation of the particles
involved, and by this virtue they cause a non-conservation of the number of right and left-handed leptons.  
Flipping rates before the electroweak transition are contributed by inverse Higgs decays, such as $e_{L} \bar{e}_{R} \leftrightarrow
\varphi^{(0)}$ and $\nu_{e L} \bar{e}_{R} \leftrightarrow \varphi^{(+)}$, with $\varphi^{(+)}$ and $\varphi^{(0)}$ forming the Higgs doublet. 
The rate of inverse Higgs decay per electron is \cite{Cline, Campbell}  
\begin{equation} \label{eq:Gamma_s}
\frac{\Gamma_{H}}{T} = \frac{\pi}{192 \zeta(3)} h_{e}^{2}  \left(\frac{m(T)}{T}\right)^2,
 \end{equation}
where $m(T)$ is the temperature-dependent effective Higgs mass, $\zeta(x)$ is the Riemann zeta function and $h_{e}$ is the Yukawa coupling for electrons. 
Another contribution is from scattering processes such as $t_{R} \bar{t}_{L} \leftrightarrow
e_{R} \bar{e}_{L}$ (where the scattering of top quark gives the dominant contribution to the reaction rate ). 
We can estimate this rate from the general expression $\Gamma = n\sigma v$, where $n$ is the particle density, $\sigma$ is the cross-section of the process, computed in Ref. \cite{Protecting} and $v$ is the velocity of the particles involved (which at high temperatures can be taken to be of order unity), allowing us to write the rate as
\begin{equation} \label{eq:ttbar}
\frac{\Gamma_{t\bar t}}{T} = \frac{(h_t h_{e})^{2}T^2}{8\pi s}\left[\frac{s^2}{(s-m_H^2)^2+\left(\pi h_t^2s/16\right)^2}+2\right],
 \end{equation}
where $h_t$ is the top Yukawa coupling and $s$ is the Mandelstam variable.
Fig.~\ref{pic:Gamma_s} depicts the chirality flipping rates \eqref{eq:Gamma_s} and \eqref{eq:ttbar} before the electroweak symmetry 
breaking. From the given plots it is visible, and in accord with findings already reported in \cite{Protecting}, that inverse Higgs decays are strongly dominant for higher 
temperatures, but for lower temperatures (corresponding to lower $m_H(T)/T$ values) they become subdominant compared to 
the $t\bar t$ processes. When the reaction rate of the $t\bar t$ processes becomes higher than the rate of the first process, the rate of Higgs inverse decay is 
already only of the order of the Hubble rate. Therefore, one could expect that the corresponding flipping rate will be negligible 
compared to the other terms in the modified MHD equations. 

\begin{figure}[tbp]
  \centering
 \includegraphics[width=0.6\textwidth]{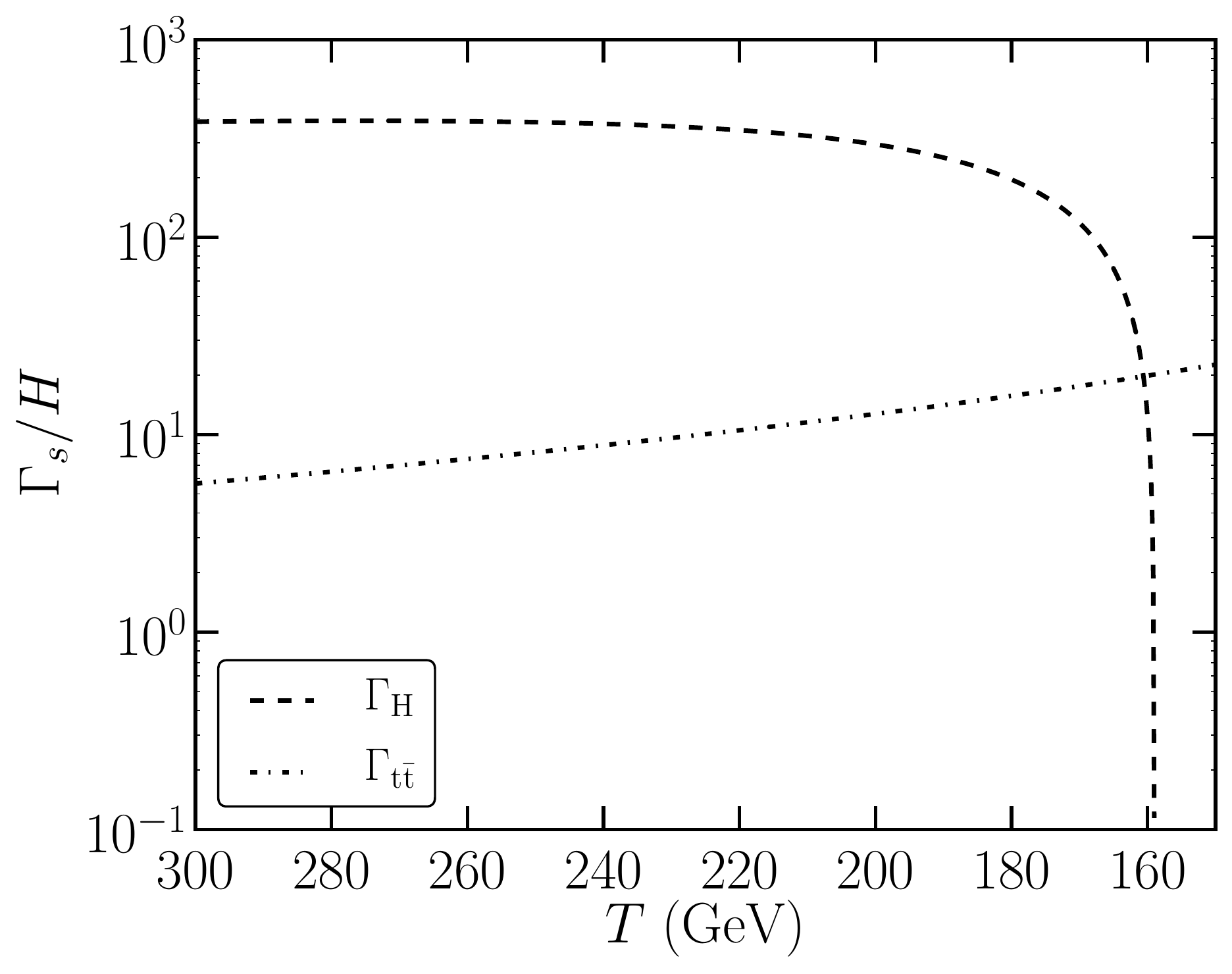}
  \caption{\label{pic:Gamma_s} The evolution of chirality flipping rates in the electroweak phase coming from the Higgs decay processes and $t\bar t$-scattering, both  normalized to the Hubble rate. Adopted from \cite{pa}}
\end{figure}

The evolution of the effective Higgs mass is not a trivial question, since its functional dependence on temperature will be related 
to the type of the phase transition and therefore on the model of elementary particles. Different additions and modifications to the Standard model, by changing 
the number of particle species -- and thus their total mass, introducing new scalar fields or variations of couplings, can change not only the numerical details of this evolution, but also
the type of the phase transition -- for instance, making it of the first order. For some recent proposals in this direction see \cite{trans1, trans2, trans3, trans4}. With a desire 
to keep our discussion as independent of assumptions characteristic for different models as possible, and being interested primarily in the electrodynamics on the electroweak scale, we 
will consider the most conservative case of the electroweak transition in the minimal Standard model. 
For this reason we follow the results of lattice simulations \cite{D'Onofrio:2014kta} regarding the order and temperature of the transition in the Standard Model, which is found to be a cross-over. 
On the other hand, in order to obtain an analytical estimate of the dependence of the thermal Higgs mass on the temperature, needed to compute the rate of Higgs inverse decays, we use the approximation of 1-loop Higgs potential below. 
The Higgs mass evolution
can be determined from the effective thermal Higgs potential in the high temperature limit to be \cite{Dine, Weinberg} 
\begin{equation} \label{eq:higgspot}
 V(\phi,T) = D(T^{2} - T_{0}^{2})\phi^{2} - ET \phi^{3} + \lambda\frac{\phi^{4}}{4}\,.
\end{equation}
 Parameters $D$, $E$ and $\lambda$ depend on the details of the particle model, where $E$ is of special importance since
it determines the order of the phase transition. 

The Higgs mass is then determined by
$m^{2}=(d^{2}V/ d \phi^{2})\mid_{ \phi=v}$, where $v$ is the Higgs expectation value obtained by
$(d V/ d\phi)\mid_{\phi=v}=0$. In the Standard Model, the parameters on Eq. \eqref{eq:higgspot} take the form
\begin{equation}
D=\frac{1}{8v_{0}^{2}}\left(2m_{W}+m_{Z}^{2} + 
2m_{t}^{2}+\frac{m_{H}^{2}}{2}\right)\,, 
\end{equation}
\begin{equation}
E=\frac{1}{4 \pi v_{0}^{3}}(2m_{W}^{3} + m_{Z}^{3})\,,
\end{equation}
\begin{equation}
 T_{0}^{2}=\frac{1}{4D}m_{H}^{2}\,,
\end{equation}
\begin{equation}
\lambda=\left(\frac{m_{H}}{2v_{0}}\right)^{2}\,,
\end{equation}
with $v_{0} \approx 246$ GeV, $m_{H} \approx 125$ GeV and $ 2D \approx 0.38$. From these expressions, 
the Higgs mass changes with temperature as $m(T)^{2}=2D\left(T^{2} - T_{0}^{2}\right)$ and smoothly goes to zero
at $T=T_{0}$. For a first order phase transition, possible for extensions
of the Standard Model, the Higgs mass is changing in the same fashion for $T>T_1$, with
\begin{equation}
 T_{1}=\frac{8 D \lambda T_{0}^{2}}{8 D \lambda - 9 E^{2} }
\end{equation}
and then instead of going to zero, reaches
\begin{equation}
 m^{2}_{T_1}= 2D(T_{1}^{2} - T_{0}^{2}) - \frac{9E^{2}T_{1}^{2}}{4 \lambda}
\end{equation}
 at $T=T_{1}$. For $T=T_{0}$ one has $m^{2}= 9E^{2}T_{0}^{2}/\lambda$ and for $T<T_1$
\begin{equation}
 m(T)^{2}=2D(T^{2}- T_{0}^{2}) - 6ETv + 3 \lambda v^{2},
\end{equation}
with 
\begin{equation}
v= \frac{3ET \pm \sqrt{9E^{2}T^{2} - 8D \lambda \left(T^{2} - T_{0}^{2}\right)}}{2 \lambda}.
\end{equation}
According to the now known value of the Higgs mass and to the results of non-perturbative
techniques, electroweak symmetry breaking in the Standard Model is of higher order than second \cite{Kajantie}. 
In some other extensions, such as in the Neutrino Minimal Standard Model, not only the order of the transition could be changed but also the physically viable values for the initial asymmetry between right- and left-handed particles.

From this dependence one can determine what are the temperature ranges at which chirality-flipping processes are negligible, demanding that the critical temperature obeys $\Gamma_{s}(T_{\rm out})/H(T_{\Gamma}) \approx 1$, where $H(T)$ is the Hubble parameter in the radiation dominated period given by $H\simeq1.08 \sqrt{ g_*/10.75}(T^2/M_{Pl})$, with $M_{Pl}$ being the Planc mass and $g_*$ the number of relativistic degrees of freedom, and we consider $\Gamma_s = \Gamma_H + \Gamma_{t \bar t}$ from here onwards. From here it follows that chirality-flipping processes are
out of equilibrium in the symmetric region for temperatures $T> T_{\rm out,1} \approx 2D \cdot 80$ TeV as negligible for temperatures $T_{0} < T<T_{\rm out, 2} \approx 159.5$ GeV.
Therefore, as the temperature in the symmetric region falls and approaches $T_{0} \approx 159$ GeV, chirality-flipping processes are becoming less significant.

\subsection{Chirality flipping rates in the broken phase}
When the Universe during its expansion reaches the temperatures around $T_{0}$ it is believed that the electroweak symmetry gets broken,
and hyperfields get replaced with ordinary electromagnetic fields, while leptons acquire finite masses. The boundary condition between the hyperfields and electromagnetic fields
is given in terms of the Weinberg angle, $\theta_{W}$
\begin{equation}
\textbf{B}=\textbf{B}_{Y} \cos \theta_{W}\,.
\label{boundary}
\end{equation}
What is also significant for our analysis is that, instead of Higgs inverse decays, the dominant contribution to chirality flipping rates
now comes from the weak and electromagnetic scattering processes, determined by the acquired mass of leptons. 
Taking into account the scaling of the respective cross-sections with the temperature gives the following rates
 \begin{equation}
\frac{\Gamma_{\rm w}}{T} \approx G_{F}^{2}T^4\left(\frac{m_{e}}{3T}\right)^{2}
\label{fliping1}
\end{equation}
\begin{equation}
\frac{\Gamma_{\rm em}}{T} \approx \alpha^{2} \left(\frac{m_{e}}{3T}\right)^{2}, 
\label{fliping2}
\end{equation}
where $G_{F}$ is the Fermi constant and $\alpha$ the fine-structure constant.
We depict both chirality flipping rates in the broken phase in Fig.~\ref{pic:Gamma_b}, and also plot the corresponding sum $\Gamma_{\rm tot} = \Gamma_{\rm em}+\Gamma_{\rm w}$, which we will be denoting as $\Gamma_b$ in the remainder of this section. 
Taking into account the discussion on the chirality flips in the symmetric and broken phase we can conclude that just before the electroweak transition, 
around  $T_0 < T<T_{\rm out, 2}$, the effect of flips should be reduced, and therefore the dissipation of the chiral assymetry should not be significant. However, for $T<T_{0}$, after the transition is completed, the total rates become relatively high. 
We can thus expect that these described consequences of the electroweak transition -- not just changing the nature of fields and masses of particles, but also the dominant chirality filipping reaction rates -- manifest in the solution of chiral MHD equations, 
and therefore cause significant changes in the evolution of chiral asymmetry.
\begin{figure}[tbp]
  \centering
  \includegraphics[width=0.6\textwidth]{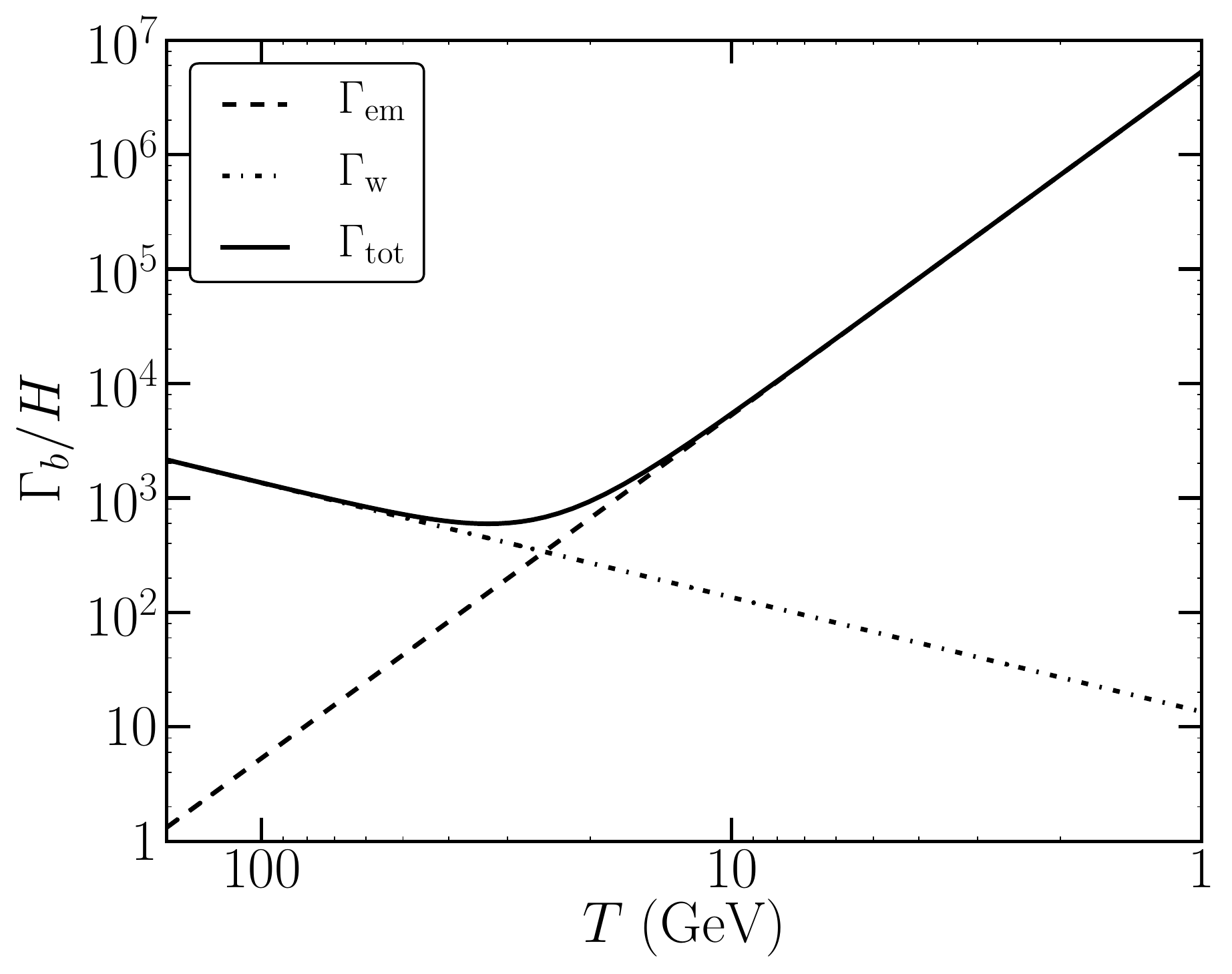}
  \caption{\label{pic:Gamma_b} The evolution of chirality flipping rates normalized to the Hubble rate in the broken phase. The total reaction rate,  $\Gamma_{\rm tot}$, is consisting of the contribution coming from electromagnetic, $\Gamma_{\rm em}$, and weak interactions $\Gamma_{\rm w}$. Adopted from \cite{pa}}
\end{figure} 
\\ \\
Before continuing the discussion of the implications of such a scenario on solutions of the MHD equations, we should stress that the 
presented picture is not entirely correct, and that the complexities of the actual processes happening have been oversimplified. This 
comes from the basic assumption that the transition of hyperfields into electromagnetic fields happens abruptly at $T_{0}$. In this picture,
the mixing angle, $\theta_{W}$, would change discontinuously from some non-vanishin value for $T>T_{0}$ to zero at $T=T_{0}$. 
That sort of discontinuous description can not essentially be in accord with the electroweak transition as a smooth crossover and the fact that gauge bosons also become massive due to the interaction with plasma, and not only due to the interactions with Higgs boson. Indeed, this
was also confirmed by analytical calculations and lattice simulations \cite{cross1, cross2}. The mixing angle will change continuously
during the transition rather than abruptly, with its values still being different from zero at temperatures lower than $T_{0}$. However,
the precise functional dependence of $\theta_{W}$ is still not proposed in the literature. This issue was addressed in \cite{kamada2},
in the context of the baryon asymmetry evolution around the electroweak crossover, where the baryogenesis is explained as a result of the change
in the baryon number that comes as a result of a coupling with hypermagnetic fields.
These changes in baryon, $Q_{B}$, and lepton number, $Q_{L}$ are determined by
\begin{equation}
\Delta Q_{B}= \Delta Q_{L}=N_{g} \Delta N_{CS} - N_{g}\frac{g'^{2}}{16 \pi^2} \Delta h_{Y}, 
\end{equation}
where $N_{g}=3$ is the number of fermionic generations, $N_{CS}$ is the Chern-Simons number and $g'$ is the gauge coupling. 
In this study simple phenomenological models were assumed
in order to model the temperature dependence of $\theta_{W}$, and it was shown that the actual effects of a continuous transition
tend to support the baryon asymmetry and lead to its higher values (since the hypermagnetic field which sources the baryon asymmetry is
active for a longer period of time). To conclude, we should have on mind that an instantaneous transition from symmetric to broken phase considered here is 
just an approximation, and the potential discontinuous properties of the obtained solutions should be understood as a relic of such assumption.  
\subsection{Analytical discussion}
Although the equations for the evolution of the magnetic field and the chiral asymmetry can not be solved analytically, their analysis in 
simplified regimes can still help us to better understand the tendencies we expect from these solutions during the electroweak crossover.
We will later see, comparing our conclusions with numerical results, that using this simplified analytical discussion it is possible to reach
some interesting conclusions. To obtain the full evolution we need to consider the set of chiral MHD equations, where we neglected the 
contributions of velocity and assumed no chiral asymmetry source terms, both before and after the crossover:
\begin{itemize}
 \item $T_{0}  <T< 10~{\rm TeV}$
\begin{equation} \label{eq:rhoks}
\frac{d\rho^{Y}_{k}}{d \tau}= -\frac{2k^{2}}{\sigma_{s}}\rho^{Y}_{k} - \frac{g'^{2}}{2 \pi^2} \frac{k^{2} \mu_{R}}{\sigma_{s}} h^{Y}_{k}
\end{equation}
\begin{equation}
\frac{d h^{Y}_{k}}{d \tau}= -\frac{2k^{2}}{\sigma_{s}} h^{Y}_{k} - \frac{2g'^{2}}{ \pi^2} \frac{ \mu_{R}}{\sigma_{s}} \rho^{Y}_{k}
\label{hiper-druga}
\end{equation}
\begin{equation} \label{eq:muRmodes}
\frac{d \mu_{R}}{d \tau}=   \frac{g'^{2}}{8 \pi^{2}}\frac{783}{88} \frac{d h^{Y}}{d \tau} - \Gamma_s \mu_{R} 
\end{equation}
\item $T<T_{0}$
\begin{equation} \label{eq:rhokb}
\frac{d \rho_{k}}{d \tau}=- \frac{2k^{2}}{\sigma_b} \rho_{k} - \frac{e^{2}}{16 \pi^2} \frac{k^{2} \mu_{5}}{\sigma_b}h_{k}
\end{equation}
\begin{equation} 
\frac{d h_{k}}{d \tau}=- \frac{2k^{2}}{\sigma_b} h_{k} - \frac{e^{2}}{4 \pi^2} \frac{ \mu_{5}}{\sigma_b} \rho_{k}
\label{druga}
\end{equation}
\begin{equation}
 \frac{d \mu_{5}}{d \tau}=  
 \frac{3e^2}{4 \pi^2}\frac{d h}{d \tau} -\Gamma_b \mu_{5}.
\label{mub}
\end{equation}
\end{itemize}
Here we used the approximation that the transition happens instantly at $T_{0}$, which was discussed in the last subsection. We need to
demand that during its evolution the chiral potential stays continuous, while fields need to satisfy the boundary condition \eqref{boundary}, while
they cross from electroweak to broken phase. One of the important physical quantities which change during the transition is the value of conductivity, changing from $\sigma_{s}$ to $\sigma_{b}$. An exact determination of conductivities which characterize the Universe
around the electroweak phase is not a simple task. The conductivity is influenced by the dominant elementary particle processes 
characteristic for each phase -- for example, the interactions between leptons, $W^{\pm}$ and $Z^{0}$ reduce the conductivity in the symmetric phase
with respect to the broken phase. In reality the conductivity should, as well as the value of the mixing angle, smoothly change during
the electroweak transition -- so the representation of a sudden change from $\sigma_{s}$ to $\sigma_{b}$ must also be understood as a simplified picture.
Still, the difference of conductivities in respective phases does not appear to be significant, 
and can be approximated by $\sigma_{s} \approx \sigma_{b} \cdot \cos^{4} \theta_W$ \cite{baym}. Due to the fact that equations before
and after the electroweak transition are essentially the same in their mathematical form, just differing in the values of coefficients
and respective functions, we can use the following notation in order to discuss them in general 
\begin{equation}
\begin{split}
c_{1}= \frac{g'^{2}}{2 \pi^2} \frac{1}{\sigma_{s}}, & \; \; c_{2}=\frac{e^{2}}{16 \pi^2} \frac{1}{\sigma_{b}}, \\  
c_{3}= \frac{g'^{2}}{8 \pi^{2}}\frac{783}{88}, & \; \; c_{4}=\frac{3 e^2}{4 \pi^2}.
\end{split}
\end{equation}
By inspecting the evolution equation for the chiral potential \eqref{chempomod} we can see that the time change of helicity of the same sign tends
to support the asymmetry growth, contrary to the effects of chirality flips. We can therefore first make an estimate of the necessary 
initial magnetic energy density in order to prevent fast damping of $\mu_{5}$ even before it reaches the electroweak crossover in its evolution. 
If $\mu_{5}$ needs to stay approximately constant then $\mu_5\cdot \Gamma_s/(c_3|d h\ d \tau|)\simeq 1$. To simplify the analysis
we assume that the initial field configuration is maximally helical, and we also approximate the spectral distribution in the following
fashion
\begin{equation}
h_{k}(\tau)=h(\tau)(\frac{k}{k_{max}})^n,
\end{equation}
for $k\leq k_{max}$, where $k_{max}$ corresponds to the shortest length scale. In reality, the time evolution of different modes will
also depend on the scale of a given moment, and it will not be possible to decouple the time and spectral dependencies. However, being
interested in obtaining the approximate expressions for the initial magnetic field strengths necessary to avoid damping of $\mu_{5}$, we
here neglect the details of the mode evolution. This is justified since we are in this place interested in the properties of total helicity
distribution, and not the detailed behavior of its spectra. Taking these assumptions and integrating \eqref{hiper-druga} we get
\begin{equation} \label{eq:h_dampthr}
h \simeq \frac{\mu_5 \Gamma_s}{c_3}\frac{|k_{max}^{n+1}-k_{min}^{n+1}|}{(n+1)\left[-\frac{2}{\sigma_s(n+3)}(k_{max}^{n+3}-k_{min}^{n+3})-\frac{g'^2 \mu_5}{\pi^2\sigma_s(n+2)}(k_{max}^{n+2}-k_{min}^{n+2})\right]}
\end{equation}
and also
\begin{equation} \label{eq:rho_dampthr}
 \rho_m \simeq \frac{(n+1)}{2(n+2)}\frac{k_{max}^{n+2}-k_{min}^{n+2}}{k_{max}^{n+1}-k_{min}^{n+1}}h.
\end{equation}
So we see that the required initial helicity and field strength needs to be proportional to the chiral asymmetry potential, flipping rates and the
difference of maximal and minimal wavenumber characterizing the spectral distribution of modes in the system. 
In a different regime, if the term related to the  change of helicity in \eqref{chempomod} is small, we have $(d h/ d \tau) \approx 0$.This can happen if the initial
value of helicity is negligible or vanishing, or on the other hand if the conductivity is very high -- close to the ideal MHD limit, suppressing any change in helicity.
Then the evolution of $\mu_{5}$ will simply be given by an exponential decay
\begin{equation} 
\mu_{5}\approx \mu^{0}_{5} \exp\left(-\int \Gamma_{s,b}(\tau) d\tau\right), 
\label{trnj}
\end{equation}
The contrary regime is achieved if the fields are strongly helical. Then $\mu_{R,5} \cdot \Gamma_{s,b}/(c_{3,4} |d h/d\tau|) \ll 1$,
and we can approximately ignore the term involving the flipping rates. In this case the chiral asymmetry and helicity are simply
related $\mu_{R,5} \approx c_{3,4} h +c$, where $c$ is a constant set by the initial values of $\mu_{R,5}$ and $h$. Proceeding with the 
same assumptions we used to obtain the expression for the minimal helicity value needed to prevent the fast decay of $\mu_{R,5}$, we can solve
the corresponding differential equations which yields
\begin{equation}
\mu_{R,5} \approx \frac{1}{2\beta}\left[\delta \cdot \tanh \left(\frac{\epsilon \tau}{2}\right) -\gamma\right],
\end{equation}
\begin{equation}
\rho_{m} \approx\frac{1}{2 c_{1,2}} \frac{\beta}{ c_{3,4}}[\mu_{R,5} - c],
\end{equation}
with the coefficients
\begin{equation}
\begin{split}
\beta=\frac{n+1}{n+2} c_{1,2} \frac{k_{max}^{n+2} -k_{min}^{n+2} }{k_{max}^{n+1} -k_{min}^{n+1}}, &\; \;\; \epsilon=\frac{2(n+1)}{(n+3) \sigma_{s,b}}  \frac{k_{max}^{n+3} -k_{min}^{n+3} }{k_{max}^{n+1} -k_{min}^{n+1}}, \\
\delta=\sqrt{(b \cdot c - e)^{2} + 4b\cdot  c \cdot e}, &\; \;\; \gamma=b \cdot c - e ,
\end{split}
\end{equation}
with $k_{min}$ denoting the largest length scale. We see that this represents a regime in which the asymmetry chemical potential grows
due to its coupling to the hypermagnetic or magnetic field. It is reasonable to assume that the conditions for the viability of such 
regime could be fulfilled near the electroweak transition and also for $T> T_{\rm out,1}$. After the electroweak transition, we expect that the evolution of $\mu_{5}$ will approach exponential damping,
given by \eqref{trnj}. However, due to the details of the dependence of flipping rates on time in the early Universe around the electroweak scale,
the regime of exponential damping will not be reached directly after the electroweak crossover. In fact, the chiral asymmetry can be expected to 
even grow at the beginning of the phase of broken electroweak symmetry. This happens due to the fact that the flipping rates we consider
actually decrease for some period of time in the broken region, after they have initially significantly increased in comparison
to the rates in the symmetric phase, until they reach their minimum. This minimum can be calculated using \eqref{fliping1} and \eqref{fliping2},
which leads to $T(\mu_{5}^{max})\approx(\alpha/G_{F})^{1/2} \sim$~GeV. 

\section{Numerical solutions}\label{numerics}
In the last section we have analytically discussed some aspects of the modified MHD equations that we expect to see around the
electroweak crossover, using various simplifications and obtaining approximate results. Complementary to that discussion, here
we present numerical solutions of the system of differential equations \eqref{eq:rhoks}-\eqref{mub}, which describe the evolution of chiral MHD 
quantities in the velocity free regime both before and after the electroweak transition. The results we present in this section were originally reported in
\cite{pa}. The main questions of interest in this study are what are the main properties of (hyper)magnetic fields
on the electroweak scale, and moreover could the electroweak crossover lead to some significant consequences on the evolution
of magnetic energy, helicity and chiral asymmetry. We will also see that numerical solutions of \eqref{eq:rhoks}-\eqref{mub} confirm 
the main conclusions derived in the last section. 

In order to present the evolution of chiral MHD quantities shortly before and after the electroweak transition, we choose
the initial temperature to be $T=300$ GeV and numerically solve the equations until the MeV scale. It needs to be stressed
that the value of the initial chiral asymmetry is a completely unknown quantity and it essentially remains as a free variable when modeling
the MHD processes on the electroweak scale. Using our current knowledge we can only conclude that on lower temperatures, for instance on the MeV scale, the chiral
asymmetry -- and thus any practical relevance of the chiral anomaly effect -- will be strongly suppressed due to the growth of flipping
rates, as we discussed in the previous section. However, this does not give any insights on the viable values of $\mu_{5}$ during
the electroweak era, and these values will be highly dependent on specific models which lead to the creation of asymmetry between right-handed 
and left-handed particles. The only definite condition we can impose on the initial values of $\mu_{5}$ is that the energy stored
in the chiral asymmetry must be smaller than the total radiation energy density, $\rho_{\rm tot}= \pi^2 g_*T^4/30$ in order that it
does not lead to any significant departures from the observationally confirmed dynamics of cosmological expansion. This leads to the 
condition $\mu_{5}/T \ll 1$ which necessarily needs to be satisfied. The concrete value we will use in our simulations was taken based on a rough estimate
linking this asymmetry with the baryon asymmetry of the Universe, assuming that similar processes could be involved in their production. 
We therefore take $\mu_{5}(300 GeV)=10^{-9} T$, which can also be regarded as a rather conservative value, which does not 
assume an unrealistically high significance of the chiral magnetic effect in the early Universe. The similar issue of arbitrariness
is also present in choosing the initial value for the energy stored in the hypermagnetic field. Although the existence of hypermagnetic fields
on cosmological scales seems rather natural -- if we can base our reasoning on the observed presence of magnetic fields on astrophysical scales today -- we
currently don't know if this was really the case, and if magnetogenesis did indeed happen before the electroweak transition, the values and 
spectral properties of these fields will strongly depend on the specific magnetogenesis model. Again, an obvious constraint can be made
requiring that the magnetic energy needs to be smaller than the total radiation energy. For parameterizing the relative strengths of
magnetic fields we introduce the ratio between the magnetic energy density and total energy density $\Omega_{\rm mag} = \rho_m/\rho_{\rm tot}$. 
We also need to assume an initial spectral distribution, which we choose to be given by a simple power-law dependence 
$\rho_k^{0}/\rho_{\rm tot}=\Omega_{\rm mag}^0 k^5/(k_{max}^5-k_{min}^5)$, with  $k_{min} < k < k_{max}$. For a continuous spectrum
the system given by \eqref{eq:rhoks}-\eqref{mub} would imply it is necessary to solve an infinite number of coupled differential equations. 
This is of course not possible to achieve numerically, so the spectrum needs to be approximated by a discrete spectrum, such that separation
between steps becomes small. Therefore, the corresponding system of chiral MHD equations was solved by approximating it to 
a discrete spectrum given by $i=1,10$ modes $k_i =k_{min} 2^{i-1}$, with the chosen $k_{min}/T=10^{-10}$. 
One of the important questions in this system is the evolution of magnetic helicity, and its interdependence with other variables.
Therefore in every simulation we have taken two different initial values for helicity, covering the opposite initial settings --
no initial helicity present (which we depict in green color) and fields which are initially maximally-helical, 
$h_{max}(k)=2 \rho_k^{(Y)}/k$ (which we depict in red color).

\subsection{Chiral asymmetry chemical potential}  \label{sec:mu5}

\begin{figure}[tbp]
  \centering
 \includegraphics[width=.53\textwidth]{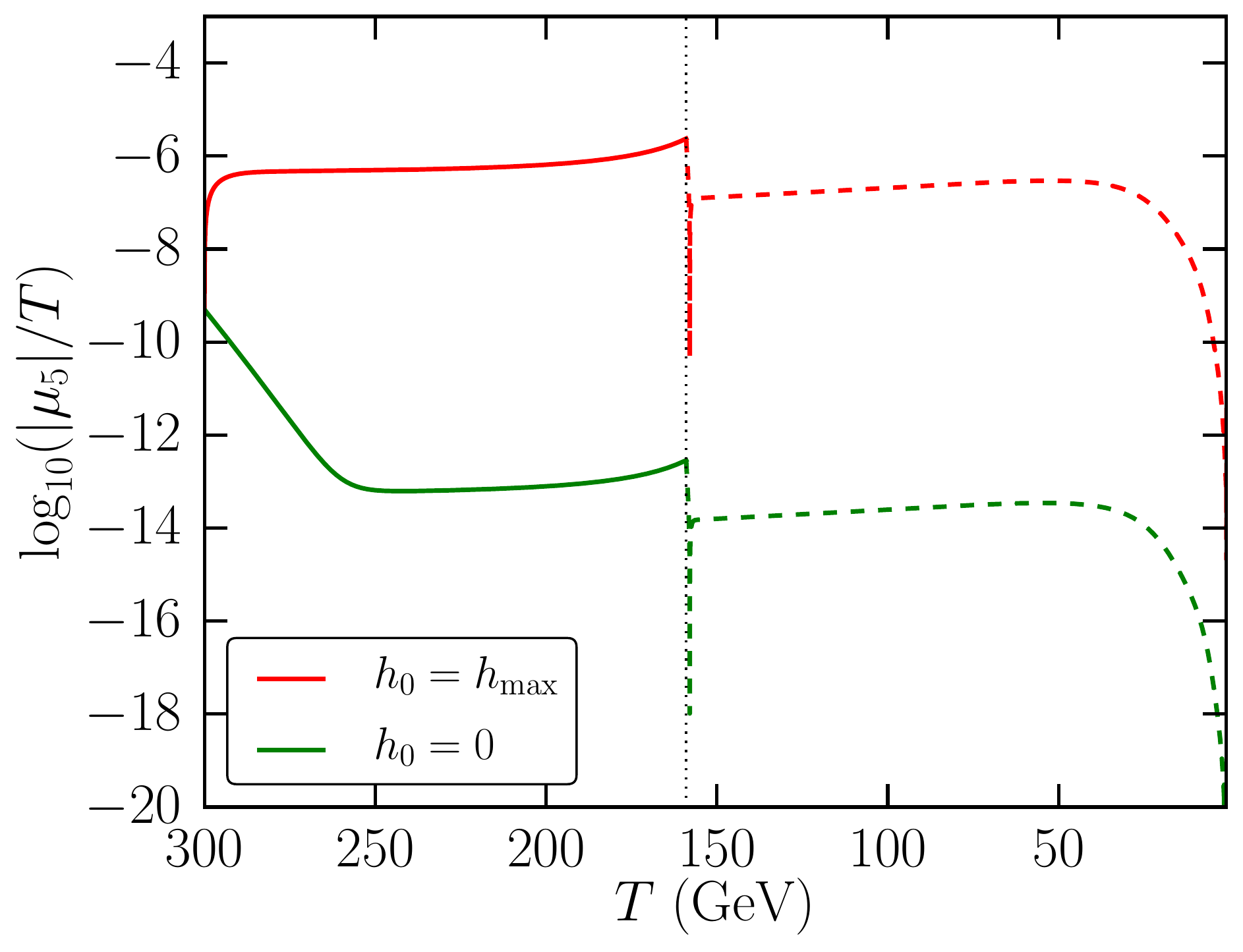}
 \hfill
 \includegraphics[width=.50\textwidth]{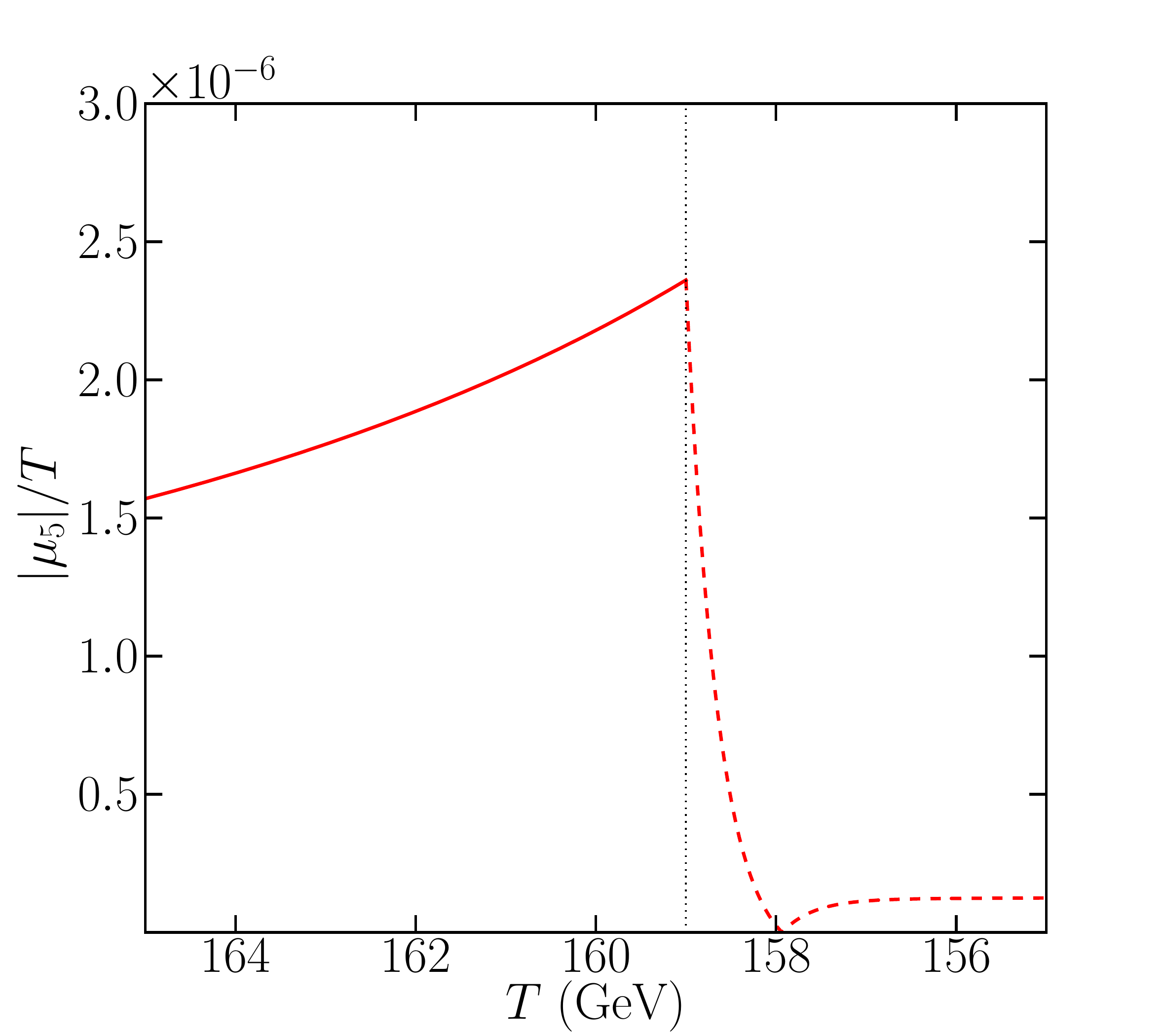}
 \caption{ \label{pic:mu5} Evolution of the logarithm of the chiral chemical potential $\log_{10}(|\mu_{5}|/T)$ with temperature, before (solid lines) and after (dashed lines) the electroweak transition, with $\Omega_{\rm mag}^0=10^{-10}$ and $\mu_5^0/T=10^{-9}$, beginning at $T=$300 GeV, for the minimal initial helicity density $h^Y_0=0$ (in green) and maximal $h^Y_0=h_{\rm max}$ (in red), on the left-hand side. Zoom around the transition for the initially maximal helical case on the right-hand side. Adopted from \cite{pa}}
\end{figure}

In Figure~\ref{pic:mu5} we show the evolution of the electron chiral chemical potential, where we have considered both the 
initially non-helical and maximal helical hypermagnetic fields. The most emphasized feature of the behavior of the chiral asymmetry
potential is its very strong decrease which happens around the transition temperature. As we noted in the last section we need to have
on mind that the apparent strong jump in $\mu_{5}$ at the transition temperature is partially due to the underlying assumption of our model, 
that the transition between the symmetric and broken phase happens suddenly, instead of over a continuous range of temperatures. This limitation will, 
however, mostly influence only the rates of changes of the considered quantities, and not the overall picture of processes. This change of the chiral
asymmetry shows how can the electroweak transition possibly lead to significant influences on the electrodynamics of the early Universe,
even if is not of the first order -- and therefore its influence is not direct (like it happens in the first order transition processes, which can
lead to turbulence and charge separation due to the formation of bubbles of the broken phase in the symmetric phase), but it manifests
via changes in the behavior of $\mu_{5}$. If we compare Figure~\ref{pic:mu5} with the analytical discussion in the last section, we
see that they are in accordance. The evolution of the chiral asymmetry potential has the following basic features, which were already predicted
in the last section. First, beginning from some temperature above the electroweak transition and 
taking $\mu_{L}=0$, the asymmetry potential, $|\mu_{5}|=|-\mu_R/2|$, grows while the temperature approaches $T_{0}\simeq 159$ GeV due to the 
decrease of the chirality flipping rate $\Gamma_s$, which can be checked in
 Fig.~\ref{pic:Gamma_s}). When the transition temperature is reached (represented by the 
vertical line), $|\mu_{5}|$ suddenly falls due to the activation of the weak interaction and electromagnetic spin flipping 
processes, determined now by the non-vanishing electron mass, as described by \eqref{fliping1} and \eqref{fliping2}. Next stage happens
after the completion of the transition, and now the chiral potential continues to grow slowly, until it reaches its local maximum in the 
broken phase, which happens around 40 GeV. Finally, due to the increase of weak and electromagnetic flipping rates, the evolution of
$\mu_{5}$ approaches to the regime of exponential decay for lower temperatures. Comparing the evolution of $\mu_{5}$ in the cases
of initially maximally helical field and initially vanishing helicity, we see how the initial field configuration has a huge impact
on the subsequent values of the chiral potential -- and this results in a difference of six orders of magnitude.

\begin{figure}[tbp]
  \centering
 \includegraphics[width=0.6\textwidth]{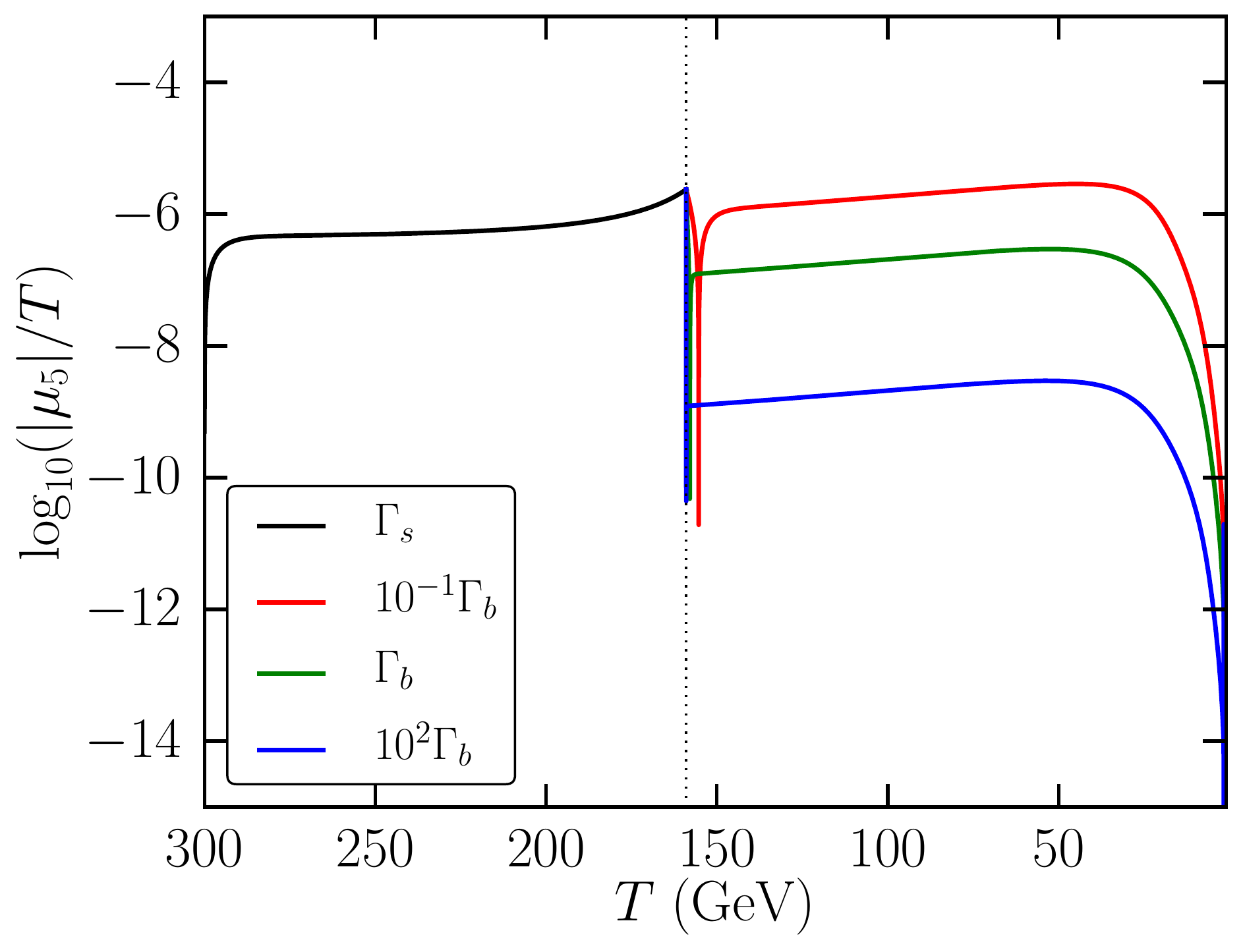}
 \caption{ \label{pic:mu5_gamma} Evolution of the logarithm of the chiral chemical potential $\log_{10}(|\mu_{5}|/T)$ with temperature, with $\Omega_{\rm mag}^0=10^{-10}$ and $\mu_5^0/T=10^{-9}$, for the maximal initial helicity density $h^Y_0=h_{\rm max}$, for different modified values of chirality flipping rates in the broken phase. Adopted from \cite{pa}}
\end{figure}

One of the potential shortcomings of the model for the chiral MHD around the electroweak transitions we considered could be that the 
estimates for the flipping rates values shortly after the transition. In order to demonstrate that our analysis does not qualitatively
and essentially depend on the values of the flipping rates just after the transition in Fig.~\ref{pic:mu5_gamma} we model this imprecision
by varying the values of flipping rates over several orders of magnitude, which shows that the presented general picture always stays the same.

\subsection{Evolution of the magnetic energy} \label{sec:rho_m}

In accord with the main conclusions of our analytical discussion in the previous section, the regime of chiral MHD evolution will be
determined by the strength of chirality flips, and initial values of magnetic energy and chiral asymmetry -- where the last two
quantities remain as free parameters. In particular, if the initial value of magnetic energy, given by its ratio to the total radiation
energy density -- $\Omega_{\rm mag}^0$ -- is too small then the $\mu_{5}$ evolution will tend to enter into the regime of fast decay, 
and therefore the subsequent difference between chiral MHD solutions and the one describing the regular MHD will be negligible. 
On the other hand, in order not to significantly influence the expansion of the Universe, magnetic fields need to satisfy 
$\Omega_{\rm mag}^0<1$ 

\begin{figure}[tbp]
  \centering
 \includegraphics[width=0.6\textwidth]{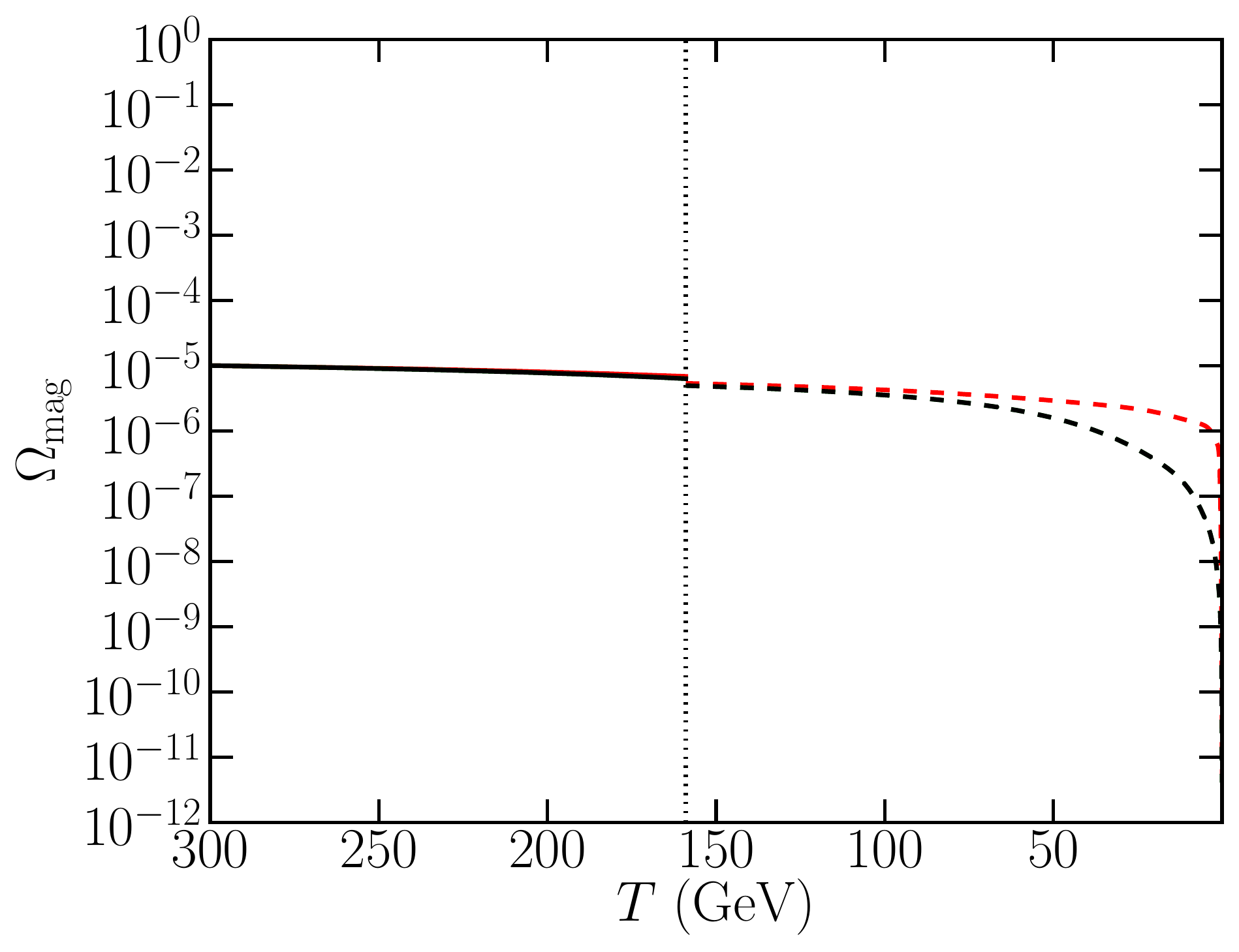}
 \caption{\label{pic:energy}  Evolution of the magnetic energy density normalized to the total energy density, $\Omega_{\rm mag}^Y$ before (solid lines) and $\Omega_{\rm mag}$ after (dashed lines) the electroweak phase transition with respect to temperature for $\Omega_{\rm mag}^0=10^{-5}$ and $\mu_5^0/T=10^{-9}$, in red. The curves in black represent the evolution of the magnetic energy density in the absence of $\mu_5$. Adopted from \cite{pa} }
\end{figure}

The red curve in Figure~\ref{pic:energy} depicts  the magnetic energy evolution in the region preceeding the electroweak phase transition. 
The curve in black shows the magnetic energy evolution in 
case the chiral asymmetry was not present, i.e. if magnetic fields would only be subject to resistive damping.  
We can note that there is no significant growth of magnetic energy in both regions, not even in the short interval 
around the phase transition. This conclusion would also follow from a very simplified estimate, since neglecting chirality flipping
processes and assuming maximal helicity we obtain $\delta h= \delta \mu_{R}/c_{3}$.  
For $\delta (|\mu_{5}|/T) \sim 10^{-6}$ around the transition, the corresponding growth of magnetic energy and helicity is negligible. 
Even if we would assume that the initial $\mu_{5}$ is much higher, but still such that $\mu_{5}/T<1$, the values of flipping rates would still
lead to a fast damping of such initial asymmetry, which could be prevented only with stronger initial magnetic fields. Therefore, we 
conclude that a significant amplification or generation of magnetic fields around the electroweak crossover is not viable in the Standard model,
even when the chiral magnetic effect is taken into account. 

In Figure~\ref{pic:energy} we show the difference between ordinary resistive damping and the evolution of magnetic energy in 
the presence of $\mu_5$, where we see that the decay is slower due to the anomaly effect, which becomes less significant as 
$\mu_{5}$ approaches the regime of exponential damping. \\ \\
Let us note that in our work \cite{pa} we also obtained numerical solutions for the evolution of helicity around the electroweak transition, 
demonstrating that -- in accord with the analytical discussion in the previous section -- helicity is necessary created from non-vanishing 
fields in a fast transient process, due to the chiral magnetic effect -- which causes the induction of helicity as a consequence of the 
change in the chiral chemical potential. However in this case, as discussed in \cite{pa}, this means that the values of helicity and 
$\mu_{R,5}$ will be significantly lower compared to the case when the fields were initially helical. 

\section{Conclusions}
In this chapter we have studied the application of chiral MHD, in the regime of negligible velocity effects, for the description 
of electrodynamics in the early Universe. Specifically, we have studied the evolution of chiral asymmetry, magnetic helicity density and 
magnetic energy around the electroweak transition in the Standard model. In the Standard model, the electroweak transition is not a first order 
phase transition, and thus it does not lead to the creation of bubbles of the broken phase in the symmetric phase -- which could cause 
the creation of gravitational waves or significant amplification of magnetic fields. Still, when the quantum anomalous effects are taken into account, 
even a higher order transition can have the consequences for the magnetohydrodynamics of the early Universe. This comes as a result of different 
interaction rates, describing the transition between chiral states in both phases, which therefore tend to suppress the chiral asymmetry. We have discussed 
the evolution of chiral MHD quantities of interest first using the simplified analytical picture and then solving the equations numerically, 
demonstrating a correspondence between the conclusion reached by both methods. Starting from the typical conditions expected in the early 
Universe around the electroweak crossover, we have demonstrated that the chiral asymmetry will slightly increase as it approaches the transition 
temperature and then significantly decrease in the broken phase, as the rates of electromagnetic and nuclear processes 
increase. The fast sudden change of $\mu_{5}$ around the transition temperature comes from the simplified treatment of the crossover 
as happening instantly, while a more realistic model would need to consider its continuous character. Due to the increase of the flipping rates, the chiral asymmetry will eventually enter the regime of exponential damping, and will 
typically be erased at the MeV scale. \\ \\
By assuming realistic initial values for the chiral chemical potential, $\mu_{5}$, such that $\mu_{5} \ll T$ we have concluded that there will be no
significant amplification of magnetic fields. Therefore, the higher order electroweak transition does not seem as a likely candidate 
for explaining the magnetogenesis problem even if anomalous contributions to the description of the electroweak MHD are considered. However, 
it was demonstrated that the decay of magnetic fields, due to the finite conductivity, will be slowed down in the chiral MHD. If the magnetic 
fields were generated by some process before the electroweak transition a study of their subsequent evolution needs to also consider 
this effect, which suppresses resistive decay. The transformation of non-helical magnetic fields into helical magnetic fields, which 
naturally happens around the electroweak scale due to the chiral anomaly, can significantly impact the growth of the magnetic field correlation 
length due to the role of helicity in the evolution of MHD turbulence. The shortcomings of magnetogenesis models which predict a correlation length that 
is not compatible with its observed values, could then in principle be fixed by considering these modifications related to chiral effects, assuming the existence 
of some non-vanishing initial value of the chiral asymmetry. Conversely, even if there is no initial chiral asymmetry, but the magnetic fields are helical, 
an asymmetry between left-handed and right-handed chiral particles will be created, which can have consequences on different baryogenesis 
models. We conclude that the proper study of electromagnetic fields and the early Universe plasma around the electroweak scales requires a complete chiral MHD 
description. The only case in which magnetic fields can be treated as governed by standard MHD equations is if the fields were 
non-helical from their creation and if there was no asymmetry of left and right handed particles around the electroweak scale.

\chapter{Chiral magnetohydrodynamic turbulence}\label{chirturb}
Turbulent phenomena were initially observed in the context of terrestrial fluid motions, such as rivers, stirred water or smoke-- as one of the 
common everyday experiences. However, today we believe - quite similarly as for electromagnetic phenomena -- that turbulent motions 
appear much more generally and are an inseparable part of physical processes in various studied structures appearing in the Universe,
including the solar wind, accretion discs, galaxies, galaxy clusters, the interstellar medium and intracluster medium \cite{t1,t2,t3,t4}. In general,
the length scales of the structures in the Universe are much larger than dissipative scales, which tends to lead to high Reynolds numbers
characterizing flows that can occur in those systems. Therefore, various forms of motion of matter in the Universe will naturally 
tend to develop into turbulent flows. Since -- as discussed in the previous chapter -- most of the cosmic medium can be described as plasma
permeated by magnetic fields, the evolution of turbulence and magnetic fields will be interconnected and coupled in a complex manner. Magnetic fields
will directly influence the development of turbulent structures via Lorentz force appearing in the Navier-Stokes equation, while the evolution
of velocity directly influences magnetic fields via corresponding term in Maxwell's equations, physically coming from the transformation of the 
electric field when the fluid is not at rest. The natural framework for the study of turbulence in the Universe is therefore a magnetohydrodynamic
description leading to magnetohydrodynamic turbulence. This is particularly the case for astrophysical objects which 
are characterized by strong magnetic fields and turbulent motions, such as neutron stars and  core collapse supernovae 
\cite{sup1, sup2, sup3, sup4, sup5}
This description can also easily be extended to the case of the early Universe. Together
with the assumed existence of magnetic fields, it seems rather convincing that matter in the conditions of the early Universe was also 
set in motion, which would --at least in some regimes -- lead to turbulent motions. For instance, if there was some first order phase transition -- such as QCD, electroweak or even some earlier
phase transition -- then the formation of bubbles of the new phase in the old phase and their collisions
typically leads to turbulent motion of the cosmological medium \cite{kamion, kosow,ratra, leitao, mark, child, iso}. After the transition, thus created MHD turbulence would be described as a 
freely decaying turbulence, but its effects would
also remain relevant for some period of time, changing the details of the evolution of magnetic fields in the Universe and their properties --
such as strength, spectral distribution and correlation length. On the other hand, as it is well known, the early Universe is believed to be 
characterized by density perturbations from which the observed astrophysical formations will subsequently develop. Density perturbations
naturally lead to movement of the primordial plasma, therefore creating some finite velocity field, which can develop into turbulence
\cite{wag, dunsby}. \\ \\
As discussed in the previous chapter, for high temperature regimes leading to systems of charged relativistic particles immersed in (hyper)magnetic
fields the standard MHD description of the plasma is not complete. In the case of massless particles, or particles that can effectively be taken
as massless in the $m \ll T$ regime, a new degree of freedom needs to be introduced which describes the number difference of left-handed and 
right-handed chiral states and which then defines the associated chemical potential, $\mu_{5}$. Due to the chiral magnetic effect this
chiral chemical potential will change in the exterior helical magnetic field, leading to the creation of an effective electric current in the 
direction of the field. We have seen how this effective current enters into Maxwell's equations, strongly changing the dynamics of magnetic
fields and leading to a number of new phenomena -- tending to produce exponential field growth, creating maximally-helical fields and 
support the transfer of energy from small to larger scales. Since the chiral current is in the direction of the magnetic field it will not contribute to the Lorentz-force, and therefore
it will not directly modify the Navier-Stokes equation and no new terms will appear in it. However, as it will modify the evolution of
magnetic fields, the modified fields now entering into Navier-Stokes equation can strongly change the evolution of velocity. On the other hand, the presence of turbulent 
velocity fields will naturally affect the solutions of the modified Maxwell's equations. Therefore, taking into account the chiral effects for high temperature
plasmas necessarily leads to the complex problem of chiral MHD turbulence. Since the origin of chiral modifications is a microscopic and 
quantum phenomenon, and furthermore the evolution of $\mu_{5}$ is influenced by the rates of particle processes which lead to spin flipping of left and right-handed states,
chiral MHD turbulence leads to the coupling between velocity, hyperfields and the particle content of the theory. The primary motivation
for the study of chiral MHD turbulence can be understood as a theoretical need to generalize the study of MHD turbulence to the case 
of high energy plasmas where the chiral anomaly effect becomes significant. Apart from this theoretical motivation, the chiral anomaly effect
can have different practical applications in different high-energy settings -- it can significantly influence the evolution of magnetic 
fields in the early Universe, magnetic field amplification in magnetars, and strongly modify the details of baryogenesis and leptogenesis models. 
While the potential role of the chiral magnetic effect was extensively studied in relation to all those problems, but ignoring the velocity effects -- as discussed in the previous chapter --
the study of those issues in the context of full chiral MHD description is still an open task for a research. Due to the complicated nature
of chiral MHD equations, and further significant difficulties that come into play when the coupling between the chiral anomaly and turbulent phenomena --
is introduced -- which still remains unsolved problem of physics on its own, this task can not be approached in a direct manner. In order to make some first steps in understanding of this complex
interplay, it is necessary to discuss simplified and approximate regimes, as well as numerical simulations. Specifically, 
it seems necessary that analytical approaches towards understanding of chiral turbulence will need to use qualitative reasoning and scaling
arguments, that are also used in the study of standard MHD turbulence. \\ \\
First systematic studies of chiral MHD turbulence started only recently. The effects of velocity field on the evolution
of magnetic fields together with chiral magnetic and chiral vortical effects were considered in \cite{hiro}, neglecting the 
back-reaction on the velocity evolution. The velocity distribution in \cite{hiro} was simply assumed to be given by a standard Kolmogorov
spectrum, with no modifications on it coming from the chiral effect, and moreover generally important advection term $\nabla \times(\mathbf{v} \times \mathbf{B})$ was neglected. In \cite{45} the system of chiral MHD equations was discussed 
using scaling symmetries, from which the general scaling laws were proposed. In \cite{novi} chiral turbulence was discussed further, analysing 
the general properties of turbulent MHD equations, and discussing how the chiral magnetic effect leads to the creation of helical magnetic fields and changes the evolution
 of magnetic energy and correlation length, while supporting the inverse cascade. Chiral turbulence was discussed for the concrete setting of the electroweak transition in \cite{sem}, where the approximation
 for the velocity field based on the simple proportionality relation between the velocity and Lorentz force  was used. Further theoretical considerations of chiral MHD, including the generalization to spatially dependent chemical potential, were recently discussed in \cite{roga}, while numerical
 simulations of chiral MHD turbulence were recently presented in \cite{brandy} and \cite{jennifer}. 
 \section{Chiral MHD numerical simulations}\label{numericki}
At the moment of writing this work there were only two numerical studies of chiral MHD turbulence reported \cite{brandy,jennifer}, both of which 
used similar methodology and approach. Both simulations extend their study to the case of spatially dependent chemical potential which then
introduces a new diffusion term in the evolution equation for $\mu_{5}$ \cite{boyflu} which is now extended to read 
\begin{equation}
\frac{d \mu_{5}}{d t}=\frac{1}{T^{2}} \frac{3e^2}{4 \pi^2}\frac{d h}{d t} + D \nabla^{2}\mu_{5},  
\end{equation}
where $D$ is a chiral diffussion coefficient. On the other hand, both simulations neglect the influence of chirality flipping rates, $\Gamma_{f}$. 
This approaximation naturally leads to the conservation of the sum of volume-averaged chiral potential and term including magnetic helicity. 
The study of chiral turbulence in \cite{brandy} proceeds from the initial configuration characterized by small seed non-turbulent fields and sufficient
amount of initial chiral asymmetry, which then leads to the strong growth of magnetic fields due to chiral anomaly effect. During this regime
the only non-linear contribution in the induction equation for magnetic field is coming from the anomaly term, since the non-linear $\mathbf{v} \times \mathbf{B}$ term, including the 
non-linear contribution of the Lorentz force in the Navier-Stokes equation, is negligible. During this regime of field growth, sourced by the 
chiral instability some small-scale velocity fluctuations will be created which contribute to nonlinearities in the velocity evolution, 
until the Lorentz-force related nonlinearity becomes comparable with the anomaly contribution. The field will then be turbulent and the 
magnetic field will grow until the value of $\mu_{5}$ becomes insignificant. When this happens helicity can no longer change in time and becomes saturated, 
while the magnetic field begins to decay. In the same work the following spectrum for the developed chiral turbulence in the inertial range 
was proposed
\begin{equation}
\rho_{m}(k,t)= C_{\mu} \tilde{\rho}\mu_{5}^{3} \eta^{2}k^{-2}, 
\end{equation}
motivated by the $k^{-2}$ spectrum for weak turbulence, expected in the case of strong stohastic magnetic fields, when the main contribution to the 
evolution of velocity comes from the Lorentz force \cite{galtier}. In the above equation $C_{\mu}$ is a new type of Kolmogorov-like 
constant characterizing the chiral turbulence, $\rho$ is the mean density of the plasma and $\eta$ is the magnetic diffusivity. This type 
of spectrum was reported to be confirmed in \cite{jennifer} where the numerical analysis was further extended. This study also presented 
the typical evolution of magnetic fields in chiral MHD turbulence consisting of three phases: 1) small scale growth due to the chiral effect, 
2) generation of chirally driven MHD turbulence and excitation of large scale dynamo instability, 3) the final stage where chiral potential 
subsequently gets damped and helicity saturates.
These works are significant for giving the first detailed numerical results of the interplay between turbulence and chiral anomaly, but 
more detailed and more general studies are necessary. It is of particular interest to study in much more details the regime where turbulence 
and anomaly effects are comparable, by discussing in depth typical temporal and spatial scaling of MHD quantities. Moreover, it is important 
to ask how would typical chirality flipping processes influence the properties of chiral MHD turbulence. The primary conceptual limitation 
of the above studies is that they assume a rather special initial configuration of small seed field and dominant chiral effect. The presented 
picture of chiral MHD turbulence develops rather naturally and as expected from this initial configuration. There are however different possible 
initial regimes -- for instance when the non-linear turbulence effects are comparable or dominant compared to the chiral contribution, 
when flipping terms become significant and have a complicated dependence -- as for instance around the electroweak transition, which can all be expected 
to evolve rather differently. 
\section{The effects of chiral anomaly on the evolution of MHD turbulence}
In order to analytically study some important modifications that come as a result of generalization of turbulence theory to the high temperature 
case of chiral MHD turbulence, in this section we will use the main approximations discussed in more detail in the previous chapter. We first assume that 
the characteristic bulk velocity of the turbulent fluid is non-relativistic, so that we can use non-relativistic MHD equations (note thet although the 
bulk plasma velocity is assumed to be much smaller than speed of light, the considered particles are themselves relativistic since $m \ll T$
in order that the chiral effect becomes relevant). We moreover assume that the chiral potential is a space-independent function (in the more 
general approach given by \eqref{chempomod} this corresponds to the approximation of negligible diffusion constant $D$) and that turbulence is 
incompressible. We also use the definitions and notations for magnetic energy and helicity spectral modes, $\rho_{k}$ $h_{k}$, as well as the magnetic correlation length, 
$\xi_{m}$, as defined in the chapter on turbulence \ref{turbo}, again assuming that magnetic fields are statistically homogeneous and isotropic. 
Combining equations \eqref{Maxwell} and \eqref{induction} in Fourier space we now need to consider the following evolution of magnetic fields
 { \begin{equation}
\label{eq:B_ev}
 \partial_t \mathbf{B_k}= - \frac{k^2}{\sigma}  \mathbf{B}_\mathbf{k} - 2 c_{1,2} \mu_5 (i\mathbf{k}\times \mathbf{B}_\mathbf{k}) + \frac{i}{(2\pi)^{3/2}}\mathbf{k}\times\int d^3q (\mathbf{v}_\mathbf{k-q}\times \mathbf{B}_\mathbf{q}) \, ,
\end{equation} 
while the evolution of velocity is given by \eqref{navsto}, and the evolution of chiral potential by \eqref{chempomod}. 

In order to obtain the time evolution of the power spectra $\rho_k$ and $h_k$ defined in \eqref{eq:rhok_ev} and \eqref{eq:hk_ev} on can multiply \eqref{eq:B_ev} and its complex conjugate 
by $\mathbf{B}_\mathbf{k}^*$ and $\mathbf{B}_\mathbf{k}$ leading to
\begin{equation} \label{eq:rhok_ev}
\partial_t  \rho_k= - \frac{2 k^2}{\sigma} \rho_k -  c_{1,2} \mu_5 k^2 h_k + I_{1}(k)\, ,
\end{equation}  
\begin{equation} \label{eq:hk_ev}
 \partial_t  h_k= - \frac{2 k^2}{\sigma} h_k - 4 c_{1,2} \mu_5 \rho_k + I_{2}(k) \, ,
\end{equation}  
where 
\begin{equation}
I_{1}(k)=  k \left[(\mathbf{k}\times \mathbf{I_k}) \cdot \mathbf{B_{k}}^{*}+(\mathbf{k}\times \mathbf{I_k}^{*}) \cdot \mathbf{B_{k}} \right],
\end{equation}
\begin{equation}
I_{2}(k)= k \left[(-i) \mathbf{I_k} \cdot \mathbf{B_k}^{*} + \mathbf{A_{k}} \cdot (\mathbf{k} \times \mathbf{I_k}^{*})\right],
\end{equation}
and 
\begin{eqnarray}
\label{integral}
\mathbf{I_k} = \frac{i }{(2\pi)^{3/2}} \int d^3q (\mathbf{v}_\mathbf{k-q}\times \mathbf{B}_\mathbf{q}) \, 
\end{eqnarray}
A particularly realistic and interesting scenario is the one which assumes that there exists a developed MHD turbulence at the time of formation 
of asymmetry between charged left and right-handed chiral particles, possibly due to some process beyond the Standard model, and that the contribution of effective chiral current does not change significantly the evolution of magnetic fields, that is $|\mathbf{J}_{5}|<|\mathbf{J}|$. Freely decaying
MHD turbulence characterized by zero helicity will stay non-helical during its subsequent evolution. However, as discussed in the previous 
chapter, the chiral anomaly effect naturally leads to the creation of helicity from initially non-helical fields. The question which directly arises 
is -- can this fact modify subsequent evolution of turbulence, even assuming that chiral modifications are suppressed with respect to the standard 
evolution of magnetic fields. We will start to address this question in the following subsection. 
\subsection{Inverse cascade and chiral anomaly}
In standard MHD magnetic helicity is a conserved quantity in the limit of infinite conductivity, and since in many cases of interest conductivity is high, 
this quantity will be important for the description of turbulent MHD flows even in the resistive regime. It can be shown that helicity is also an ideally conserved
quantity in the chiral MHD case. Using the definition of helicity we have $\dot{h}=(1/V)\partial_{t}\int d^3 r\, \mathbf{A} \cdot \mathbf{B}$ 
and with \eqref{Maxwell} and \eqref{induction}, we get
\begin{equation} \label{eq:hel_ev}
\frac{dh}{dt}=-\frac{2}{V} \int d^3{r} \left[\frac{1}{ \sigma}\left(\nabla \times \textbf{B}\right)\cdot  \textbf{B} +2c_{1,2} \mu_{5,R}|\textbf{B}|^{2}\right]. 
\end{equation}
From this expression we see that in the limit $\sigma \rightarrow \infty $, there can be no change of $\mu_{5}$ due to the chiral anomaly since helicity is 
conserved. Therefore, high conductivities tend to support the conservation of magnetic helicity and suppress the significance of the chiral magnetic effect. 
In a similar manner we can discuss the influence of the chiral effect on the conservation of the magnetic flux. Using the same equations,
together with definition for magnetic flux, $\Phi= \int_{S} \mathbf{B} \cdot d\mathbf{S}$ we obtain
\begin{equation}
\frac{d \Phi}{dt}=-\frac{1}{ \sigma} \oint_{\ell} \mathbf{J} \cdot d\mathbf{\ell} - 2c_{1,2} \mu_{R, 5}\oint_{\ell} \mathbf{B} \cdot d\mathbf{\ell} . 
\end{equation}
As in the case of standard MHD, in the case of ideal chiral MHD the magnetic flux will also be conserved, while in the resistive regime, 
the flux changes -- corresponding to cutting and reconnecting of field lines -- will be enhanced 
by the chiral anomaly. \\ \\ 
One of the important properties of magnetic helicity in MHD turbulence is that it can support an inverse energy transfer. It is well known
that MHD turbulence differs from the hydrodynamical turbulence by the fact it can -- beside the direct cascade, that is a transfer of energy 
from the large inertial scale down to the small scales where viscosity becomes significant -- also lead to an inverse cascade in the 3D case, that is 
transfer of energy from small to large scales. This inverse energy transfer has an important role in changing the evolution of magnetic fields
and leading to the formation of organized turbulent structures from random initial conditions \cite{po, shiromizu}. The existence of such an
inverse transfer seems to be strongly connected with the existence of finite helicity in such turbulent flows. As we discussed, for high conductivities magnetic helicity can be treated as an effectively conserved quantity, which prevents the decay of helicity in short-scale modes and 
supports their transfer to large scale modes, and -- together with the non-linear interactions between different lengths scales -- leads to 
the effect of inverse cascade. While it seems clear that the existence of a non-vanishing magnetic helicity, and its conservation in the 
high-conductivity limit, is closely related to the inverse energy transfer, it is however still not clear if the presence of magnetic helicity
is a necessary condition for the development of an inverse cascade. Indeed, in some recent numerical simulations it is reported that inverse
cascades were observed in MHD turbulence where magnetic fields were non-helical \cite{zrake, tev}. This is in contradiction with some 
earlier works which concluded that inverse cascade is not possible for non-helical fields \cite{campanelli, son, christenson, ban-jed}. 
It is therefore necessary to wait for new and more detailed studies in order to solve this question. However, even if the existence of 
magnetic helicity in MHD turbulence is not a necessary condition for the development of inverse cascades it will strongly support it, and therefore 
have a strong impact on the turbulent evolution. This was recently confirmed in \cite{repibane}, were the presented numerical results
suggested that non-helical inverse transfer is much less efficient in transferring magnetic energy to large scales, and that for non helical
fields this efficiency depends on the Prandtl number,$Pr=\nu/ \eta$, being more emphasized for the low Prandtl numbers. \\ \\
In order to start discussing the effects of chiral effect on the properties of MHD turbulence let us assume we initially, starting 
from a time $t_{0}$, have a developed non-helical turbulence. To describe the evolution of magnetic energy we use the following 
initial scaling
\begin{equation}
\rho_{k}(t)=\sqrt{\frac{t_{0}}{t}}k \rho_{k}\left(k\sqrt{\frac{t}{t_{0}}},t_{0}\right) ,
\label{nohel}
\end{equation}
which was proposed in \cite{Olesen3}, and which  seem to be consistent with the result of numerical simulations in Refs. \cite{zrake, tev}
It then follows that the total magnetic energy density, $\rho_{m}$, scales like $\rho_{m} \sim 1/t$ and the correlation length as $\xi_{m} \sim \sqrt{t}$. 
We now assume that at some later time $t_{i}$, a finite $\mu_{5}$ is created due to some particle processes which become active. The chiral anomaly will, according to \eqref{eq:hk_ev}, then lead to a creation of helicity. 
After some short time interval, $\Delta t \equiv t -t_{i}$ there will be a finite helicity created according to
\begin{equation}
h_{k}=-4c_{1,2} \int_{t_{i}}^{t} \mu_{5}(t) \sqrt{\frac{t_{0}}{\tau}} k \rho_{k}\left(k\sqrt{\frac{\tau}{t_{0}}},t_{0}\right) d \tau,
\label{helind}
\end{equation}
where we concentrate on the case of strong magnetic fields such that $\Gamma=\rho_{k}/\rho_{mag} \ll 1$, allowing us to ignore the contribution of 
$\int_{t_{i}}^{t} I_{2} d \tau$ during this short time sequence (also, since fields are initially non-helical $I_{2}$ is initially vanishing). 
We furthermore assume that $\mu_{5}$(t) is a smooth function and can therefore be written as $\mu_{5}(t)= \sum_{n} c_{n}t^{n}$ on a small time interval $\Delta t$. The values of the expansion coefficients, $c_{n}$ will be given
by the explicit relations for the flipping and source terms entering in \eqref{chempomod}, and they will naturally be different for different physical systems. 
The induced magnetic helicity density is then given by
\begin{eqnarray} \label{eq:hinduced}
h_{in}(t) &=& \int h_{k} d\ln k \nonumber \\
&=& - 4 c_{1,2} t_{0} \sum_{n} c_{n} \frac{(t^n - t_{i}^n)}{n} \int_{0}^{\infty} \rho_{k}(x, t_{0})dx,
\end{eqnarray}
and it can be seen that the time evolution of the induced helicity, $h(t) \sim t^{n}$, is guided by the evolution of $\mu_{5}(t)$.
in the special case of $\mu_{5}=const.$ one gets a logarithmic scaling with time
\begin{equation}
 h_{in}^{sta}= -4 K c_{1,2}t_{0}  \log\left( \frac{t}{t_{i}}\right) \int_{0}^{\infty} \rho_{k}(x,t_{0})dx .
\end{equation}
The growth of magnetic helicity will approximately follow these relations as long as the term in Eqs. \eqref{eq:hk_ev} and \eqref{eq:rhok_ev} containing $h_{k}$ remains much smaller than the term containing the energy density.
From this we may conclude that the induction of helicity in a turbulent MHD plasma is a short transient phenomenon, after which turbulence 
will continue to evolve as helical MHD turbulence. The field evolution, and therefore implicitly also the velocity evolution, will now be 
additionally modified by the effective chiral current, $\mathbf{J}_{5}$. In this regime it is necessary to consider the full set of coupled differential equations for energy end helicity density, 
and even a qualitative understanding of their properties is not a simple task. From the conclusions reached so far we can see that chiral 
and turbulent effects tend to support each other since both of them tend to establish the inverse energy transfer. Their physical mechanisms
are however completely different: while the chiral effect is based on the small-scale quantum instability, turbulence is believed to stem from  
purely classical non-linear effects coming from interaction of different length scales, which exists macroscopically -- only in the range 
of scales sufficiently higher than the scales on which viscosity effects become dominant. Moreover, while the inverse cascade is actually supported by 
the conservation of magnetic helicity the evolution of chiral asymmetry is determined by the change of magnetic helicity, and chiral effects will thus be 
suppressed if the changes of helicity are negligible, even more if chirality flipping processes are active and source terms negligible. From this we conclude that the regime in which 
turbulent and chiral phenomena coexist as equally important is likely to be only temporary in its character, and that a system will tend to approach the state where one of the effects determines 
the main features of its dynamics, while the other one has only a minor role, which can be treated as a correction. \\ \\
We first consider a rather conservative scenario in which chiral asymmetry, $\mu_{5}$, during the evolution of chiral MHD turbulence stays 
so small that the only significant effect induced by the anomaly is the afore discussed creation of helicity, after which helical magnetic fields
follow the usual non-chiral MHD evolution. This could indeed by a likely scenario in the early Universe due to several reasons. First of all, it is
questionable how high the values of $\mu_{5}$ could be, since the mechanisms creating the chiral asymmetry are not properly understood. It is very possible that these values, 
compared to the characteristic temperatures around the electroweak scale, could be very modest. Moreover, high conductivities of the plasma 
suppress the change of total helicity and the level of chiral asymmetry would be further reduced by electromagnetic, nuclear and Higgs processes 
which reduce $\mu_{5}$ (see previous chapter \ref{chir}). To analytically discuss this case we will generalize the usual scaling arguments discussed
in standard textbooks \cite{biskamp} -- which, despite their approximate nature and simplicity, still lead to a satisfactory matching with the results of numerical simulations. 
In this regime we use the approximate scaling 
\begin{equation}
\rho_{m} \xi_{m} \sim \frac{h_{in}}{2} \approx const.
\label{magintsc}
\end{equation}
The use of this scaling is based on the fact that helicity will be approximately conserved during the phase of developed turbulence 
when the inverse cascade occurs, where the definition of correlation length in \eqref{corr} was used. 
Assuming a constant energy transfer rate proportional to the eddy-turnover rate in the inertial interval we write
\begin{equation}
\frac{d \rho_{tot}}{dt} \sim - \rho_{tot} \frac{ \rho_{K}^{1/2}}{\xi_{m}},
\end{equation}
where $\rho_{tot}=\rho_{m} + \rho_{K}$. 
It then follows that
\begin{equation} \label{eq:scaling}
\frac{d}{dt}\left[\rho_{m}(1+ \Gamma)\right] \sim - \frac{\rho_{m}^{5/2}\Gamma^{1/2}(1+ \Gamma)}{h_{in}}. 
\end{equation}
Numerical simulations typically show that the ratio between kinetic and magnetic energy asymptotically approaches a constant value in the case of standard MHD turbulence \cite{ban-jed}. Therefore, we can now treat $\Gamma$ as independent of time, and then from \eqref{eq:scaling} we obtain
that scaling of magnetic energy follows $\rho_{m} \sim t^{-2/3}$. We thus conclude that even in this very conservative scenario the presence
of a weak chiral anomaly effect will eventually tend to modify the decay of the magnetic field from $\rho_{m} \sim 1/t$ to $\rho_{m} \sim t^{-2/3}$. 
During the later course of the evolution of such a system this helicity, initially induced by the anomaly, will tend to organize the field spectrum in the direction of maximally helical 
configuration. This comes from the observation that standard MHD turbulence tends to organize magnetic fields which were initially fractional helical to become maximally helical 
\cite{andrey, campanelli2}. We thus conclude that even a comparatively small chiral asymmetry, leading to anomaly terms that are small compared to turbulent and magnetic field terms, can significantly 
change the subsequent turbulent evolution on initially non/helical turbulence.\\ \\
Now we can turn to another regime, starting from the same initial setting, that is initially non-helical developed MHD turbulence, but now with chiral effects strong enough
such that they can not be ignored after the transient induction of helicity, i.e. $|\mathbf{J}| \approx |\mathbf{J}_{5}|$. Assuming high conductivities, typical for the early universe, any change 
in helicity will be highly suppressed, and we can make an approximation that helicity would be exactly conserved in the $\mu_{5}=0$ case, and the complete contribution to magnetic helicity
conservation comes only from the associated change in the chiral chemical potential: $\dot{h} \sim \kappa_{T} \dot{\mu}_{5}$, where $\kappa_{T}=T^{2}/c_{3,4}$. Here we assume that 
source and helicity rates are either small or compensating each other -- in the more general case the arguments we will present here can be easily extended using the complete equation
\eqref{chempomod}. We can again define the integral scale as in \eqref{magintsc}, but now it will depend on time, since helicity will also change due to the anomaly. Neglecting 
the temperature changes we have
\begin{equation}
\frac{d}{dt}\left[\rho_{m}(1+ \Gamma)\right] \sim - \frac{\rho_{m}^{5/2}\Gamma^{1/2}(1+ \Gamma)}{\kappa_{T} \mu_{5} } \, . 
\end{equation} 
Again assuming that $\Gamma$ approaches a constant value and $\mu_{5}(t)= \sum_{n} c_{n}t^{n}$, we get 
\begin{equation}
\rho_{m} \sim \left(\int\frac{dt}{k_{T}\sum_n c_{n} t^{n}}\right)^{-2/3} .
\end{equation}
If the evolution of $\mu_{5}$ is such that one can approximate it with a power-law solution, $\mu_{5}=Kt^{n}$, which is of interest in several practical contexts,
this leads to the following scaling  $\rho_{m} \sim  t^{2(n-1)/3}$. At the time when the regime of maximal helicity is reached it follows
$\xi_{m} \approx  \kappa_{T} \mu_{5}/(2 \rho_{B}) \approx t^{(n+2)/3} $. We see that the overall evolution of magnetic field and the correlation length is determined by the evolution
of the chiral chemical potential, which for a fast enough growth of $\mu_{5}$,  with $n>1$, leads to the total growth of magnetic energy. In realistic systems this phase can naturally 
happen only during a limited span of time, since the amount of magnetic energy that can be converted from the chiral potential is constrained by its initial value, and this regime will 
stop when $\mu_{5}$ starts to deplete. This type of regime essentially corresponds to the initial phases of chiraly driven MHD turbulence in the simulations \cite{jennifer, brandy}. In the more
general case of arbitrary flipping and sourcing terms, one needs to solve the following equation, which will then be dependent on their particular form in a given system
\begin{equation}
\frac{d\rho_{m} }{dt} \sim - \frac{\rho_{m}^{5/2}}{k_{T}\sum c_{n} t^{n} - \int\left(\Pi_{sr} - \Gamma_{f} \sum_n c_{n} t^{n} \right)dt}\, .
\end{equation}
\\ \\
Another type of transition to the chiral MHD turbulence can happen if the turbulent cosmological magnetic fields (created for instance during the inflationary 
expansion of the Universe) are already helical, while there is initially no chiral asymmetry present. The cration of chiral asymmetry will happen in such systems 
even if there are no active particle processes which could create some chiral asymmetry. Due to the fact that helicity is not 
an exactly conserved quantity, these changes in helicity will induce a difference in the number of left and right-handed charged particles 
corresponding to the chemical potential
\begin{equation}
\frac{d \mu_{5}}{dt}=-c_{3,4}\frac{2}{V T^{2}} \int d^3{r} \left[\frac{1}{ \sigma}\left(\nabla \times \textbf{B}\right)\cdot  \textbf{B}\right] - \Gamma_{f}
\label{ascre}
\end{equation}
and during the later evolution this created asymmetry will lead to an effective chiral current, $\mathbf{J}$, entering into Maxwell's equations 
and changing the evolution of magnetic and velocity field. The approximate relation \eqref{ascre} is valid only if $|\mathbf{J_{5}}| \ll |\mathbf{J}| $, and when 
this is no longer true the total contribution to the change in the chiral potential needs to be considered, as described in \eqref{chempomod} and 
\eqref{helchange}. This type of evolution leading to chiral MHD turbulence is more robust in the sense that it does not require any additional 
processes to create the chiral asymmetry. However, it is less probable to lead to some significant deviations from the standard MHD since 
this creation of the chiral asymmetry is strongly suppressed by the high conductivity. It could therefore be of interest in high-temparature 
systems with lower conductivities. In the case of the early Universe, as we discussed in the previous chapter, we can take that the conductivity on the electroweak scale is given by $ \sigma \propto T$, 
so the induction of the chiral asymmetry is proportional to $1/T^{3}$. From this follows that such a process is more significant for lower 
temperatures after the electroweak crossover. However as the temperature decreases, the rates of electromagnetic processes, causing the 
spin flips, increase proportionally to $1/T$, eventually leading to an exponential damping of the anomaly. So the most favorable region for such creation 
of asymmetry seems to be just after the electroweak transition -- around the minimal values of flipping rates, as discussed in the previous chapter. 
\section{Consideration of specific regimes of chiral MHD turbulence}
\subsection{Weak anomaly regime}
In this section we will focus our attention on the regime where the chiral anomaly contributions are small with respect to the standard 
MHD terms, i.e. when $ |2 c_{1,2}  \mu_{R, 5} \textbf{B}| \ll |\textbf{v} \times \textbf{B}+\textbf{E}|$. This regime was also briefly 
discussed in the previous section using qualitative scaling arguments, and here our aim will be to study it in more detail and to obtain 
some concrete solutions demonstrating its properties. These resulted were initially reported by us in \cite{novi}. In this setting, we expect that the dynamics will be determined by the terms
present in the MHD equations in the $\mu_{5}=0$ case, and that the chiral anomaly will just lead to small corrections to these background solutions. 
Under these assumptions we can write the anomaly contribution to magnetic energy and helicity as a small perturbation to the standard background MHD solutions, which we label as
 $\rho^{\rm bg}$ and $h^{\rm bg}$, and subsequently ignore all the terms higher than the first order in the perturbation. We therefore write
 \begin{equation}
\rho_{B}=\rho^{\rm bg} + \rho^{\mu},
\end{equation}
and 
\begin{equation}
h=h^{\rm bg} + h^{\mu}. 
\end{equation}
Adding those perturbations to \eqref{eq:rhok_ev} - \eqref{eq:hk_ev} around $\rho_{B}=\rho^{\rm bg}$ and $\mu_{5}=0$
we obtain in the zeroth order
\begin{equation}
\partial_t  \rho^{\rm bg}_k= - \frac{2 k^2}{\sigma} \rho^{\rm bg}_k  + I_{1}(k)\, ,  
\end{equation}
\begin{equation}
\partial_t h^{\rm bg}_k= - \frac{2 k^2}{\sigma}  h^{\rm bg}_k + I_{2}(k) ,
\end{equation}
and to first order 
\begin{equation}
\partial_t  \rho^{\mu}_k= - \frac{2 k^2}{\sigma} \rho^\mu_{k} -  c_{1,2} \mu_5 k^2 h^{\rm bg}_k \, ,
\end{equation}
\begin{equation}
\partial_t  h^\mu_{k} = - \frac{2 k^2}{\sigma} h^\mu_{k} - 4 c_{1,2} \mu_5 \rho^{\rm bg}_k \, .
\end{equation}
In these equations an additional assumption was made, namely that the coupling between the chiral asymmetry, given by $\mu_{5}$, and velocity 
can be neglected. This is justified if the kinetic energy is significantly smaller than magnetic energy, that is $\Gamma \ll 1$, and these coupled 
terms are by definition a negligible product of two small quantities. We will therefore in the following focus on the case of magnetically 
dominant MHD turbulence with the addition of small chiral effects. The presented perturbative equations have simple analytical solutions
\begin{equation}
\rho^\mu_{k}= \frac{\int e^{\frac{2 k^{2}}{\sigma}t} f(t) dt}{e^{\frac{2 k^{2}}{\sigma}t}} + const.
\label{pert1}
\end{equation}
and
\begin{equation}
h^\mu_{k}=\frac{\int e^{\frac{2 k^{2}}{\sigma}t} g(t) dt}{e^{\frac{2 k^{2}}{\sigma}t}} + const. \, , 
\label{pert2}
\end{equation}
where $f(t)= - c_{1,2} \mu_5 k^2 h^{\rm bg}_k $ and $g(t)=- 4 c_{1,2} \mu_5 \rho^{\rm bg}_k$. 
In order to obtain concrete solutions in these regime we need to specify the background evolution of magnetic field and helicity given 
by spectral modes $\rho^{\rm bg}_k$ and $h^{\rm bg}_k$. The problem of course appears in the fact that no such simple solutions, which could 
properly and in general portray the properties of turbulent MHD flows, actually exist. In fact, the existence of such solutions in terms 
of simple functions would contradict the essentially irregular character of turbulence. The only way to continue analytically 
is to use a simplified analytical model which at least manifests some aspects of MHD turbulence in some limited regime of applicability. 
With this aim we will use the analytical model proposed in Refs.~\cite{campanelli, campanelli2}. This approximation for the background solution 
giving the magnetic field and helicity evolution requires some further assumptions to be made. First of all, the dissipation term
$\nu \nabla^{2}\textbf{v}$ in the Navier-Stokes equation is neglected -- which can be taken as justified as long as
 we are focusing on the dynamics well inside the inertial interval, where dissipation effects are small. In \cite{campanelli} it is further assumed that the Navier-Stokes equation \eqref{navsto} can be
quasi-linearized neglecting the term $(\textbf{v} \cdot \nabla) \textbf{v}$. This is consistent with our already stated assumption that turbulence 
is magnetically dominated, $\Gamma<1$, and this term is then a non-linear contribution of a small quantity. Next, the approximation done in 
\cite{campanelli} is that the Lorentz-force is the only significant mechanism driving the turbulence, and that velocity can be further 
approximated using the fluid-response time per unit density $\zeta_L$ \cite{Sigltime} so that $\mathbf{v}\approx\zeta_L \mathbf{F}_L$, with 
$\mathbf{F}_L = \mathbf{J}\times \mathbf{B}$. This approximation should be taken with care, and properly speaking one can expect its validity 
only in the weak anomaly regime, where fields are close to the maximally-helical configuration and when the anomaly is not significantly driving 
the evolution of magnetic fields, since then the anomaly growth scale should give the characteristic time scale of the system, and not 
just the fluid-response time parameter \cite{novi}. Following \cite{campanelli, novi} the drag time is determined by $\zeta_L \approx \Gamma(0)[\tau_{\rm eddy}(0) + \gamma t]$, where $\tau_{\rm eddy}(0)$
is the initial eddy turnover time on a a specific scale $l$, $\tau_{\rm eddy}=l/v$ and $\gamma$ is some constant. As stated, magnetic fields are in this 
approximation not taken as maximally helical, but initially characterized by some fraction of magnetic helicity, such that $h^{\rm bg}_k(0)= \epsilon \rho^{\rm bg}_k(0)/(2k)$, with $0 \leq \epsilon \leq 1$.
Using the approximations described here, and introducing the defined quantities, one can derive the following ansatz for the background 
evolution of magnetic energy end helicity, leading to an inverse cascade evolution \cite{campanelli}
\begin{eqnarray}
\rho^{\rm bg}_k(t)=\rho^{\rm bg}_k(0) e^{-2k^{2} l_{\rm diss}^{2}} [\cosh(2k l_{\alpha})+\epsilon \sinh(2kl_{\alpha})],\,
\label{background1}
\end{eqnarray}
\begin{eqnarray}
h^{\rm bg}_k(t)=&&h^{\rm bg}_k(0) e^{-2k^{2} l_{\rm diss}^{2}}[\sinh(2k l_{\alpha})+\epsilon \cosh(2kl_{\alpha})],\,
\label{background2}
\end{eqnarray}
where we introduced $\eta_{\rm eff}= (\sigma)^{-1} + 4 \rho^{\rm bg} \zeta_{L}/3$, $\alpha_{B}=- \dot{h}^{\rm bg} \sigma \zeta_{L}/ 3$, $ l_{\rm diss}^{2}= \int_0^\tau d\tau \eta_{\rm eff}$ and $l_{\alpha}= \int_0^\tau d\tau \alpha_{B}$.
When there is no helicity present $\epsilon=0$, we see that $\alpha=0=l_{\alpha}$. Therefore, initially non-helical fields stay non-helical, 
and the evolution of magnetic energy modes is given by exponential decay, physically coming from the resistive damping. As discussed 
in \cite{campanelli} the evolution of helical magnetic fields described by this approximation will follow first resistive damping, and then inverse cascade. 
Here we concentrate only on the regime where inverse transfer takes place since it is of more physical interest in the context of questions 
discussed in this chapter. To obtain specific solutions in the weak anomaly approximation of chiral MHD turbulence it is also necessary to determine the evolution 
of $\mu_{5}$. Rather than going into details of specific settings determining the concrete form of flipping terms, potential source terms 
and initial conditions we will here consider some archetypal cases which determine the evolution of $\mu_{5}$. Note that by definition of an inverse cascade 
regime, helicity is approximately conserved, and the anomaly induced helicity is much smaller than the background helicity, thus $dh/dt \approx 0$ and 
$\mu_{5} \approx \Pi_{sr} - \Gamma_{f}$. \\ \\
Since departures from standard MHD turbulence are by definition small in the weak anomaly regime, the most dramatic change in the 
plasma evolution in this regime is the one happening if magnetic fields are initially non-helical. According to the standard MHD description they 
will, in this approximation, just decay and there will be no inverse transfer of energy. If there is initially some chiral asymmetry present it will lead to a fast induction of helicity 
at early times, as described by \eqref{pert2}. This induced helicity will then appear in \eqref{background1} and \eqref{background2}, after 
a short transient period, playing the role of the background helicity and eventually leading to an inverse energy transfer. The comparison of two solutions for magnetic 
energy density spectrum, one starting from $\mu_{5}(0)=0$ and other from $\mu_{5}=const.$, for the case where source and flipping terms 
are both negligible and are canceling each other, is shown in Fig.~\ref{pic:seldec}

\begin{figure}
\centering
\includegraphics[width=0.6\textwidth]{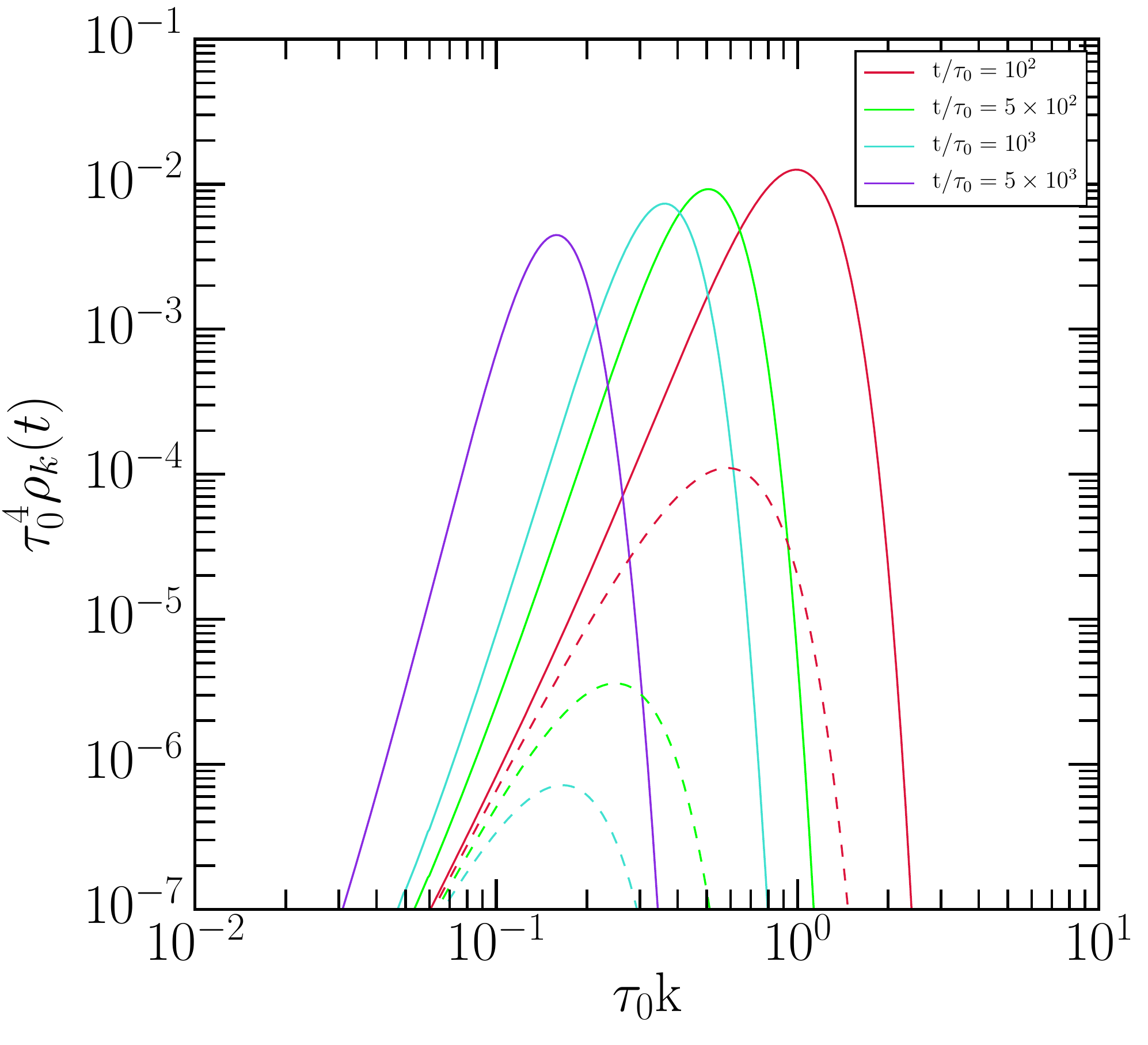}

\caption{\label{pic:seldec} Total magnetic energy density spectrum in the case of an initially vanishing helicity. Dashed lines, computed from \eqref{background1} in the absence of $\mu_5$, show resistive damping of modes in time. Solid lines in the presence of $\mu_5^0=-10^3\tau_0^{-1}$ for $\Gamma_f\mu_5 = \Pi_{sr}$, which induces a finite helicity, are illustrating an inverse cascade in time. Adopted from \cite{novi}}
\end{figure}
Another example of a significant departure from the non-chiral tubulence happens in the case when there is a source term active for some time, caused by reactions such as electron capture (for instance in neutron stars), which 
creates a chiral asymmetry at rate $\Pi_{sr}$, such that $\Gamma_{f} \ll \Pi_{sr}$. In this case, the amount of energy transferred from 
the chiral potential to magnetic fields leads to significant growth of magnetic fields during the inverse transfer. However, when the 
amount of anomaly induced magnetic energy becomes comparable to the background magnetic energy the system leaves the weak anomaly regime 
and the discussed treatment can no longer be applied. This case is demonstrated in Fig.~\ref{pic:3hels}, where a constant source term was taken, described by: $\Pi_{sr}=10^{-2}\tau_0^{-1}$, If the source term is active even longer, then the system can even enter in the opposite regime of strong anomaly, where the overall  
dynamics is ruled dominantly by effects related to the chiral current. 

\begin{figure}
\centering
\includegraphics[width=0.6\textwidth]{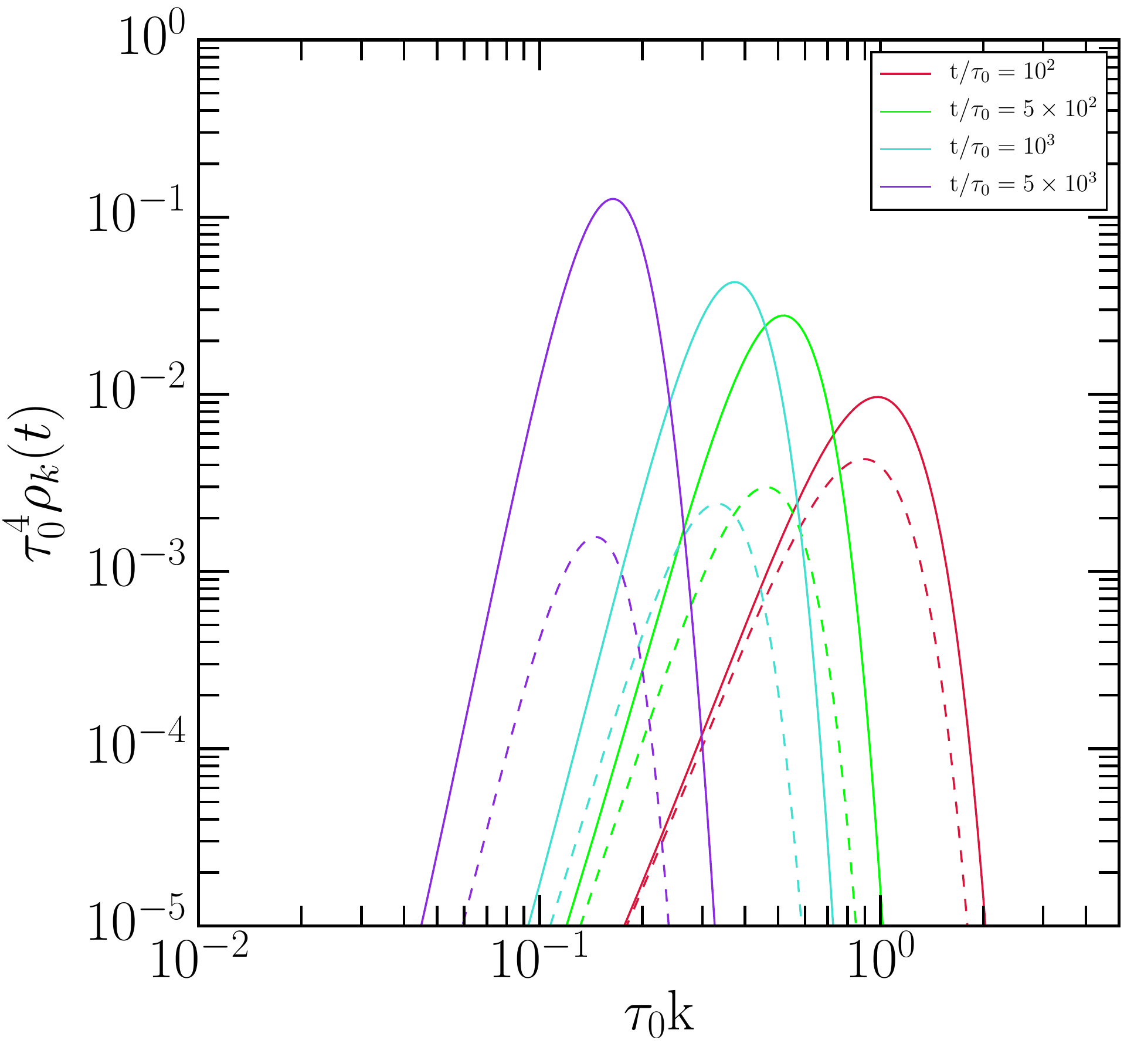}
\caption{\label{pic:3hels} Total magnetic energy density spectrum (solid) and background magnetic energy density spectrum (dashed) at different times, computed from \eqref{pert1} and \eqref{chempomod}, and \eqref{background1}, respectively, with $\mu_5^0=-10^3\tau_0^{-1}$ and $\epsilon=7\times 10^{-3}$, for $\Gamma_{f}=0$ and $\Pi_{sr}=const.$ Adopted from \cite{novi}}
\end{figure}

In the case where helical turbulence is already in the inverse cascade regime, the contributions induced by the chiral magnetic effect will remain to be a small perturbation, as it visible in the solution depicted in
Fig.~\ref{pic:4hels}, where we show the case of no active source term and flipping rates corresponding to the weak interaction rate, 
 $\Gamma_{f}(t)=(G_F^2m_e/3t)^2\tau_0^{-1}$, where $G_F$ is the Fermi coupling and $m_e$ is the electron mass.
 \begin{figure}
\centering
\includegraphics[width=0.6\textwidth]{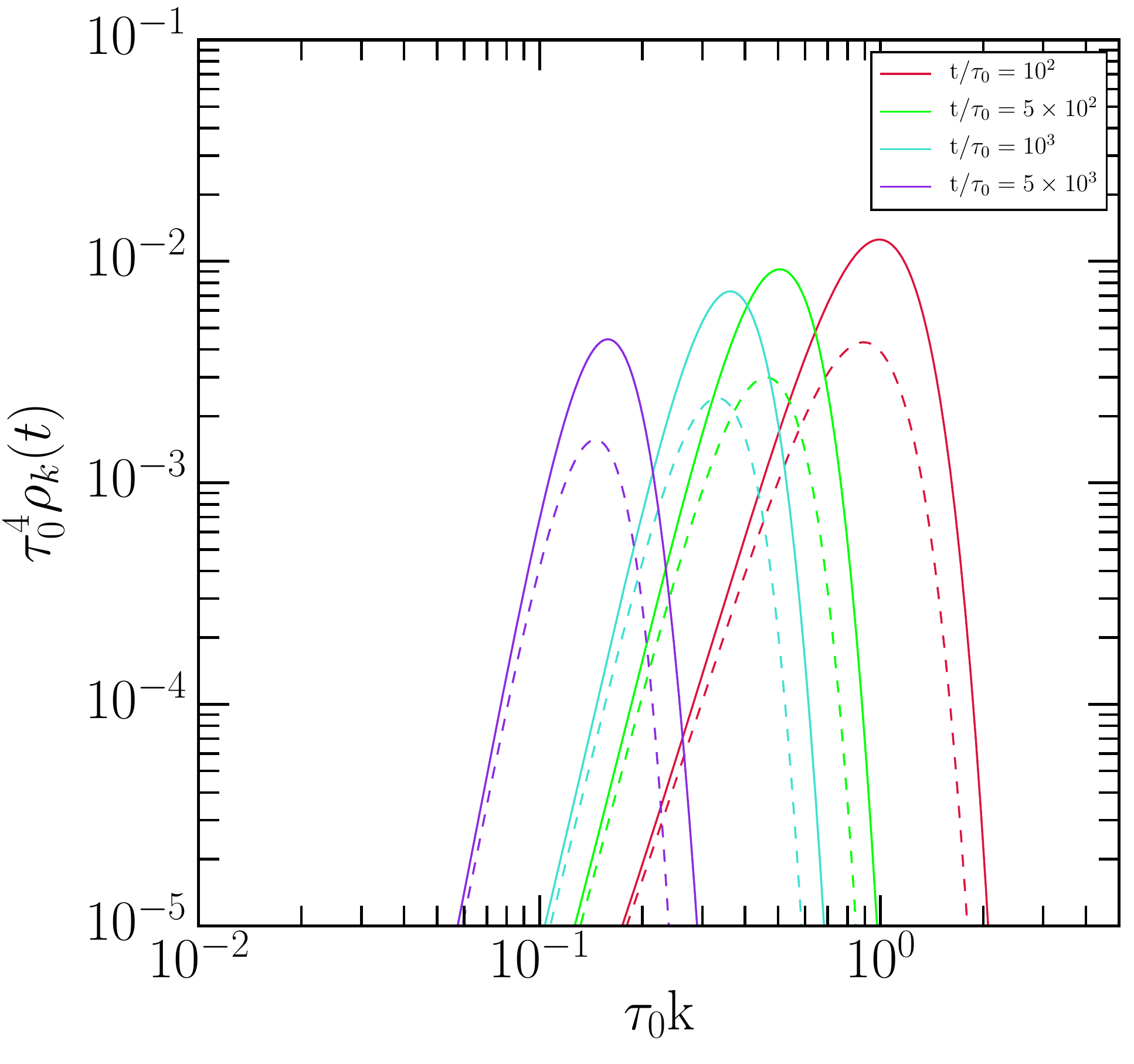}
\caption{\label{pic:4hels} Total magnetic energy density spectrum (solid) and background magnetic energy density spectrum (dashed) at different times, computed from \eqref{pert1} and \eqref{chempomod}, and \eqref{background1}, respectively, with $\mu_5^0=-10^3\tau_0^{-1}$ and $\epsilon=7\times 10^{-3}$, for $\Gamma_{f}\propto t^{-2}\tau_0^{-1}$ and $\Pi_{sr}=0$. Adopted from \cite{novi}}
\end{figure}

\subsection{Strong anomaly regime}
Let us now discuss the specific regime of chiral MHD turbulence, where the anomaly effects are so big initially that the effective chiral current dominates over other terms governing 
the evolution of magnetic fields, $|2 c_{1,2} \mu_{5} \nabla \times \mathbf{B}|\gg |\nabla \times (\mathbf{v} \times \mathbf{B})+ \nabla^{2} \mathbf{B}/\sigma|$. In the first 
approximation, it is easy to see that the magnetic field will exponentially grow in time (see the discussion in the previous chapter \ref{chir}). Assuming moreover that MHD turbulence is magnetically dominated,
$\Gamma \ll 1 $, we can approximate the Navier-Stokes equation taking that the evolution of velocity is dominated by the Lorentz force, $\mathbf{F}_{L}=(\nabla \times \mathbf{B}) \times \mathbf{B}$. 
Since the chiral current is in the direction of magnetic field it gives no contribution to the evolution of velocity under this approximation. The dominant change of the magnetic field
will then be given by the exponential growth in time, and ignoring spatial variations during this evolution, which are negligible in the regime of a dominant anomaly contribution and a spatial-independent chiral potential, 
the velocity is effectively determined by the following equation
\begin{equation}
  \rho \frac{\partial \textbf{v}}{\partial t} \approx (\nabla \times \mathbf{B}(0))\times \mathbf{B}(0) e^{2 Ct},
\end{equation}
where the anomaly governed growth of magnetic fields was approximated as $\mathbf{B}= \mathbf{B}(0)e^{C t}$, and $C$ is a constant. Therefore, in the anomaly dominated regime the velocity will initially 
tend to also grow exponentially in time, assuming that the magnetic field is initially not a force-free field, $(\nabla \times \mathbf{B}(0))\times \mathbf{B}(0) \neq 0$. This corresponds to the assumption that the magnetic field is initially not close to the maximally-helical configuration. 
However, the growth of anomaly dominated fields will tend to organize magnetic fields as maximally-helical, since the assumptions of 
this regime dictates that $\nabla \times \mathbf{B} \approx -2 \sigma c_{1,2} \mu_{R,5} \mathbf{B}$ and this dominant contribution to the Lorentz force will then decrease. However, this growth 
of magnetic field will induce small scale fluctuations which will then make a contribution to the Lorentz force. In the same time, 
as the magnetic and velocity field grow, the term $\mathbf{v} \times \mathbf{B}$ will also become more significant, until finally making 
the initial approximation of exponential field growth invalid. This signifies the transition from the anomaly dominated regime to the regime 
where turbulence and anomaly effects are comparable. An understanding of the subsequent evolution then requires the analysis of complete system of equations 
\eqref{Maxwell} - \eqref{chempomod}. Moreover, the initial exponential growth of magnetic fields is supported by the conversion of energy stored in the 
chiral potential into magnetic energy, and $\mu_{5}$ will eventually be depleted. In this later stage the chiral MHD turbulence will approach 
the limit of the standard MHD turbulence, since due to the conversion into magnetic energy, the chiral asymmetry potential will approach $\mu_{5} \approx 0$. This analysis is in agreement 
with the results of recently reported numerical simulations \cite{jennifer,brandy} discussed in section \ref{numericki}, explaining the observed phases of: small scale growth,
chiraly driven turbulence, and a transition to standard turbulent evolution. The initial parameters used in these studies
correspond to the strong anomaly regime analysed here. The analytical approximations presented above can not however be directly applied
to the cases considered in \cite{brandy, jennifer} since the fields are there initially near to the force-free configuration and the main component determining 
the velocity evolution will be force effects coming from small scale fluctuations. 
\section{Minimal energy configurations}\label{minencon}
In complex non-linear systems, such as the one given by chiral MHD turbulence, it is of special interest to use the global properties 
of evolution given by energy considerations. In this way it is hopefully possible to obtain a better understanding of its universal aspects, which 
is much more difficult to see by using the straightforward computation. This becomes even more pronounced when direct analytical computation
is not possible, as is the case in the problem of chiral MHD turbulence. When dealing with energy considerations a special interest lies in the study of minimal energy 
configurations characterizing some system, since during its evolution there will naturally exist a tendency towards relaxation to such a state. 
This approach is used in plasma physics, in the study of standard MHD turbulence, where the relaxation towards a force-free minimal energy 
state was observed in various settings \cite{plazma3, plazma5, plazma6}. It is therefore of special interest to discuss the minimal energy states of 
chiral MHD turbulence, and to understand configurations which would naturally be approached during its relaxation. We will also show that, as 
well as in the previous discussions presented in this chapter, the chiral magnetic effect leads to some interesting difference in the field 
properties of such configurations, although the resulting minimal-energy field is still force-free. \\ \\
Let us first review the invariant and ideally invariant quantities of chiral MHD turbulence. Since we are interested 
in discussing the relaxation states of chiral MHD turbulence, in this section we assume that there are no source terms active, $\Pi_{sr}=0$, 
which would disturb the tendency of approaching a minimal energy state. In the case of chiral MHD turbulence we have 
three types of energy contributions -- the magnetic energy density:
\begin{equation} 
\rho_{m}=\frac{1}{2V} \int d^{3} r \mathbf{B}^{2}(\mathbf{r},t),
\label{magnetic}
\end{equation}
the kinetic energy density:
\begin{equation}
\rho_{K}=\frac{1}{2V}\int d^{3} r \rho  v^{2},
\label{kinetic}
\end{equation}
and the energy associated with $\mu_{5}$
 \begin{equation} 
\rho_{5}=\frac{\mu_{5}^{2}T^{2}}{6} . 
\label{chirale}
\end{equation}
The total energy will clearly by given by $\rho_{tot}=\rho_{m} + \rho_{K} + \rho_{5}$. The ideal-invariants are given by magnetic and 
cross-helicity, which are conserved in the limit of ideal MHD. Even in the case of resistive MHD, when the characteristic conductivities 
are high, they remain to be important quantities and their suppressed decay, that comes from this property, leads to important effects on the properties of magnetized turbulence, as we will soon discuss more. We repeat their definitions for the convenience of the reader:
\begin{equation}
h_{c}= \frac{1}{V}\int  \mathbf{v} \cdot \mathbf{B} d^{3} r
\end{equation}
\begin{equation}
h=\frac{1}{V}\int_{V} \mathbf{A} \cdot \mathbf{B} d^{3} r
\label{heldef}
\end{equation}
We now derive the expressions for the time change of quantities defined above. We first consider the equation giving the evolution of 
non-chiral energy contributions, that is dissipation of the sum of magnetic and kinetic energy. First we multiply the Navier-Stokes equation \eqref{navsto} with $\mathbf{v}$ and Eq. \eqref{induction} with $\mathbf{B}$, and add them using \eqref{magnetic}, 
\eqref{kinetic} and for simplicity use the Alfv\'{e}n time units \cite{biskamp}. The pressure divergence can then be written as a function of other quantities in the incompressible limit, $\nabla^2 p=-\nabla \cdot \left[ (\mathbf{v} \cdot \nabla)\mathbf{v}- (\mathbf{B} \cdot \nabla)\mathbf{B} \right] $,
and ignoring the surface terms during the integration it follows
\begin{equation}
\frac{d}{dt}(\rho_{m}+\rho_{K})= - \frac{1}{V} \int\left[\frac{(\nabla \times \mathbf{B})^{2}}{\sigma} + \nu \cdot \omega^{2}+ 2c_{1,2} \mu_{5} \mathbf{B} \cdot (\nabla \times \mathbf{B} +\mathbf{J}_{5})\right] d^{3}r,  
\label{maghyen}
\end{equation}
where $\mathbf{\omega}$ is the vorticity defined as $\mathbf{\omega}= \nabla \times \mathbf{B}$, and
we introduced the effective chiral current $\mathbf{J}_{5}=2 \sigma c_{1,2} \mu_{5}  \mathbf{B}$. We can see that the change of magnetohydrodynamic 
energy consists of two parts. The first contribution is given by resistive and viscous processes and it is strictly negative. The second 
contribution to the change of magnetohydrodynamic energy comes from the energy associated with the effective chiral current, and its 
contribution can be both positive and negative. This corresponds to transforming the energy stored in the chiral potential 
into magnetic energy, and to the energy transfer from the magnetic field to the chiral potential in the opposite case. We see that if the chiral 
current is significantly strong its effects can completely suppress the significance of resistive and damping terms and determine the evolution of energy. 
We can also use \eqref{Maxwell} and \eqref{induction}, together with \eqref{heldef} to compute the time change of magnetic helicity
\begin{equation}
\frac{dh}{dt}=-\frac{2}{V \sigma} \int d^3{r}
(\mathbf{\nabla \times \mathbf{B}}+ \mathbf{J}_{5}) \cdot \mathbf{B},
\label{helchange}
\end{equation}
which apart from the standard violation of helicity conservation, which comes as a result of finite conductivity, also contains an additional 
contribution coming from the anomaly effect. 
To obtain the equation for time evolution of the total chiral MHD energy, $\rho_{tot}$, we use \eqref{chirale}, together with \eqref{chempomod}. By virtue
of \eqref{helchange} we note that the change of energy stored in chiral potential cancels the anomaly created change in \eqref{maghyen}. This is indeed necessary 
for the conservation of energy. We obtain: 
\begin{equation}
\frac{d \rho_{tot}}{dt} =  - \frac{1}{V} \int\left[\frac{(\nabla \times \mathbf{B})^{2}}{\sigma} + \nu \cdot \omega^{2}\right] d^{3}r -\frac{T^{2}}{3} \mu_{5}^{2} \Gamma_{f} .
\end{equation}
It can be seen that dissipation of total energy happens due to the effects of viscosity, resistivity and chirality flipping processes (which 
change only the energy of the chiral sector, acting in analogy with macroscopic dissipative coefficients, $\nu$ and $\eta$), which are all 
strictly negative. If there are no flipping processes, the total chiral MHD energy has the same evolution as the sum of kinetic and magnetic 
energy in the standard MHD turbulence -- meaning that all changes in energy can happen only as a redistribution of energy between the magnetic 
and chiral sector. 
We can also derive the equation for cross-helicity evolution
\begin{equation}
\frac{dh_{c}}{dt}= -\frac{1}{V} \int \left[(\nu + \frac{1}{\sigma})(\nabla \times \mathbf{B}) \cdot \mathbf{\omega} + 2c_{1,2} \mu_{5} \cdot (\nabla \times\mathbf{B} +\mathbf{J}_{5})\cdot \mathbf{v}\right]d^{3}r ,
\end{equation}
From these equations we can argue that the evolution towards minimal energy state will not proceed with the same dynamic for the total energy 
on the one hand, and magnetic and cross-helicity on the other. The main difference lies in the fact that the energy change needs to be strictly negative,
while the change in magnetic and cross-helicity can be both positive and negative, since $(\nabla \times \mathbf{B}) \cdot \mathbf{B}$ and 
$(\nabla \times \mathbf{B} \cdot \mathbf{\omega}$ can take both signs. Like in the non-chiral case those quantities are described by dissipative 
terms which involve different order of spatial derivatives (like $\nabla \times \mathbf{B}$ for helicity and $(\nabla \times \mathbf{B})^{2}$ for magnetic energy).
From the approximate scalling following from \eqref{helchange} $h \sim (1/l) \int \rho_{tot} d^{3}r$, with $l$ being a characteristing length, we can see that the helicity is in general 
expected to decay slower. 
Moreover, while both helicities can additionaly increase in time due to the anomaly contribution, the total energy needs to strictly
decrease with time. During the transition towards a minimal energy state, where the temporary growth of both helicities needs to stop 
due to the depletion of initial $\mu_{5}$, we can therefore treat the total energy as decaying much faster than magnetic and cross-helicity, 
and thus take the minimal energy state as characterized by (approximately) constant magnetic or cross-helicity. This type of selective decay argument 
also plays an important role in the study of non-chiral MHD. \\ \\
Following our conclusions we note that there will be two possible relaxation processes - the relaxation towards a minimal energy state characterized by a
constant magnetic helicity or the minimal energy state characterized by a constant cross-helicity. Transition towards a minimal energy state with constant helicity of course
means that at the later stages of chiral turbulence the chiral anomaly contribution will become negligible, since it is directly related to the change in helicity. This type of evolution, ending
in a phase where the magnetic helicity saturates, was also recently reported in numerical studies of chiral MHD turbulence \cite{jennifer, brandy}. We will first consider this type of selective decay process. 
Requiring that the minimal energy configuration is established with respect to the constraint of constant helicity we need to perform the variational principle of the following expression
\begin{equation}
J= \frac{1}{V}\int \left[\frac{1}{2}  (v^{2} +(\nabla \times \mathbf{A})^{2})  +  \frac{\mu_{5}^{2}T^{2}}{6}\right] d^{3} r -  \frac{\alpha}{2V} \int \mathbf{A} \cdot (\nabla \times \mathbf{A}) d^{3}r,
\label{variational}
\end{equation}
where $\mathbf{A}$ is the vector potential, and $\alpha$ is a Lagrangian multiplier. Varying \eqref{chirale} with respect to velocity, while requiring the extremal configuration,
simply yields $\mathbf{v}=0$, which shows that velocity will dissipate approaching the zero velocity case, as a minimal energy state. Ignoring the rates of flipping reactions, 
$\Gamma_{f}=0$, and using \eqref{chempomod} assuming a constant temperature $T$ \footnote{We note that this assumption does not present any limitation for the application
of this approach to the case of the early Universe -- where temperature clearly needs to change, since introducing the description in terms of conformal quantities, temperature does not appear explicitly in  \eqref{chirale}, and the procedure is mathematically the same, after replacing the ordinary time with conformal time and variables with their conformal counterparts  (for instance see \cite{pa}) }, while varying \eqref{chirale} with respect to vector potential, using the identity 
$\mathbf{A} \cdot (\nabla \times \delta \mathbf{A})=\delta \mathbf{A} \cdot(\nabla \times \mathbf{A}) - \nabla \cdot (\mathbf{A} \times \delta \mathbf{A})$, we obtain
\begin{equation}
\nabla \times \mathbf{B} + \left[\frac{2c_{3,4}}{3}(\frac{c_{3,4}}{T^{2}}h +C)- \alpha\right] \mathbf{B}=0, 
\label{varhel}
\end{equation}
where
\begin{equation}
 C=\mu_{5}^{in} - \frac{1}{T^{2}}c_{3,4} h^{in},
 \label{constant}
\end{equation}
and $\mu_{5}^{in}$, $h^{in}$ are the initial values of the chiral asymmetry potential and magnetic helicity. Since the relaxation state is achieved with respect to the constraint of
constant magnetic helicity, the helicity value appearing in \eqref{varhel} is the value towards which helicity saturates during this selective decay. Therefore we see that during this relaxation process the system approaches
a force-free field configuration, which is a solution of the equation \eqref{varhel}, and which is determined by the parameters describing the chiral anomaly effect. In the case of standard MHD turbulence, i.e. for $c_{3,4}=0$ we obtain the condition 
$\nabla \times \mathbf{B} - \alpha \mathbf{B}=0$, that is the curl of magnetic field can only vanish if the field itself vanishes for $\alpha \neq 0$, while for the chiral case
the curl of $\mathbf{B}$ can also vanish for non-vanishing $\mathbf{B}$ if $(2c_{3,4}/3)(c_{3,4}/T^{2}h +C)= \alpha$. Also, in the non-chiral case the orientation of curl of $\mathbf{B}$ is either parallel (for $\alpha >0$)
or anti-parallel ($\alpha <0$) for all field configurations, while in the chiral case both orientations are possible for single $\alpha$, depending on the values of helicity and
and integration constant \eqref{constant}. The constant $\alpha$ can be expressed in terms of the magnetic energy characterizing the minimal configuration, $\rho_{m}^{min}$, magnetic helicity, 
and the remaining chiral parameters -- if we take a dot product of \eqref{varhel} with $\mathbf{A}$, integrate both sides over volume, and use definitions of magnetic and helicity energy densities. 
We then obtain the following condition
\begin{equation}
\alpha = \frac{2\rho_{min}}{h}- \frac{2c_{3,4}}{3}(\frac{c_{3,4}}{T^{2}}h +C)
\end{equation}
\\
We can now discuss the main properties of the chiral MHD solutions in the later phase of their evolution, 
assuming that the velocity and magnetic field have already evolved significantly close to their minimal-energy configurations. As the velocity has significantly decreased, 
approaching to $\mathbf{v}=0$, the second-order quantities in velocity appearing in Navier-Stokes equation can be ignored, and the field configuration will then also be close
to the vanishing force state described above. Then the evolution of velocity will be determined by the following equation
\begin{equation}
\frac{\partial \textbf{v}}{ \partial t}  - \nu \nabla^{2}\textbf{v}= 0,
 \label{san}
\end{equation}
and therefore in this approximation decoupled from the evolution of magnetic fields. 
Using the Fourier decomposition
\begin{equation}
\mathbf{v}(\mathbf{r},t)=\int \frac{d^{3} q}{(2 \pi)^{3}}e^{i\mathbf{q} \cdot{\mathbf{r}}}\mathbf{v}(\mathbf{q},t)  
\end{equation}
we obtain the solutions for the velocity modes
\begin{equation}
\mathbf{v}_{\mathbf{q}}(t)=\mathbf{v_{q}}(0) e^{-\nu q^2 t},
\label{veli}
\end{equation}
were $\mathbf{v_{q}}(0)$ is an integration constants. 
Using \eqref{integral} we also obtain the velocity contribution of the following form
\begin{equation}
\frac{i}{(2\pi)^{3/2}}\int d^3q e^{-\nu q^2 t} \left[k_{a}v^{i}_{\mathbf{k}-\mathbf{q}}(0)<B^{a}_{\mathbf{q}}B^{i}_{\mathbf{k}}> - 
k_{a}v^{a}_{\mathbf{k}-\mathbf{q}}(0)<B^{i}_{\mathbf{q}}B^{i}_{\mathbf{k}}>\right]
\end{equation}
Focusing our attention on the evolution of statistically homogeneous and isotropic magnetic fields we use the condition 
\begin{equation}
<B_{i}(\mathbf{k},t) B_{j}(\mathbf{q},t)>=\frac{(2 \pi)^3}{2} \delta(\mathbf{k}+\mathbf{q})[(\delta_{ij}-\frac{k_{i} k_{j}}{k^2})S(k,t) + i \epsilon_{ijk}\frac{k_{k}}{k}A(k,t)], 
\label{correlator}
\end{equation}
where $S$ and $A$ denote the symmetric and anti-symmetric parts of the correlator. Applying this condition and using the assumption of incompressibility, it follows that the contribution of the last term vanishes, and we are left only with the velocity-free chiral MHD equations:
 \begin{equation} 
\partial_t  \rho_k= - \frac{2 k^2}{\sigma} \rho_k - 4 c_{1,2} \mu_5 k^2 h_k \, ,
\label{rhok}
\end{equation}  
\begin{equation} 
 \partial_t  h_k= - \frac{2 k^2}{\sigma} h_k - 16 c_{1,2} \mu_5 \rho_k  \, .
 \label{hok}
 \end{equation}
We have therefore demonstrated that if the evolution of chiral MHD turbulence proceeds towards a minimal energy configuration under the constraint of constant helicity, during the later stages
of this evolution, when the magnetic field approaches a force-free configuration, the evolution of chiral magnetic fields and turbulence will tend to effectively 
decouple. In general, we thus expect that the strongest interplay between the chiral effect and turbulence will happen in the regimes which are far away from this minimal energy state. \\ \\
We now focus on the second type of relaxation process, where the transition towards minimal energy state is determined by the constraint of constant cross-helicity, again justifying this
assumption with the already discussed difference of dissipation rates of energy and cross-helicity. Which of the two processes will actually happen in a given system seems to be determined by
the specific configuration of initial conditions -- if the system is strongly helical on the one hand, or on the other -- if there is a significant initial alignment between the 
velocity and the magnetic field \cite{biskamp,tel}. In the case of cross-helicity dictated selective decay we need to vary the following expression
\begin{equation}
J{'}= \frac{1}{V}\int \left[\frac{1}{2}  (v^{2} +(\nabla \times \mathbf{A})^{2}))  +  \frac{\mu_{5}^{2}T^{2}}{6}\right] d^{3} r -  \frac{\alpha'}{V} \int \mathbf{v} \cdot (\nabla \times \mathbf{A}) d^{3}r.
\label{cross}
\end{equation}
Varying \eqref{cross} with respect to velocity while demanding $\delta J'=0$ we get
\begin{equation}
\mathbf{v}= \alpha' \mathbf{B},  
\label{jedn1}
\end{equation}
while varying with respect to the vector potential leads to
\begin{equation}
\nabla \times \mathbf{B} + \frac{2}{3}c_{3,4}(\frac{c_{3,4}}{T^{2}}h + C)\mathbf{B}= \alpha' (\nabla \times \mathbf{v}).  
\label{jedn2}
\end{equation}
In the special case when $\alpha'=\pm 1$ we see that if $\mathbf{B}\neq 0$ then equations \eqref{jedn1} - \eqref{jedn2} have a solution only if the chiral effect is not present, that is if 
$c_{3,4}=0$, so this state is not achievable in chiral MHD turbulence. On the other hand, when $|\alpha'|\neq 1$ then the field obviously satisfies 
\begin{equation}
\nabla \times \mathbf{B}= -\frac{2}{3} \frac{c_{3,4}(\frac{c_{3,4}}{T^{2}}h + C)\mathbf{B}}{1-\alpha'^{2}} 
\end{equation}
Multiplying this equation with $\mathbf{A}$ and integrating over volume, we can express the coefficient $\alpha'$ in terms of the magnetic energy density characterizing the minimal energy
configuration, $\rho_{min}$ and helicity, $h$, as
\begin{equation}
\alpha'=\pm \sqrt{1+\frac{c_{3,4}}{3 \rho_{min}} {(\frac{c_{3,4}}{T^{2}}h + C)h}}. 
\end{equation}
We can see the physical implication of this result better if we write the Navier-Stokes equation \eqref{navsto} in the Els\"{a}sser variables, $\mathbf{z}^{\pm}=\mathbf{v} \pm \mathbf{B}$
\begin{equation}
\frac{\partial z^{\pm}}{\partial t}+ z^{\mp}\cdot \nabla z^{\pm}= - \nabla p + \frac{1}{2}(\nu + \eta) \nabla^{2} z^{\pm} + \frac{1}{2}(\nu - \eta) \nabla^{2} z^{\mp}
\end{equation}
We see that in the non-chiral MHD case, leading to $\mathbf{v}= \pm \mathbf{B}$, the non-linear interaction term is necessarily zero, since it couples the Els\"{a}sser variables of different
signs -- and therefore the evolution of turbulence is guided only by dissipation processes. However, in the case of chiral MHD turbulence the interaction term will in general 
still be present. Therefore, the introduction of chiral anomaly effects leads to the presence of non-linear interactions even in the minimal energy configuration. 
\section{Conclusions}
Some first attempts in the study of the complex phenomenon of chiral MHD turbulence -- which leads to a non-trivial interdependence between the evolution of magnetic fields, quantum 
anomalous effects and particle processes -- have been presented in this chapter. We have here demonstrated how the anomalous modified
MHD turbulence can lead to a significantly different time evolution of magnetic fields. Due to the high mathematical complexity of this problem, numerical simulations are an important 
tool in its study. Unfortunately, to our knowledge, only two reports on the results of numerical study of chiral MHD turbulence were presented, which are based on a similar set of assumptions 
and proceed from the same type of initial conditions, causing a fast anomaly driven magnetic field growth until the saturation. Therefore, there is a need for a larger number of numerical 
studies, investigating different regimes and focusing more on the complicated dynamics of the regime in which both anomalous and turbulent effects are of comparable importance. 
We have further discussed how the anomalous turbulence can support the existence of inverse cascades and also lead to the creation of maximally helical fields from  magnetic fields 
with initially vanishing helicity. Using qualitative and scaling arguments, the modifications of the evolution of magnetic energy were first discussed, and then supplemented with 
some concrete solutions obtained in the specific regimes of chiral MHD turbulence. First studying the case when anomaly  effects  are  small  compared  to
the standard MHD terms, we argued that the decrease of magnetic energy with time will proceed slower, $\rho_{m} \sim t^{-2/3}$, with respect to an initially non-helical and non-chiral
turbulence. In the case when the anomaly effects stay significant after helicity was induced, a Kolmogorov-like reasoning was used to obtain the scaling of magnetic energy 
and correlation length with time. It was shown that if the chiral potential can be approximated as $\mu_{5} \approx t^{n}$, we then have $\rho_{m} \sim t^{2(n-1)/3}$ and 
$\xi_{m} \sim t^{(n+2)/3}$. We then focused on the analysis of the weak anomaly regime. In this regime the anomaly contribution to magnetic energy and helicity was treated as a small
perturbation to the standard MHD background. The solutions thus obtained  demonstrate how chiral effects support the inverse cascade and the growth of the correlation length in 
the weak anomaly regime. The opposite regime, in which the anomaly effects dominate over the turbulence terms was also briefly discussed and it was argued how it can be approximately 
explained with an exponential growth of magnetic field, caused by chiral magnetic effects, which then induces a further change of the velocity, causing a backreaction 
of the velocity field, and finally leading to depletion of $\mu_{5}$ and the saturation of the magnetic helicity. Finally, chiral MHD turbulence was discussed by studying the relaxation towards a minimal energy state. We first discussed the dissipation of ideal MHD invariants and showed how the magnetic helicity and cross-helicity can be taken as approximately conserved even in the resistive 
MHD turbulence, due to their slower decay with respect to magnetic energy. It was also argued that this effect can become even more pronounced in the case of chiral turbulence. Using this approximation, 
we derived minimal energy configurations using a variational procedure. As in the standard MHD case a minimal energy configuration is given by a force-free field. In general, 
there will be two possible relaxation states -- with respect to approximately conserved magnetic helicity and approximately conserved
cross-helicity.  The precise form of the field corresponding to such minimal energy configurations is given by the coefficients 
related to chiral anomaly effects, and it was shown that non-linear interactions can still be present even in the minimal energy configuration due to the anomalous effects.

\chapter{Summary and outlook}

In this work we have studied quantum anomalous effects on the description  
of
high energy electrodynamics, reporting our recent results, and reviewing  
some
new advances in this topic. These effects should become important on  
temperatures
characterizing the electroweak scale, and could therefore influence
processes happening in the early Universe, and potentially in astrophysical
objects such as proto-neutron stars. For charged massless fermions the  
quantum
effect of the chiral anomaly leads to a non-vanishing divergence of the  
chiral current,
related to the number of left-handed and right-handed chiral particles. In
the exterior magnetic field, characterized by a non-vanishing magnetic  
helicity,
the difference in number of left-handed and right-handed particles will not
be conserved, and this will lead to an effective chiral current in the  
direction
of magnetic field. When this current is introduced into the Maxwell  
equations,
describing the macroscopic dynamics of the electromagnetic field, it will  
lead to
new effects and important modifications of the field evolution. It is  
expected
that such anomalous effects will also play a role after the electroweak  
crossover,
when leptons become massive, but temperatures are still high enough so  
that their
masses can be effectively ignored. In such cases, the effects coming from
finite masses can be added perturbatively, in terms of the reactions rates  
causing
the changes in the chirality of the considered particles.
\\ \\
After discussing the main assumptions of the magnetohydrodynamic (MHD)  
approximation
for the evolution of electrodynamic fields, and justifying its application
to the case of the early Universe and similar systems, we have analysed
the main properties of MHD equations modified by the chiral magnetic  
effect,
and coupled to the particle processes via the flipping reaction rates. It  
was then
discussed how these equations can lead to solutions exhibiting such  
properties as
the creation of helicity from initially non-helical fields, the creation  
of the chiral asymmetry by helical magnetic fields, exponential magnetic  
field amplification
and an inverse transfer of magnetic energy.
This system of chiral MHD equations was then considered in the case of the  
evolution of magnetic fields around the electroweak crossover, while  
ignoring velocity effects.
It was demonstrated that even a transition that is not of the first order  
has direct consequences on the evolution of the asymmetry between left and  
right-handed leptons. If the existence of magnetic fields before the  
electroweak transition is assumed, as well
as some initial chiral asymmetry, the results of our computations suggest  
that
the chiral asymmetry typically first grows in time, while approaching the  
crossover,
and then undergoes a strong decrease around the symmetry breaking, caused
by the activation of flipping rates connected with the newly acquired  
leptonic
masses. Afterwards, the asymmetry gets subsequently damped reaching
lower temperatures in the phase of the broken electroweak symmetry. In this
setting of an higher order electroweak transition, we have argued why it  
is not
possible to obtain any significant amplification of magnetic fields for  
realistic
values of the chiral asymmetry.
\\ \\
When velocity effects are considered in systems characterized by
large Reynolds numbers, the electrodynamics of the plasma in the  
magnetohydrodynamic
approximation leads to the theory of MHD turbulence. We
have argued that in the conditions pertaining to temperatures comparable
to the electroweak energy scale, this description needs to be generalized  
to a theory of chiral MHD turbulence. The systematic study of this phenomenon
started only recently. After reviewing some recent numerical findings, we  
have presented a discussion showing that chiral effects can lead to the  
creation of maximally-helical fields from initially non-helical fields, support the  
inverse energy transfer, enhance the growth of the correlation length and suppress the
decay of magnetic energy. Using a qualitative reasoning based on the  
considerations of energy transfer and approximate scalings between MHD quantities, a
discussion regarding the change in the time evolution of the MHD  
quantities were presented. The specific regimes of weak and strong anomaly were then
presented, obtaining specific solutions and supporting the previous  
qualitative discussion. At the end, we focused on the relaxation states of chiral MHD
turbulence, showing that - like in the case of non-chiral MHD turbulence -
the minimal energy state is given by a force-free field configuration. Two  
possible relaxation states - with respect to approximately conserved magnetic  
helicity and approximately conserved cross-helicity - were considered, and it was  
shown that the details of the field configurations are now determined by the  
coeffcients related to chiral anomaly effects. We have furthermore argued that during  
the approach towards these minimal-energy states the evolution of magnetic  
fields and chiral asymmetry will effectively decouple. These results seem to be  
consistent with the recently reported findings of numerical simulation modeling
chiral MHD turbulence. The work on understanding the consequences of quantum anomalous effects
on elecromagnetic fields is far from being completed. Further developments  
in this topic will require a synthesis of different results and conclusions  
derived by
using analytical, numerical and observational methods, in such different  
research felds as the early Universe, neutron stars and quark-gluon plasma physics.  
A particularly demanding task is given by the need for a proper  
understanding of
chiral MHD turbulence, where only some initial steps have been made so far.






\chapter*{Acknowledgments}
I first need to thank my supervisor, Prof. Günter Sigl, for taking me as a PhD student and thus giving me the opportunity to fully 
devote myself to research and to the study of physics. I am also thankful to him for introducing me to the topics presented in this work and 
without him this work would not be possible. I would also like to thank dr. Thomas Konstandin for being my co-supervisor and for interesting discussions 
which helped us preparing our first article on the chiral anomaly. My thanks go to all the members of the theoretical
Astroparticle group of the II. Institute for Theoretical Physics that assisted me with various advice and help when needed. 
A special thanks in this respect needs to go to the old man among postdocs and PhD students in our group, Andrey Saveliev, always ready 
to assist young lost newcomers in private and professional matters. \\ \\
The results presented in this thesis would not be here without my collaborator and comrade, Natacha, without her presence and assistance (and the proof reading of this thesis is just a small 
part of it!), without our talks, investigations
and ideas born; but to further and properly express the joy and beauty of our trips to various spheres of natural philosophy and metaphysics is simply not possible. \\ \\
I'm thankful to Peter Niksa for many rich discussions on turbulence, for reading my thesis, giving many suggestions and hunting for the errors with a fanatic determination. \\ \\
I thank Claudia for her constant help with the practical matters related to survival in this unusual country, for making the time spent 
around DESY much nicer, and for kindly assisting me with writing the abstract in German. \\ \\
My comrade Andrej, the soldier who shared the same bunker during this war, had very difficult tasks of living with me, and -- as if that was not enough -- many new and old ideas, starting from physics 
and ending in yet uncharted territories, were permanently tested and reshaped, crafted and thrown to the dust in the discussions on everyday basis, making a tremendous influence 
on the course of this phase of my journey. \\ \\
My stay in Hamburg was made much nicer thanks to my gentle companions -- I'm thankful to the proud river Elbe and many animals, plants 
and stones of Blankenesse for all the common moments. \\ \\
It is a common practice to express the gratitude to your parents in the acknowledgments. I'm not doing it because it is common, nor 
for respecting the form and expectations (I will always be against such principles), but out of the fact that I had the luck to be raised and supported in such conditions where I had 
the full freedom to develop as I want and follow my own goals and paths, not connected with any practical and useful considerations. My gratitude can not be expressed with words. \\ \\
Thinking that I'm the only child of my parents, it was so incredible to find out I have a sister and to actually meet her for the first time 
in Hamburg -- and while I got lost because of all of that, she also joyfully helped me to liberate some suppressed parts of myself and to bring them to the light, while walking through 
the forest. \\ \\
I thank Andrea for being impatient, for casting her radiance all up to Hamburg, for binding my thoughts, for being an axis of cycles of arrivals and departures
and for everything which can not be mentioned. \\ \\
I bow my head in gratitude in front of my dear teachers, that introduced me to the principles of physics and investigations of Nature, nourishing my love for these questions on various 
stages of my development and rewarding me with a special attention -- Jura Štranjger during the elementary school (with so much humor), Tanja Mihaljević during the high school days (with so much charm), and Nevenko Bilić and Dubravko Horvat
that guided me in a relaxed fashion during my first steps in independent research in theoretical physics. \\ \\
I'm thankful to all the great immortal masters of knowledge, the masters I could not meet 
directly, but that were teaching and guiding me through their books and works. I'm especially indebted to Nikola Tesla, who stimulated my fascination for the
electromagnetic phenomena and who will in this matters always remain to be my greatest teacher. \\ \\
Finishing this work I focus my mind on the goal of physics and the final destination of all searches, the infinite and unlimited Nature, which from itself manifests as matter in a myriad of forms, undergoing the constant 
change and development without end, establishing relations between all of its infinite parts, and brings them together to unity.

\end{document}